\documentclass[prl,aps,superscriptaddress,footinbib,twocolumn]{revtex4-2}
\usepackage[latin9]{inputenc}
\usepackage{color}
\usepackage{bbm}
\usepackage{amsmath}
\usepackage{amssymb}
\usepackage{stmaryrd}
\usepackage{threeparttable}
\usepackage[ruled]{algorithm2e}

\usepackage{multirow} 
\usepackage{makecell}
\usepackage{color}
\usepackage{listings}
\usepackage[unicode=true,
 bookmarks=true,bookmarksnumbered=false,bookmarksopen=false,
 breaklinks=false,pdfborder={0 0 1},backref=false,colorlinks=true]
 {hyperref}
\hypersetup{linkcolor=magenta, urlcolor=blue, citecolor=blue, pdfstartview={FitH}, hyperfootnotes=false, unicode=true}

\makeatletter

\usepackage{ntheorem}
\newtheorem*{proof}{Proof}

\newtheorem{proposition}{Proposition}
\newtheorem{lemma}{Lemma}
\newtheorem{corollary}{Corollary}
\newtheorem{definition}{Definition}
\newtheorem{assumption}{Assumption}

\usepackage{amsfonts}\usepackage{tabularx}\usepackage{dcolumn}\usepackage{bm}\usepackage{graphicx}\usepackage{epstopdf}
\usepackage{times}
\hypersetup{urlcolor=blue}

\def\ket#1{\left|#1\right\rangle}
\def\bra#1{\left\langle#1\right|}

\makeatother

\begin{document}
\title{Factoring integers with sublinear resources on a superconducting quantum processor}

\author{Bao Yan}
 \thanks{These authors contributed equally to this work.}
\affiliation{State Key Laboratory of Mathematical Engineering and Advanced Computing, Zhengzhou 450001, China}
\affiliation{State Key Laboratory of Low-Dimensional Quantum Physics and Department of Physics, Tsinghua University, Beijing 100084, China}

\author{Ziqi Tan}
 \thanks{These authors contributed equally to this work.}
\affiliation{School of Physics, ZJU-Hangzhou Global Scientific and Technological Innovation Center, Interdisciplinary Center for Quantum Information, and Zhejiang Province Key Laboratory of
Quantum Technology and Device, Zhejiang University, Hangzhou 310000, China}

\author{Shijie Wei}
 \thanks{These authors contributed equally to this work.}
\affiliation{Beijing Academy of Quantum Information Sciences, Beijing 100193, China}

\author{Haocong Jiang}
\affiliation{Institute of Information Technology, Information Engineering University, Zhengzhou 450001, China}
\author{Weilong Wang}
\affiliation{State Key Laboratory of Mathematical Engineering and Advanced Computing, Zhengzhou 450001, China}
\author{Hong Wang}
\affiliation{State Key Laboratory of Mathematical Engineering and Advanced Computing, Zhengzhou 450001, China}
\author{Lan Luo}
\affiliation{State Key Laboratory of Mathematical Engineering and Advanced Computing, Zhengzhou 450001, China}
\author{Qianheng Duan}
\affiliation{State Key Laboratory of Mathematical Engineering and Advanced Computing, Zhengzhou 450001, China}
\author{Yiting Liu}
\affiliation{State Key Laboratory of Mathematical Engineering and Advanced Computing, Zhengzhou 450001, China}
\author{Wenhao Shi}
\affiliation{State Key Laboratory of Mathematical Engineering and Advanced Computing, Zhengzhou 450001, China}
\author{Yangyang Fei}
\affiliation{State Key Laboratory of Mathematical Engineering and Advanced Computing, Zhengzhou 450001, China}
\author{Xiangdong Meng}
\affiliation{State Key Laboratory of Mathematical Engineering and Advanced Computing, Zhengzhou 450001, China}
\author{Yu Han}
\affiliation{State Key Laboratory of Mathematical Engineering and Advanced Computing, Zhengzhou 450001, China}
\author{Zheng Shan}
\affiliation{State Key Laboratory of Mathematical Engineering and Advanced Computing, Zhengzhou 450001, China}
\author{Jiachen Chen}
\affiliation{School of Physics, ZJU-Hangzhou Global Scientific and Technological Innovation Center, Interdisciplinary Center for Quantum Information, and Zhejiang Province Key Laboratory of
Quantum Technology and Device, Zhejiang University, Hangzhou 310000, China}
\author{Xuhao Zhu}
\affiliation{School of Physics, ZJU-Hangzhou Global Scientific and Technological Innovation Center, Interdisciplinary Center for Quantum Information, and Zhejiang Province Key Laboratory of
Quantum Technology and Device, Zhejiang University, Hangzhou 310000, China}
\author{Chuanyu Zhang}
\affiliation{School of Physics, ZJU-Hangzhou Global Scientific and Technological Innovation Center, Interdisciplinary Center for Quantum Information, and Zhejiang Province Key Laboratory of
Quantum Technology and Device, Zhejiang University, Hangzhou 310000, China}
\author{Feitong Jin}
\affiliation{School of Physics, ZJU-Hangzhou Global Scientific and Technological Innovation Center, Interdisciplinary Center for Quantum Information, and Zhejiang Province Key Laboratory of
Quantum Technology and Device, Zhejiang University, Hangzhou 310000, China}
\author{Hekang Li}
\affiliation{School of Physics, ZJU-Hangzhou Global Scientific and Technological Innovation Center, Interdisciplinary Center for Quantum Information, and Zhejiang Province Key Laboratory of
Quantum Technology and Device, Zhejiang University, Hangzhou 310000, China}
\author{Chao Song}
\affiliation{School of Physics, ZJU-Hangzhou Global Scientific and Technological Innovation Center, Interdisciplinary Center for Quantum Information, and Zhejiang Province Key Laboratory of
Quantum Technology and Device, Zhejiang University, Hangzhou 310000, China}
\author{Zhen Wang}
 \email{2010wangzhen@zju.edu.cn}
\affiliation{School of Physics, ZJU-Hangzhou Global Scientific and Technological Innovation Center, Interdisciplinary Center for Quantum Information, and Zhejiang Province Key Laboratory of
Quantum Technology and Device, Zhejiang University, Hangzhou 310000, China}
\author{Zhi Ma}
\email{ma\_zhi@163.com}
\affiliation{State Key Laboratory of Mathematical Engineering and Advanced Computing, Zhengzhou 450001, China}
\author{H. Wang}
\affiliation{School of Physics, ZJU-Hangzhou Global Scientific and Technological Innovation Center, Interdisciplinary Center for Quantum Information, and Zhejiang Province Key Laboratory of
Quantum Technology and Device, Zhejiang University, Hangzhou 310000, China}
\author{Gui-Lu Long}
\email{gllong@tsinghua.edu.cn}
\affiliation{State Key Laboratory of Low-Dimensional Quantum Physics and Department of Physics, Tsinghua University, Beijing 100084, China}
\affiliation{Beijing Academy of Quantum Information Sciences,  Beijing 100193, China}
\affiliation{ Beijing National Research Center for Information Science and Technology and School of Information Tsinghua University, Beijing 100084, China}
\affiliation{Frontier Science Center for Quantum Information, Beijing 100084, China}

\begin{abstract}
\textbf{Shor's algorithm has seriously challenged information security based on public key cryptosystems. However, to break the widely used RSA-2048 scheme, one needs millions of physical qubits, which is far beyond current technical capabilities. Here, we report a universal quantum algorithm for integer factorization by combining the classical lattice reduction with a quantum approximate optimization algorithm (QAOA). The number of qubits required is $O(\text{log} N/\text{loglog} N)$, which is sublinear in the bit length of the integer $N$, making it the most qubit-saving factorization algorithm to date. We demonstrate the algorithm experimentally by factoring integers up to 48 bits with 10 superconducting qubits, the largest integer factored on a quantum device. We estimate that a quantum circuit with 372 physical qubits and a depth of thousands is necessary to challenge RSA-2048 using our algorithm. Our study shows great promise in expediting the application of current noisy quantum computers, and paves the way to factor large integers of realistic cryptographic significance.}
\end{abstract}

\maketitle
Quantum computing has entered the era of noisy intermediate scale quantum (NISQ)~\cite{preskill2018quantum,arute2019quantum}. A milestone in the NISQ era is to prove that NISQ devices can surpass classical computers in problems with practical significance, that is, to achieve practical quantum advantage. Low-resource algorithms, which harness only limited available qubits and circuit depths to perform classically challenging tasks, are of great significance. Variational quantum algorithms, adopting a ``classical+quantum" hybrid computing framework, hold great promise for a meaningful quantum advantage in the NISQ era~\cite{cerezo2021variational,peruzzo2014variational,farhi2014quantum,Wang2022}. One  representative is the quantum approximate optimization algorithm (QAOA)~\cite{farhi2014quantum}, which was proposed to solve eigenvalue problems, and has subsequently been widely used in various fields such as chemical simulation~\cite{mcardle2020quantum,wei2020full}, machine learning~\cite{biamonte2017quantum}, and engineering applications~\cite{wang2018quantum,harrigan2021quantum}.

Integer factorization has been one of the most important foundations of modern information security~\cite{rivest2019method}. The exponential speedup of integer factorization by  Shor's algorithm~\cite{shor1994algorithms} is a great manifestation of the superiority of quantum computing. However, running  Shor's algorithm on a fault-tolerant quantum computer is quite resource-intensive~\cite{gidney2021factor,gouzien2021factoring}. Up to now, the largest integer factorized by Shor's algorithm in current quantum systems is 21~\cite{vandersypen2001experimental,monz2016realization,martin2012experimental}. Alternatively, integer factorization can be transformed into an optimization problem, which can be solved by adiabatic quantum computation (AQC)~\cite{farhi2001quantum,schaller2010role,borders2019integer,yan2021adiabatic} or QAOA~\cite{anschuetz2019variational}. Larger numbers have been factored using these approaches, in various physical systems~\cite{xu2017experimental,jiang2018quantum,li2017high,karamlou2021analyzing}. The maximum integers factorized are 291311~(19-bit) in NMR system~\cite{li2017high}, 249919~(18-bit) in D-Wave quantum annealer~\cite{jiang2018quantum},  1099551473989~(41-bit) in superconducting  device~\cite{karamlou2021analyzing}. However, it should be noted that some of the factored integers have been carefully selected with special structures~\cite{mosca2022factoring}, thus the largest integer factored by a general method in a real physical system by now is 249919~(18-bit). 

\begin{figure*}
\centering
\includegraphics[width=0.98\textwidth]{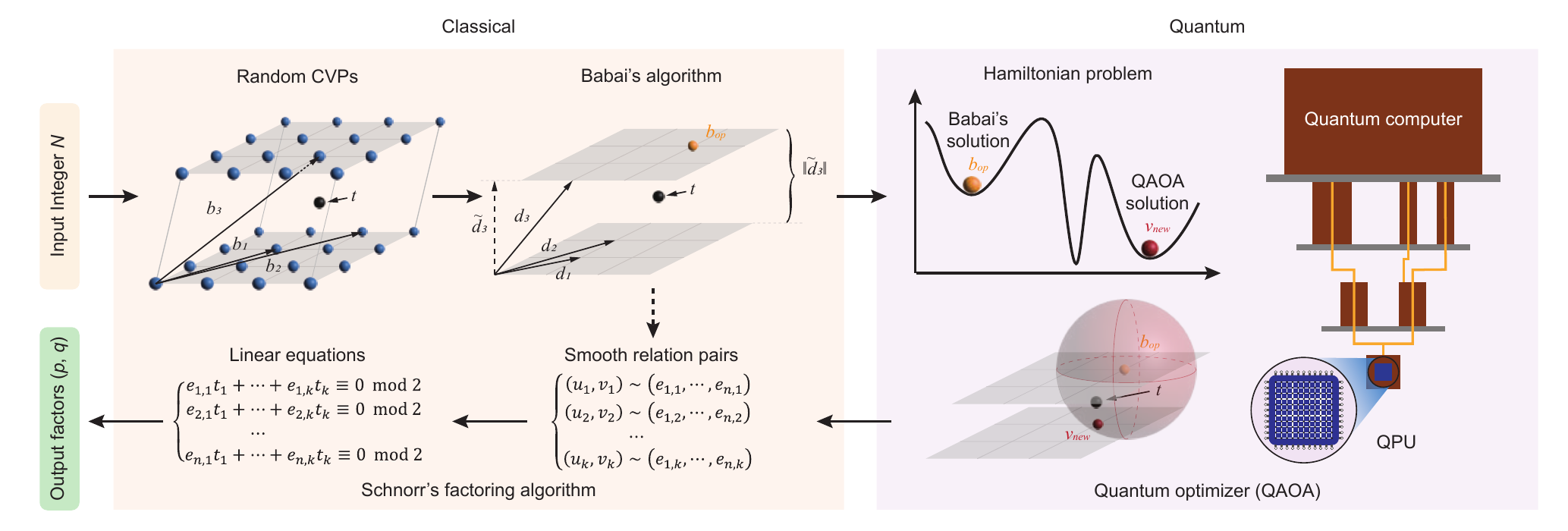}
\caption{\textbf{Workflow of the sublinear-resource quantum integer factorization (SQIF) algorithm.} The algorithm adopts a ``classical+quantum" hybrid framework where a quantum optimizer QAOA is used to optimize the classical Schnorr's factoring algorithm. First, the  problem is preprocessed as a closest vector problem (CVP) on a lattice. Then, the  quantum computer works as an optimizer to refine the classical vectors computed by Babai's algorithm, and this step can find a higher quality (closer) solution of CVP. The optimized results will feedback to the procedure in Schnorr's algorithm. After post-processing, finally output the factors $p$ and $q$. }\label{flow}
\end{figure*}

In this paper, we propose a universal quantum algorithm for integer factorization that requires only sublinear quantum resources. The algorithm is based on the classical Schnorr's  algorithm~\cite{schnorr2013factoring,schnorr2021fast}, which uses lattice reduction to factor integers. We take advantage of QAOA to optimize the most time-consuming part of Schnorr's algorithm to speed up the overall computing of the factorization progress. For an $m$-bit integer $N$, the number of qubits needed for our algorithm is $O(m/\text{log}m)$, which is sublinear in the bit length of $N$. This makes it the most qubit-saving quantum algorithm for integer factorization compared with the existing algorithms, including Shor's algorithm.  Using this algorithm, we have successfully factorized the integers 1961~(11-bit), 48567227~(26-bit) and 261980999226229~(48-bit), with 3, 5 and 10 qubits in a superconducting quantum processor, respectively. The 48-bit integer, 261980999226229, also refreshes the largest integer factored by a general method in a real quantum device. We proceed by estimating the quantum resources required to factor RSA-2048. We find that a quantum circuit with 372 physical qubits and a depth of thousands is necessary to challenge RSA-2048 even in the simplest 1D-chain system. Such a scale of quantum resources is most likely to be achieved on NISQ devices in the near future.

\vspace{.5cm}
\noindent\textbf{\large{}The framework of the algorithm }{\large\par}

\noindent  
The workflow of the sublinear-resource quantum integer factorization (SQIF) algorithm is summarized in Fig.~\ref{flow}, which essentially manifests itself as a ``classical+quantum" hybrid framework. The core idea is to utilize the quantum optimizer QAOA to optimize the most time-consuming part of Schnorr's algorithm, as a result, improving the whole efficiency
of the factoring process. As illustrated in the left panel of Fig.~\ref{flow}, Schnorr's algorithm involves two substantial steps, finding enough smooth relation pairs (sr-pairs for short) and solving the resulted linear equation system.  Generally, finding sr-pairs is the most important and consuming part of the algorithm while solving  equation system can be done in polynomial time. In Schnorr's algorithm \cite{se}, the sr-pair problem is converted to the closest vector problem (CVP) on a lattice, and resolved by lattice reduction algorithms such as Babai's algorithm~\cite{babai1986lovasz}. Based on the fact that CVP is a famous NP-hard problem~\cite{micciancio2001hardness}, we are supposed to have only the approximate other than the severe solution of CVP in polynomial time or other acceptable time consuming. Meanwhile, the probability of getting an sr-pair is proportional to the quality of the CVP solution~\cite{schnorr2013factoring}. Namely, the closer the solution vector of CVP, the more efficient the sr-pair acquaintance. Based on the facts mentioned above, we propose a scheme which utilizes QAOA to further optimize the CVP solution obtained by Babai's algorithm. The whole process of the SQIF algorithm is presented by detailed examples in \cite{se}. We mainly focus on the quantum procedures of the algorithm in the following part.

 \begin{figure*}
\centering
\includegraphics[width=0.98\textwidth]{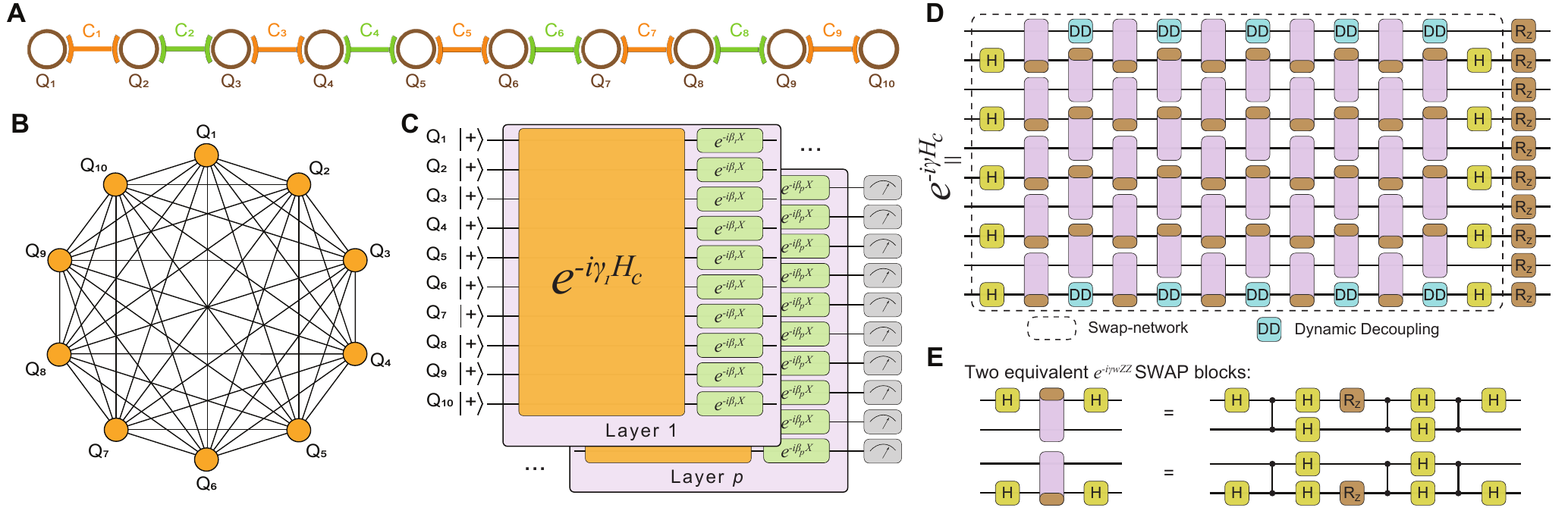}
\caption{\textbf{Experimental setup and the QAOA circuit of the SQIF algorithm.} \textbf{A}, The 10 qubits selected on a superconducting quantum processor, with each qubit coupled to its nearest neighbors mediated by frequency-tunable couplers. \textbf{B}, Native interaction topology of the problem Hamiltonian for the 10-qubit factoring case, mapped into a chain topology depicted in \textbf{A}. \textbf{C}, Circuit diagram of a $p$-layer QAOA. All qubits are initialized into $\ket{+}$, followed by $p$ layers of repeated application of the problem Hamiltonian (orange) and the mixing Hamiltonian (green), finished by population measurements (gray). Note that the variational parameters $\{\gamma, \beta\}$ are different for all layers. \textbf{D}, Routing circuit for the 10-qubit all-to-all Hamiltonian into the linear nearest neighbor topology, built by a brickwork of two similar SWAP blocks with two layers of Hardamard gates (H) applied at the start and end,  followed by a layer of R$_z$($\theta$) gates. Here, the rotation angle is omitted. The depth of the circuit is proportional to the number of qubits used. \textbf{E}, Detailed compilation of the quantum circuit into the native gates of the superconducting quantum processor.}\label{experiment}
\end{figure*}

We combine  Babai's algorithm with QAOA to solve the CVP on a lattice. 
Given a lattice $\Lambda$ with  a group of basis $B=[\mathbf b_1,...,\mathbf b_n]\in\mathbb R^{(n+1)\times n}$ and a target vector $\mathbf t\in\mathbb R^{n+1}$, Babai's algorithm can find a vector $\mathbf b_{op}\in \Lambda$ which is approximately closest to the target vector $\mathbf t$ via two steps. First, perform LLL-reduction with parameter $\delta$ for the given basis  $B=[\mathbf b_1,...,\mathbf b_n]$.  Consequently, we have a set of LLL-reduced basis denoted by $D=[\mathbf d_1,...,\mathbf d_n],$ and the corresponding Gram-Schmidt orthogonal basis denoted by $ \tilde{D}=[\tilde{\mathbf d_1},...,\tilde{\mathbf d_n}]$. The second step is a ``size-reduction" of the target  vector $\mathbf t$ using the LLL-reduced basis. Then we have the approximate closest vector, denoted by
\begin{equation}
\mathbf b_{op}=(b_{op}^1,...,b_{op}^{n+1})'=\sum_{i=1}^{n}{c_i\mathbf d_i},
\end{equation}
where the coefficient $c_i=\lceil \mu_i\rfloor=\lceil\langle {\mathbf d,\tilde {\mathbf d}}_i\rangle/\langle\tilde{\mathbf d}_i,\tilde{\mathbf d}_i\rangle\rfloor$ is obtained by rounding to the nearest integer to the Gram-Schmidt coefficient $\mu_i$. Here, we notice that the round-to-nearest function takes only one approximation at a time. In fact, if the values of the two rounding functions can be taken into the calculation simultaneously, a higher-quality solution can be obtained \cite{se}. This process will exponentially increase the amount of classical operations, which is unaffordable for a classical computer. Here we adopt
the idea of quantum computing, using the superposition effect of qubits to encode the coefficient values obtained by the two rounding functions at the same time.  Then we construct the optimization problem based on the Euclidean distance between the new lattice vector and the target vector. The details of the construction are as follows.

Let $\mathbf v_{new}$ be the new vector obtained by randomly floating $x_i\in \{0,\pm 1\}$ on the coefficient $c_i$, satisfying
\begin{equation}
\mathbf v_{new}=\sum_{i=1}^{n}{(c_i+x_i)\mathbf d_i}=\sum_{i=1}^{n}{x_i\mathbf d_i+\mathbf b_{op}}.
\end{equation}
We  construct the loss function of the optimization problem as follows
\begin{equation}\label{lost}
F(x_1,...,x_n)=\lVert\mathbf t- \mathbf v_{new}\rVert^2=\lVert \mathbf t-\sum_{i=1}^{n}{x_i\mathbf d_i}-\mathbf b_{op}\rVert^2.
\end{equation}
The function value $ \lVert\mathbf t-\mathbf v_{new}\rVert^2$ represents the squared Euclidean distance from the new vector to the target vector. The lower the loss function value, the closer the new vector is to the target vector $\mathbf t$, and the higher the quality of the solution. When all variables $x_{i,i=1,...,n}$ take $0$, the optimal solution based on Babai's algorithm is obtained. 

By mapping the variable $x_i$ to the Pauli-Z terms, the problem Hamiltonian corresponding to Eq.~\ref{lost} can be constructed as
\begin{equation}\label{hc}
Hc=\lVert \mathbf t- \sum_{i=1}^{n}{\hat x_i\mathbf d_i-\mathbf b_{op}}\rVert^2
=\sum_{j=1}^{n+1}|t_j-{\sum_{i=1}^{n}\hat x_id_{i,j}-b_{op}^j|^2},
\end{equation}
where $\hat x_i$ is a quantum operator mapped to the Pauli-Z basis according to the single-qubit encoding rules, which can be found in \cite{se}.

In this case, the number of qubits needed for the quantum procedure to optimize Babai's algorithm is equal to the dimension of the lattice. According to the analysis in \cite{se}, the lattice dimension satisfies $n\sim 2c\text{log}N/\text{loglog}N$, with $c$ a lattice parameter close to 1. Therefore, to factorize an $m$-bit integer $N$, the number of qubits required in the algorithm is $O(m/\text{log}m)$, which is a sublinear scale of $m$, compared to $O(m)$ qubits in Shor's algorithm~\cite{shor1994algorithms} and $O(m^2)$ qubits in the product table method~\cite{jiang2018quantum}. This makes our algorithm the most qubit-saving method to date, and it is also the first general quantum factoring algorithm with sublinear qubit resources.

\vspace{.5cm}
\noindent \textbf{\large{}The experiment and results}{\large\par}

\noindent 
We demonstrate the algorithm by experimentally factoring three integers on a superconducting quantum processor, where ten qubits and nine couplers arranged in a chain topology are selected. All qubits and couplers are frequency-tunable transmons, with single-qubit rotations around the $x$- or $y$-axis of the Bloch sphere realized by applying drive signals with gate information encoded in the amplitude and phase of the microwave pulses. We adopt virtual-z gates to implement single-qubit rotations around $z$-axis. Two-qubit controlled-Z (CZ) gates can be achieved by swapping the joint states $\ket{11}$ and $\ket{02}$ (or $\ket{20}$) of the neighboring qubits, when the interaction mediated by the coupler is activated~\cite{zhang2022digital}. Cross-entropy benchmarkings (XEB) in parallel yield average fidelities close to 99.9\% and 99.5\% for the single-qubit rotations and the CZ gates, respectively. More details of the experimental setup and characteristics of the quantum processor in \cite{se}. 

We factorize the 11-bit integer 1961, 26-bit integer 48567227 and 48-bit integer 261980999226229 with 3, 5 and 10 superconducting qubits, respectively. Here we demonstrate the process of obtaining one sr-pair by quantum method in each group of experiments. The calculations of other sr-pairs are similar and will be obtained by numerical method.  The details of all the sr-pairs and the corresponding linear equation systems are presented in \cite{se}.

The topology of the ZZ-items in the problem Hamiltonian is an $n$-order complete graph (Kn) according to Eq. \ref{hc} \cite{se}. An example for the 10-qubit case is shown in Fig.~\ref{experiment}\textbf{B}. To make the Kn-type Hamiltonian work on the 1D-chain of physical qubits, we have adopted a routing method based on the classical parallel bubble sort algorithm, in which the all-to-all qubits interactions can be mapped into the nearest-neighbor two-qubit interactions on a chain through elaborate swap networks, as shown in Fig.~\ref{experiment}\textbf{D}. In fact, the routing method is optimal with only a linear increase of circuit depth overhead. The swap networks are further complied into the native gates (Fig.~\ref{experiment}\textbf{E}), which can be directly executed on the quantum processor. Notably, a tiny skill has been used by an up-down combination of the ZZ-SWAP block in the even and odd layers of swap networks. As a result, a linear depth of H gates can be reduced.

QAOA can find the approximate ground state of the Hamiltonian system by updating the parameters (Fig.~\ref{experiment}\textbf{C}, a detailed description can be found in \cite{se}). The parameter optimization process of QAOA can be understood through the landscape of the energy function $E(\mathbf{\gamma,\beta})$. The comparison between  the theoretical and the experimental landscapes is a qualitative diagnostic for the application of QAOA to real hardware. For the hyperparameter $p=1$, we can visualize the energy landscape as a function of the parameters $(\gamma,\beta)$ in a three-dimensional plot in Fig.~\ref{landscape}. Here, the energy function values are normalized by $E^*=(E-E_{min})/(E_{max}-E_{min})$. Fig.~\ref{landscape} shows the noiseless simulated (left) and experimental (right) energy maps for the 3, 5 and 10 qubits cases, respectively. The different colors of the pixel blocks in the figure represent different function values. We overlay the convergence path of the classical optimization procedure, as the red curve shown in Fig.~\ref{landscape}. To optimize the parameters, we use the model gradient descent method, which performs well both numerically and experimentally on some variational quantum ansatzes. We find that the algorithm can converge to the region of global minimum within 10 steps in all three cases. We can see that the convergence paths of the experiments differ from those of the theoretical results, however, converged to the optimum in comparable steps. This indicates that the algorithm is robust to certain noise.
\begin{figure}[htb]
\centering
\includegraphics[width=0.48\textwidth]{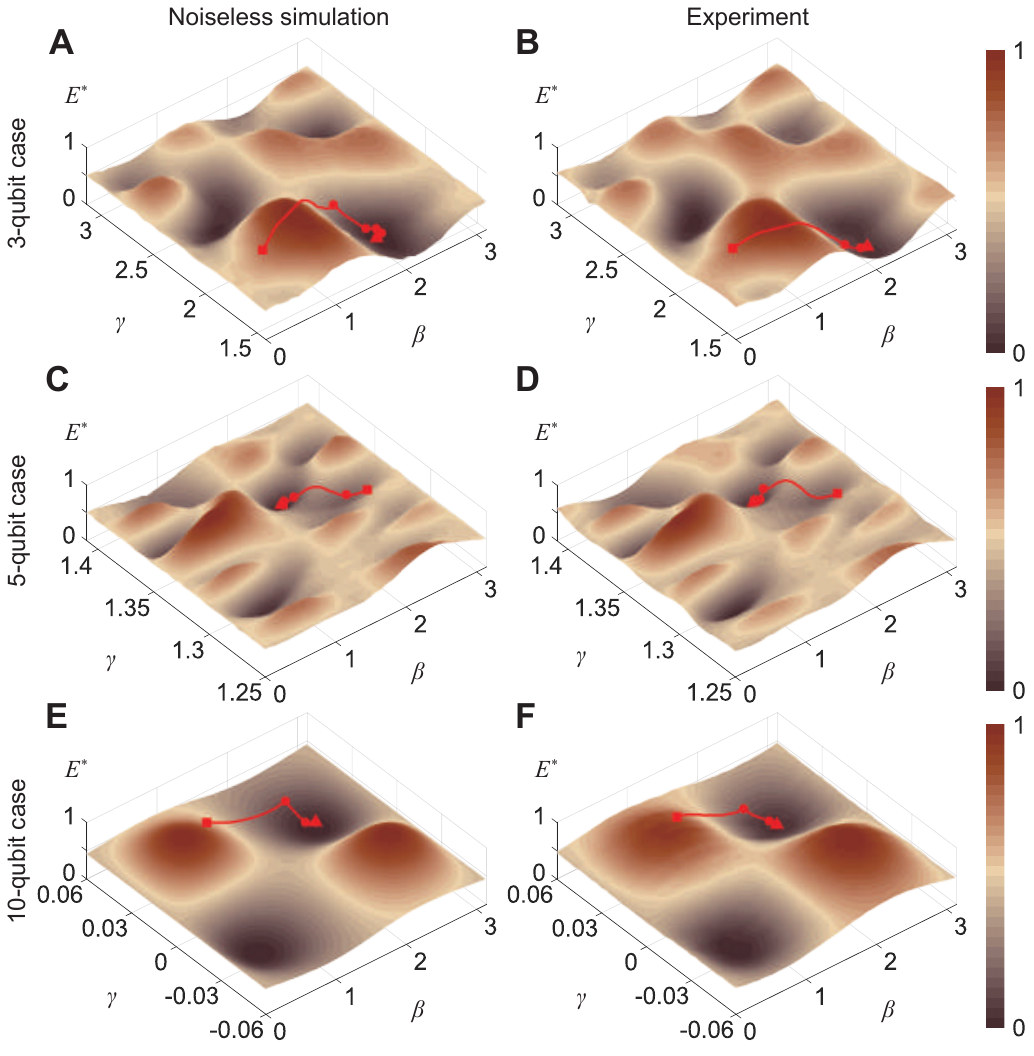}
\caption{\textbf{Energy landscapes and convergence paths of QAOA for $p=1$.} \textbf{A, B}, Numerical and experimental landscapes for the 3-qubit case, \textbf{C, D} 5-qubit case, and \textbf{E, F} 10-qubit case. In each group of the experiment, $41\times 41$ combinations of $(\gamma,\beta)$ have been evaluated, which are evenly distributed grid points in a sub-zone of the entire 2-dimensional parameter space. For each grid point, the expectation value is estimated using 30,000 circuit repetitions. The comparison of the experimental and numerical landscapes shows a clear correspondence of landscape features. An overlaid optimization trace (red, initialized from the square marker and converged into the triangle) demonstrates the ability of a classical optimizer to find optimal parameters.}\label{landscape}
\end{figure}

\begin{figure*}[htb]
\centering
\includegraphics[width=0.98\textwidth]{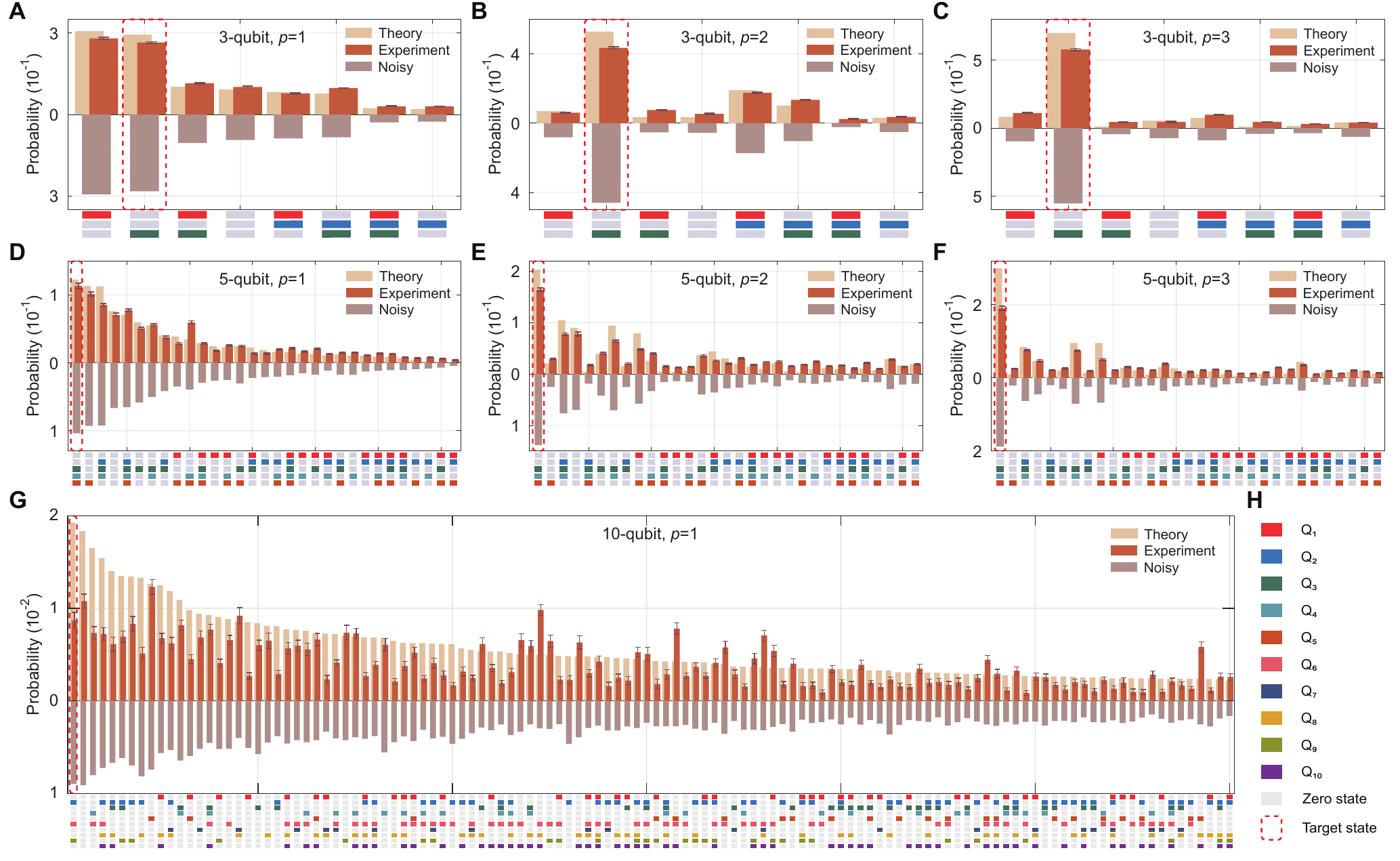}
\caption{\textbf{Experimental performance of QAOA for the three factoring cases.}  \textbf{A}-\textbf{C}, QAOA performance of the 3-qubit case with $p=1$, $p=2$ and $p=3$, respectively. \textbf{D}-\textbf{F}, QAOA performance of the 5-qubit case with $p=1$, $p=2$ and $p=3$, respectively. \textbf{G}, $p=1$ performance of QAOA for the 10-qubit case . The experimental results shown in orange are averaged over 20 repeated experiments with error bars giving a confidence interval of one standard deviation. The theory(yellow) and 0.01-noise(taupe) results are also given for comparison. It can be observed that all the three groups of experimental results on the superconducting quantum processor are in good agreement with the theoretical and 0.01-noise values. \textbf{H}, Representations of the color blocks that are basis states of different qubits in x-tick labels.}\label{results}
\end{figure*}

In QAOA, the core work of the quantum computer is to prepare the quantum states according to the given variational parameters. The performance of QAOA will be improved by increasing the depth of hyperparameter $p$ in theory. However, the errors are accumulated during the increasing of circuit depth and the bonus of the computation can be counteracted. Here we report the performance of the superconducting quantum processor on  running circuits at the optimal $\beta,\gamma$ parameters.  We show QAOA layers up to $p=3$ for the cases of 3 and 5 qubits, and a single-layer QAOA for the 10-qubit case. The results of $p=3$ for the 10-qubit case have also been performed and are apparently  better than random guess, however, not as good as that of $p=1$ \cite{se}. We can observe in Fig.~\ref{results}\textbf{A-C} that the probability of the target state (red dashed box) increases as the hyperparameter $p$ grows. Although the increase is not as large as the theoretical value, it is in good agreement with the noise simulation. Similar results can be found in the 5-qubit experiment, see Fig.~\ref{results}\textbf{D-F}. The results for the 10-qubit case with $p=1$ are shown in Fig.~\ref{results}\textbf{G}. We only show the most significant 120 states according to the theoretical results for illustration. We can find that the theoretical probability of the target state is 0.02~(the highest), while the experimental result is around 0.008, which is close to the noise result 0.009. The experimental results are significantly larger than that of random guess 0.001, which means the computation bonus of QAOA is still considerable. In addition, the shape of the probability distribution of each quantum state is symmetric with that of the simulation results, which shows that the experimental results are in good agreement with the theoretical values. 

\vspace{.5cm}
\noindent \textbf{\large{}The quantum resource estimation}{\large\par}
\noindent 
Here we report the quantum resources needed to challenge some real-life RSA numbers based on the SQIF algorithm in this paper. The main quantum resources mentioned include the number of qubits and the quantum circuit depth of QAOA in one layer. Usually, quantum circuits cannot be directly executed on quantum computing devices, as their design does not consider the qubits connectivity characteristics of actual physical systems. The execution process often requires additional quantum resources such as ancilla qubits and extending circuit depths.  We have discussed the quantum resources required in quantum systems under three typical topologies, including all connected system (Kn), 2D-lattice system (2DSL), and 1D-chain system (LNN). We demonstrate with specific schemes that the embedding process needs no extra qubits overhead and the circuit depths of QAOA  in one layer are $O(n)$ for all three systems. As a result, a sublinear quantum resource is necessary for factoring integers using our algorithm. Taking RSA-2048 as an example, the number of qubits required is $n= 2*2048/\text{log}2048\sim 372$. The quantum circuit depth of QAOA with a single layer is $1118$ in Kn topology system, $1139$ in 2DSL system and $1490$ in the simplest LNN system, which is achievable for the NISQ devices in the near future. The quantum resources required for different lengths of RSA numbers are shown in Table \ref{quantum}. The detailed analysis can be found in \cite{se}.

\begin{table}[b]
\caption{\label{tab:table1} Resource estimation for RSA numbers. The main quantum resources mentioned are the number of qubits, the quantum circuit depth of QAOA with a single iteration in three typical topologies, including all connected system (Kn), 2D-lattice system (2DSL) and 1D-chain system (LNN).  The results are obtained without considering the native compilation of the ZZ-basic module (or ZZ-SWAP basic module) in a specific physical system. }
\begin{ruledtabular}
\begin{tabular}{lcccc}
RSA number & Qubits  & Kn-depth & 2DSL-depth & LNN-depth \\
\hline
RSA-128 & 37 & 113 &121  & 150 \\
RSA-256 & 64 & 194 & 204 & 258 \\
RSA-512 & 114 & 344 & 357 & 458 \\
RSA-1024 & 205 & 617 & 633 & 822 \\
RSA-2048 & 372 & 1118 & 1139 & 1490 \\
\end{tabular}
\end{ruledtabular}\label{quantum}

\end{table}

\vspace{.5cm}

\noindent \textbf{\large{}Conclusion}{\large\par}

\noindent The integer factorization problem is the security cornerstone of the widely used RSA public key cryptography nowadays. In this paper, we have proposed a general quantum algorithm for integer factorization based on the classical lattice reduction method. To factor an  $m$-bit integer $N$, the number of qubits needed for the algorithm is $O(m/\text{log}m)$, which is a sublinear scale of the bit length of $N$. This quantum factoring algorithm uses the least qubits compared with previous methods, including Shor's algorithm. We have demonstrated the factoring principle for the algorithm on a superconducting quantum processor. The 48-bit integer 261980999226229 in our work is the largest integer factored by the general method in a real quantum system to date. We have analyzed the quantum resources required to factor RSA-2048 in quantum systems under three typical topologies. We find that a quantum circuit with 372 physical qubits and a depth of thousands is necessary to challenge RSA-2048 even in the simplest 1D-chain system. Such a scale of  quantum resources is most likely to be achieved on NISQ devices in the near future. It should be pointed out that the quantum speedup of the algorithm is unclear due to the ambiguous convergence of QAOA. However, the idea of optimizing the ``size-reduce" procedure in Babai's algorithm through QAOA can be used as a subroutine in a large group of widely used lattice reduction algorithms. Further on, it can help to analyze the quantum-resistant cryptographic problems based on lattice.

\bibliography{sample, exp, SI/sn-bib}

\begin{thebibliography}{59}%
\makeatletter
\providecommand \@ifxundefined [1]{%
 \@ifx{#1\undefined}
}%
\providecommand \@ifnum [1]{%
 \ifnum #1\expandafter \@firstoftwo
 \else \expandafter \@secondoftwo
 \fi
}%
\providecommand \@ifx [1]{%
 \ifx #1\expandafter \@firstoftwo
 \else \expandafter \@secondoftwo
 \fi
}%
\providecommand \natexlab [1]{#1}%
\providecommand \enquote  [1]{``#1''}%
\providecommand \bibnamefont  [1]{#1}%
\providecommand \bibfnamefont [1]{#1}%
\providecommand \citenamefont [1]{#1}%
\providecommand \href@noop [0]{\@secondoftwo}%
\providecommand \href [0]{\begingroup \@sanitize@url \@href}%
\providecommand \@href[1]{\@@startlink{#1}\@@href}%
\providecommand \@@href[1]{\endgroup#1\@@endlink}%
\providecommand \@sanitize@url [0]{\catcode `\\12\catcode `\$12\catcode
  `\&12\catcode `\#12\catcode `\^12\catcode `\_12\catcode `\%12\relax}%
\providecommand \@@startlink[1]{}%
\providecommand \@@endlink[0]{}%
\providecommand \url  [0]{\begingroup\@sanitize@url \@url }%
\providecommand \@url [1]{\endgroup\@href {#1}{\urlprefix }}%
\providecommand \urlprefix  [0]{URL }%
\providecommand \Eprint [0]{\href }%
\providecommand \doibase [0]{https://doi.org/}%
\providecommand \selectlanguage [0]{\@gobble}%
\providecommand \bibinfo  [0]{\@secondoftwo}%
\providecommand \bibfield  [0]{\@secondoftwo}%
\providecommand \translation [1]{[#1]}%
\providecommand \BibitemOpen [0]{}%
\providecommand \bibitemStop [0]{}%
\providecommand \bibitemNoStop [0]{.\EOS\space}%
\providecommand \EOS [0]{\spacefactor3000\relax}%
\providecommand \BibitemShut  [1]{\csname bibitem#1\endcsname}%
\let\auto@bib@innerbib\@empty
\bibitem [{\citenamefont {Preskill}(2018)}]{preskill2018quantum}%
  \BibitemOpen
  \bibfield  {author} {\bibinfo {author} {\bibfnamefont {J.}~\bibnamefont
  {Preskill}},\ }\bibfield  {title} {\bibinfo {title} {Quantum computing in the
  {NISQ} era and beyond},\ }\href
  {https://quantum-journal.org/papers/q-2018-08-06-79/} {\bibfield  {journal}
  {\bibinfo  {journal} {Quantum}\ }\textbf {\bibinfo {volume} {2}},\ \bibinfo
  {pages} {79} (\bibinfo {year} {2018})}\BibitemShut {NoStop}%
\bibitem [{\citenamefont {Arute}\ \emph {et~al.}(2019)\citenamefont {Arute},
  \citenamefont {Arya}, \citenamefont {Babbush}, \citenamefont {Bacon},
  \citenamefont {Bardin}, \citenamefont {Barends}, \citenamefont {Biswas},
  \citenamefont {Boixo}, \citenamefont {Brandao}, \citenamefont {Buell} \emph
  {et~al.}}]{arute2019quantum}%
  \BibitemOpen
  \bibfield  {author} {\bibinfo {author} {\bibfnamefont {F.}~\bibnamefont
  {Arute}}, \bibinfo {author} {\bibfnamefont {K.}~\bibnamefont {Arya}},
  \bibinfo {author} {\bibfnamefont {R.}~\bibnamefont {Babbush}}, \bibinfo
  {author} {\bibfnamefont {D.}~\bibnamefont {Bacon}}, \bibinfo {author}
  {\bibfnamefont {J.~C.}\ \bibnamefont {Bardin}}, \bibinfo {author}
  {\bibfnamefont {R.}~\bibnamefont {Barends}}, \bibinfo {author} {\bibfnamefont
  {R.}~\bibnamefont {Biswas}}, \bibinfo {author} {\bibfnamefont
  {S.}~\bibnamefont {Boixo}}, \bibinfo {author} {\bibfnamefont {F.~G.}\
  \bibnamefont {Brandao}}, \bibinfo {author} {\bibfnamefont {D.~A.}\
  \bibnamefont {Buell}}, \emph {et~al.},\ }\bibfield  {title} {\bibinfo {title}
  {Quantum supremacy using a programmable superconducting processor},\ }\href
  {https://doi.org/10.1038/s41586-019-1666-5} {\bibfield  {journal} {\bibinfo
  {journal} {Nature}\ }\textbf {\bibinfo {volume} {574}},\ \bibinfo {pages}
  {505} (\bibinfo {year} {2019})}\BibitemShut {NoStop}%
\bibitem [{\citenamefont {Cerezo}\ \emph {et~al.}(2021)\citenamefont {Cerezo},
  \citenamefont {Arrasmith}, \citenamefont {Babbush}, \citenamefont {Benjamin},
  \citenamefont {Endo}, \citenamefont {Fujii}, \citenamefont {McClean},
  \citenamefont {Mitarai}, \citenamefont {Yuan}, \citenamefont {Cincio} \emph
  {et~al.}}]{cerezo2021variational}%
  \BibitemOpen
  \bibfield  {author} {\bibinfo {author} {\bibfnamefont {M.}~\bibnamefont
  {Cerezo}}, \bibinfo {author} {\bibfnamefont {A.}~\bibnamefont {Arrasmith}},
  \bibinfo {author} {\bibfnamefont {R.}~\bibnamefont {Babbush}}, \bibinfo
  {author} {\bibfnamefont {S.~C.}\ \bibnamefont {Benjamin}}, \bibinfo {author}
  {\bibfnamefont {S.}~\bibnamefont {Endo}}, \bibinfo {author} {\bibfnamefont
  {K.}~\bibnamefont {Fujii}}, \bibinfo {author} {\bibfnamefont {J.~R.}\
  \bibnamefont {McClean}}, \bibinfo {author} {\bibfnamefont {K.}~\bibnamefont
  {Mitarai}}, \bibinfo {author} {\bibfnamefont {X.}~\bibnamefont {Yuan}},
  \bibinfo {author} {\bibfnamefont {L.}~\bibnamefont {Cincio}}, \emph
  {et~al.},\ }\bibfield  {title} {\bibinfo {title} {Variational quantum
  algorithms},\ }\href {https://doi.org/10.1038/s42254-021-00348-9} {\bibfield
  {journal} {\bibinfo  {journal} {Nat. Rev. Phys.}\ }\textbf {\bibinfo {volume}
  {3}},\ \bibinfo {pages} {625} (\bibinfo {year} {2021})}\BibitemShut {NoStop}%
\bibitem [{\citenamefont {Peruzzo}\ \emph {et~al.}(2014)\citenamefont
  {Peruzzo}, \citenamefont {McClean}, \citenamefont {Shadbolt}, \citenamefont
  {Yung}, \citenamefont {Zhou}, \citenamefont {Love}, \citenamefont
  {Aspuru-Guzik},\ and\ \citenamefont {O'brien}}]{peruzzo2014variational}%
  \BibitemOpen
  \bibfield  {author} {\bibinfo {author} {\bibfnamefont {A.}~\bibnamefont
  {Peruzzo}}, \bibinfo {author} {\bibfnamefont {J.}~\bibnamefont {McClean}},
  \bibinfo {author} {\bibfnamefont {P.}~\bibnamefont {Shadbolt}}, \bibinfo
  {author} {\bibfnamefont {M.-H.}\ \bibnamefont {Yung}}, \bibinfo {author}
  {\bibfnamefont {X.-Q.}\ \bibnamefont {Zhou}}, \bibinfo {author}
  {\bibfnamefont {P.~J.}\ \bibnamefont {Love}}, \bibinfo {author}
  {\bibfnamefont {A.}~\bibnamefont {Aspuru-Guzik}},\ and\ \bibinfo {author}
  {\bibfnamefont {J.~L.}\ \bibnamefont {O'brien}},\ }\bibfield  {title}
  {\bibinfo {title} {A variational eigenvalue solver on a photonic quantum
  processor},\ }\href {https://doi.org/10.1038/ncomms5213} {\bibfield
  {journal} {\bibinfo  {journal} {Nat. Commun.}\ }\textbf {\bibinfo {volume}
  {5}},\ \bibinfo {pages} {1} (\bibinfo {year} {2014})}\BibitemShut {NoStop}%
\bibitem [{\citenamefont {Farhi}\ \emph {et~al.}(2014)\citenamefont {Farhi},
  \citenamefont {Goldstone},\ and\ \citenamefont {Gutmann}}]{farhi2014quantum}%
  \BibitemOpen
  \bibfield  {author} {\bibinfo {author} {\bibfnamefont {E.}~\bibnamefont
  {Farhi}}, \bibinfo {author} {\bibfnamefont {J.}~\bibnamefont {Goldstone}},\
  and\ \bibinfo {author} {\bibfnamefont {S.}~\bibnamefont {Gutmann}},\
  }\bibfield  {title} {\bibinfo {title} {A quantum approximate optimization
  algorithm},\ }\href {https://arxiv.org/abs/1411.4028} {\bibfield  {journal}
  {\bibinfo  {journal} {arXiv:1411.4028}\ } (\bibinfo {year}
  {2014})}\BibitemShut {NoStop}%
\bibitem [{\citenamefont {Wang}\ \emph {et~al.}(2022)\citenamefont {Wang},
  \citenamefont {Wei}, \citenamefont {Long},\ and\ \citenamefont
  {Hanzo}}]{Wang2022}%
  \BibitemOpen
  \bibfield  {author} {\bibinfo {author} {\bibfnamefont {Z.}~\bibnamefont
  {Wang}}, \bibinfo {author} {\bibfnamefont {S.}~\bibnamefont {Wei}}, \bibinfo
  {author} {\bibfnamefont {G.-L.}\ \bibnamefont {Long}},\ and\ \bibinfo
  {author} {\bibfnamefont {L.}~\bibnamefont {Hanzo}},\ }\bibfield  {title}
  {\bibinfo {title} {Variational quantum attacks threaten advanced encryption
  standard based symmetric cryptography},\ }\href
  {https://doi.org/10.1007/s11432-022-3511-5} {\bibfield  {journal} {\bibinfo
  {journal} {Sci. China Inf. Sci.}\ }\textbf {\bibinfo {volume} {65}},\
  \bibinfo {pages} {1} (\bibinfo {year} {2022})}\BibitemShut {NoStop}%
\bibitem [{\citenamefont {McArdle}\ \emph {et~al.}(2020)\citenamefont
  {McArdle}, \citenamefont {Endo}, \citenamefont {Aspuru-Guzik}, \citenamefont
  {Benjamin},\ and\ \citenamefont {Yuan}}]{mcardle2020quantum}%
  \BibitemOpen
  \bibfield  {author} {\bibinfo {author} {\bibfnamefont {S.}~\bibnamefont
  {McArdle}}, \bibinfo {author} {\bibfnamefont {S.}~\bibnamefont {Endo}},
  \bibinfo {author} {\bibfnamefont {A.}~\bibnamefont {Aspuru-Guzik}}, \bibinfo
  {author} {\bibfnamefont {S.~C.}\ \bibnamefont {Benjamin}},\ and\ \bibinfo
  {author} {\bibfnamefont {X.}~\bibnamefont {Yuan}},\ }\bibfield  {title}
  {\bibinfo {title} {Quantum computational chemistry},\ }\href
  {https://journals.aps.org/rmp/abstract/10.1103/RevModPhys.92.015003}
  {\bibfield  {journal} {\bibinfo  {journal} {Rev. Mod. Phys.}\ }\textbf
  {\bibinfo {volume} {92}},\ \bibinfo {pages} {015003} (\bibinfo {year}
  {2020})}\BibitemShut {NoStop}%
\bibitem [{\citenamefont {Wei}\ \emph {et~al.}(2020)\citenamefont {Wei},
  \citenamefont {Li},\ and\ \citenamefont {Long}}]{wei2020full}%
  \BibitemOpen
  \bibfield  {author} {\bibinfo {author} {\bibfnamefont {S.}~\bibnamefont
  {Wei}}, \bibinfo {author} {\bibfnamefont {H.}~\bibnamefont {Li}},\ and\
  \bibinfo {author} {\bibfnamefont {G.}~\bibnamefont {Long}},\ }\bibfield
  {title} {\bibinfo {title} {A full quantum eigensolver for quantum chemistry
  simulations},\ }\href {https://doi.org/10.34133/2020/1486935} {\bibfield
  {journal} {\bibinfo  {journal} {Research}\ }\textbf {\bibinfo {volume}
  {2020}} (\bibinfo {year} {2020})}\BibitemShut {NoStop}%
\bibitem [{\citenamefont {Biamonte}\ \emph {et~al.}(2017)\citenamefont
  {Biamonte}, \citenamefont {Wittek}, \citenamefont {Pancotti}, \citenamefont
  {Rebentrost}, \citenamefont {Wiebe},\ and\ \citenamefont
  {Lloyd}}]{biamonte2017quantum}%
  \BibitemOpen
  \bibfield  {author} {\bibinfo {author} {\bibfnamefont {J.}~\bibnamefont
  {Biamonte}}, \bibinfo {author} {\bibfnamefont {P.}~\bibnamefont {Wittek}},
  \bibinfo {author} {\bibfnamefont {N.}~\bibnamefont {Pancotti}}, \bibinfo
  {author} {\bibfnamefont {P.}~\bibnamefont {Rebentrost}}, \bibinfo {author}
  {\bibfnamefont {N.}~\bibnamefont {Wiebe}},\ and\ \bibinfo {author}
  {\bibfnamefont {S.}~\bibnamefont {Lloyd}},\ }\bibfield  {title} {\bibinfo
  {title} {Quantum machine learning},\ }\href
  {https://doi.org/10.1038/nature23474} {\bibfield  {journal} {\bibinfo
  {journal} {Nature}\ }\textbf {\bibinfo {volume} {549}},\ \bibinfo {pages}
  {195} (\bibinfo {year} {2017})}\BibitemShut {NoStop}%
\bibitem [{\citenamefont {Wang}\ \emph {et~al.}(2018)\citenamefont {Wang},
  \citenamefont {Hadfield}, \citenamefont {Jiang},\ and\ \citenamefont
  {Rieffel}}]{wang2018quantum}%
  \BibitemOpen
  \bibfield  {author} {\bibinfo {author} {\bibfnamefont {Z.}~\bibnamefont
  {Wang}}, \bibinfo {author} {\bibfnamefont {S.}~\bibnamefont {Hadfield}},
  \bibinfo {author} {\bibfnamefont {Z.}~\bibnamefont {Jiang}},\ and\ \bibinfo
  {author} {\bibfnamefont {E.~G.}\ \bibnamefont {Rieffel}},\ }\bibfield
  {title} {\bibinfo {title} {Quantum approximate optimization algorithm for
  {M}axcut: A fermionic view},\ }\href
  {https://journals.aps.org/pra/abstract/10.1103/PhysRevA.97.022304} {\bibfield
   {journal} {\bibinfo  {journal} {Phys. Rev. A}\ }\textbf {\bibinfo {volume}
  {97}},\ \bibinfo {pages} {022304} (\bibinfo {year} {2018})}\BibitemShut
  {NoStop}%
\bibitem [{\citenamefont {Harrigan}\ \emph {et~al.}(2021)\citenamefont
  {Harrigan}, \citenamefont {Sung}, \citenamefont {Neeley}, \citenamefont
  {Satzinger}, \citenamefont {Arute}, \citenamefont {Arya}, \citenamefont
  {Atalaya}, \citenamefont {Bardin}, \citenamefont {Barends}, \citenamefont
  {Boixo} \emph {et~al.}}]{harrigan2021quantum}%
  \BibitemOpen
  \bibfield  {author} {\bibinfo {author} {\bibfnamefont {M.~P.}\ \bibnamefont
  {Harrigan}}, \bibinfo {author} {\bibfnamefont {K.~J.}\ \bibnamefont {Sung}},
  \bibinfo {author} {\bibfnamefont {M.}~\bibnamefont {Neeley}}, \bibinfo
  {author} {\bibfnamefont {K.~J.}\ \bibnamefont {Satzinger}}, \bibinfo {author}
  {\bibfnamefont {F.}~\bibnamefont {Arute}}, \bibinfo {author} {\bibfnamefont
  {K.}~\bibnamefont {Arya}}, \bibinfo {author} {\bibfnamefont {J.}~\bibnamefont
  {Atalaya}}, \bibinfo {author} {\bibfnamefont {J.~C.}\ \bibnamefont {Bardin}},
  \bibinfo {author} {\bibfnamefont {R.}~\bibnamefont {Barends}}, \bibinfo
  {author} {\bibfnamefont {S.}~\bibnamefont {Boixo}}, \emph {et~al.},\
  }\bibfield  {title} {\bibinfo {title} {Quantum approximate optimization of
  non-planar graph problems on a planar superconducting processor},\
  }\href@noop {} {\bibfield  {journal} {\bibinfo  {journal} {Nature Physics}\
  }\textbf {\bibinfo {volume} {17}},\ \bibinfo {pages} {332} (\bibinfo {year}
  {2021})}\BibitemShut {NoStop}%
\bibitem [{\citenamefont {Rivest}\ \emph {et~al.}(1978)\citenamefont {Rivest},
  \citenamefont {Shamir},\ and\ \citenamefont {Adleman}}]{rivest2019method}%
  \BibitemOpen
  \bibfield  {author} {\bibinfo {author} {\bibfnamefont {R.~L.}\ \bibnamefont
  {Rivest}}, \bibinfo {author} {\bibfnamefont {A.}~\bibnamefont {Shamir}},\
  and\ \bibinfo {author} {\bibfnamefont {L.}~\bibnamefont {Adleman}},\
  }\bibfield  {title} {\bibinfo {title} {A method for obtaining digital
  signatures and public-key cryptosystems},\ }\href
  {https://doi.org/10.1145/359340.359342} {\bibfield  {journal} {\bibinfo
  {journal} {Commun. ACM}\ }\textbf {\bibinfo {volume} {21}},\ \bibinfo {pages}
  {120} (\bibinfo {year} {1978})}\BibitemShut {NoStop}%
\bibitem [{\citenamefont {Shor}(1994)}]{shor1994algorithms}%
  \BibitemOpen
  \bibfield  {author} {\bibinfo {author} {\bibfnamefont {P.}~\bibnamefont
  {Shor}},\ }\bibfield  {title} {\bibinfo {title} {Algorithms for quantum
  computation: discrete logarithms and factoring},\ }in\ \href
  {https://doi.org/10.1109/SFCS.1994.365700} {\emph {\bibinfo {booktitle}
  {Proc. 35th Ann. Symp. on Foundations of Computer Science}}}\ (\bibinfo
  {year} {1994})\ pp.\ \bibinfo {pages} {124--134}\BibitemShut {NoStop}%
\bibitem [{\citenamefont {Gidney}\ and\ \citenamefont
  {Eker{\aa}}(2021)}]{gidney2021factor}%
  \BibitemOpen
  \bibfield  {author} {\bibinfo {author} {\bibfnamefont {C.}~\bibnamefont
  {Gidney}}\ and\ \bibinfo {author} {\bibfnamefont {M.}~\bibnamefont
  {Eker{\aa}}},\ }\bibfield  {title} {\bibinfo {title} {How to factor 2048 bit
  {RSA} integers in 8 hours using 20 million noisy qubits},\ }\href
  {https://quantum-journal.org/papers/q-2021-04-15-433/} {\bibfield  {journal}
  {\bibinfo  {journal} {Quantum}\ }\textbf {\bibinfo {volume} {5}},\ \bibinfo
  {pages} {433} (\bibinfo {year} {2021})}\BibitemShut {NoStop}%
\bibitem [{\citenamefont {Gouzien}\ and\ \citenamefont
  {Sangouard}(2021)}]{gouzien2021factoring}%
  \BibitemOpen
  \bibfield  {author} {\bibinfo {author} {\bibfnamefont {E.}~\bibnamefont
  {Gouzien}}\ and\ \bibinfo {author} {\bibfnamefont {N.}~\bibnamefont
  {Sangouard}},\ }\bibfield  {title} {\bibinfo {title} {Factoring 2048-bit
  {RSA} integers in 177 days with 13 436 qubits and a multimode memory},\
  }\href {https://journals.aps.org/prl/abstract/10.1103/PhysRevLett.127.140503}
  {\bibfield  {journal} {\bibinfo  {journal} {Phys. Rev. Lett.}\ }\textbf
  {\bibinfo {volume} {127}},\ \bibinfo {pages} {140503} (\bibinfo {year}
  {2021})}\BibitemShut {NoStop}%
\bibitem [{\citenamefont {Vandersypen}\ \emph {et~al.}(2001)\citenamefont
  {Vandersypen}, \citenamefont {Steffen}, \citenamefont {Breyta}, \citenamefont
  {Yannoni}, \citenamefont {Sherwood},\ and\ \citenamefont
  {Chuang}}]{vandersypen2001experimental}%
  \BibitemOpen
  \bibfield  {author} {\bibinfo {author} {\bibfnamefont {L.~M.}\ \bibnamefont
  {Vandersypen}}, \bibinfo {author} {\bibfnamefont {M.}~\bibnamefont
  {Steffen}}, \bibinfo {author} {\bibfnamefont {G.}~\bibnamefont {Breyta}},
  \bibinfo {author} {\bibfnamefont {C.~S.}\ \bibnamefont {Yannoni}}, \bibinfo
  {author} {\bibfnamefont {M.~H.}\ \bibnamefont {Sherwood}},\ and\ \bibinfo
  {author} {\bibfnamefont {I.~L.}\ \bibnamefont {Chuang}},\ }\bibfield  {title}
  {\bibinfo {title} {Experimental realization of {S}hor's quantum factoring
  algorithm using nuclear magnetic resonance},\ }\href
  {https://www.nature.com/articles/414883a} {\bibfield  {journal} {\bibinfo
  {journal} {Nature}\ }\textbf {\bibinfo {volume} {414}},\ \bibinfo {pages}
  {883} (\bibinfo {year} {2001})}\BibitemShut {NoStop}%
\bibitem [{\citenamefont {Monz}\ \emph {et~al.}(2016)\citenamefont {Monz},
  \citenamefont {Nigg}, \citenamefont {Martinez}, \citenamefont {Brandl},
  \citenamefont {Schindler}, \citenamefont {Rines}, \citenamefont {Wang},
  \citenamefont {Chuang},\ and\ \citenamefont {Blatt}}]{monz2016realization}%
  \BibitemOpen
  \bibfield  {author} {\bibinfo {author} {\bibfnamefont {T.}~\bibnamefont
  {Monz}}, \bibinfo {author} {\bibfnamefont {D.}~\bibnamefont {Nigg}}, \bibinfo
  {author} {\bibfnamefont {E.~A.}\ \bibnamefont {Martinez}}, \bibinfo {author}
  {\bibfnamefont {M.~F.}\ \bibnamefont {Brandl}}, \bibinfo {author}
  {\bibfnamefont {P.}~\bibnamefont {Schindler}}, \bibinfo {author}
  {\bibfnamefont {R.}~\bibnamefont {Rines}}, \bibinfo {author} {\bibfnamefont
  {S.~X.}\ \bibnamefont {Wang}}, \bibinfo {author} {\bibfnamefont {I.~L.}\
  \bibnamefont {Chuang}},\ and\ \bibinfo {author} {\bibfnamefont
  {R.}~\bibnamefont {Blatt}},\ }\bibfield  {title} {\bibinfo {title}
  {Realization of a scalable shor algorithm},\ }\href
  {https://doi/10.1126/science.aad9480} {\bibfield  {journal} {\bibinfo
  {journal} {Science}\ }\textbf {\bibinfo {volume} {351}},\ \bibinfo {pages}
  {1068} (\bibinfo {year} {2016})}\BibitemShut {NoStop}%
\bibitem [{\citenamefont {Martin-Lopez}\ \emph {et~al.}(2012)\citenamefont
  {Martin-Lopez}, \citenamefont {Laing}, \citenamefont {Lawson}, \citenamefont
  {Alvarez}, \citenamefont {Zhou},\ and\ \citenamefont
  {O'brien}}]{martin2012experimental}%
  \BibitemOpen
  \bibfield  {author} {\bibinfo {author} {\bibfnamefont {E.}~\bibnamefont
  {Martin-Lopez}}, \bibinfo {author} {\bibfnamefont {A.}~\bibnamefont {Laing}},
  \bibinfo {author} {\bibfnamefont {T.}~\bibnamefont {Lawson}}, \bibinfo
  {author} {\bibfnamefont {R.}~\bibnamefont {Alvarez}}, \bibinfo {author}
  {\bibfnamefont {X.-Q.}\ \bibnamefont {Zhou}},\ and\ \bibinfo {author}
  {\bibfnamefont {J.~L.}\ \bibnamefont {O'brien}},\ }\bibfield  {title}
  {\bibinfo {title} {Experimental realization of {S}hor's quantum factoring
  algorithm using qubit recycling},\ }\href
  {https://doi.org/10.1038/nphoton.2012.259} {\bibfield  {journal} {\bibinfo
  {journal} {Nat. Photon.}\ }\textbf {\bibinfo {volume} {6}},\ \bibinfo {pages}
  {773} (\bibinfo {year} {2012})}\BibitemShut {NoStop}%
\bibitem [{\citenamefont {Farhi}\ \emph {et~al.}(2001)\citenamefont {Farhi},
  \citenamefont {Goldstone}, \citenamefont {Gutmann}, \citenamefont {Lapan},
  \citenamefont {Lundgren},\ and\ \citenamefont {Preda}}]{farhi2001quantum}%
  \BibitemOpen
  \bibfield  {author} {\bibinfo {author} {\bibfnamefont {E.}~\bibnamefont
  {Farhi}}, \bibinfo {author} {\bibfnamefont {J.}~\bibnamefont {Goldstone}},
  \bibinfo {author} {\bibfnamefont {S.}~\bibnamefont {Gutmann}}, \bibinfo
  {author} {\bibfnamefont {J.}~\bibnamefont {Lapan}}, \bibinfo {author}
  {\bibfnamefont {A.}~\bibnamefont {Lundgren}},\ and\ \bibinfo {author}
  {\bibfnamefont {D.}~\bibnamefont {Preda}},\ }\bibfield  {title} {\bibinfo
  {title} {A quantum adiabatic evolution algorithm applied to random instances
  of an {NP}-complete problem},\ }\href
  {https://www.science.org/doi/abs/10.1126/science.1057726} {\bibfield
  {journal} {\bibinfo  {journal} {Science}\ }\textbf {\bibinfo {volume}
  {292}},\ \bibinfo {pages} {472} (\bibinfo {year} {2001})}\BibitemShut
  {NoStop}%
\bibitem [{\citenamefont {Schaller}\ and\ \citenamefont
  {Sch{\"u}tzhold}(2010)}]{schaller2010role}%
  \BibitemOpen
  \bibfield  {author} {\bibinfo {author} {\bibfnamefont {G.}~\bibnamefont
  {Schaller}}\ and\ \bibinfo {author} {\bibfnamefont {R.}~\bibnamefont
  {Sch{\"u}tzhold}},\ }\bibfield  {title} {\bibinfo {title} {The role of
  symmetries in adiabatic quantum algorithms},\ }\href
  {https://dl.acm.org/doi/abs/10.5555/2011438.2011447} {\bibfield  {journal}
  {\bibinfo  {journal} {Quantum Info. Comput.}\ }\textbf {\bibinfo {volume}
  {10}},\ \bibinfo {pages} {109} (\bibinfo {year} {2010})}\BibitemShut
  {NoStop}%
\bibitem [{\citenamefont {Borders}\ \emph {et~al.}(2019)\citenamefont
  {Borders}, \citenamefont {Pervaiz}, \citenamefont {Fukami}, \citenamefont
  {Camsari}, \citenamefont {Ohno},\ and\ \citenamefont
  {Datta}}]{borders2019integer}%
  \BibitemOpen
  \bibfield  {author} {\bibinfo {author} {\bibfnamefont {W.~A.}\ \bibnamefont
  {Borders}}, \bibinfo {author} {\bibfnamefont {A.~Z.}\ \bibnamefont
  {Pervaiz}}, \bibinfo {author} {\bibfnamefont {S.}~\bibnamefont {Fukami}},
  \bibinfo {author} {\bibfnamefont {K.~Y.}\ \bibnamefont {Camsari}}, \bibinfo
  {author} {\bibfnamefont {H.}~\bibnamefont {Ohno}},\ and\ \bibinfo {author}
  {\bibfnamefont {S.}~\bibnamefont {Datta}},\ }\bibfield  {title} {\bibinfo
  {title} {Integer factorization using stochastic magnetic tunnel junctions},\
  }\href {https://doi.org/10.1038/s41586-019-1557-9} {\bibfield  {journal}
  {\bibinfo  {journal} {Nature}\ }\textbf {\bibinfo {volume} {573}},\ \bibinfo
  {pages} {390} (\bibinfo {year} {2019})}\BibitemShut {NoStop}%
\bibitem [{\citenamefont {Yan}\ \emph {et~al.}(2021)\citenamefont {Yan},
  \citenamefont {Jiang}, \citenamefont {Gao}, \citenamefont {Duan},
  \citenamefont {Wang},\ and\ \citenamefont {Ma}}]{yan2021adiabatic}%
  \BibitemOpen
  \bibfield  {author} {\bibinfo {author} {\bibfnamefont {B.}~\bibnamefont
  {Yan}}, \bibinfo {author} {\bibfnamefont {H.}~\bibnamefont {Jiang}}, \bibinfo
  {author} {\bibfnamefont {M.}~\bibnamefont {Gao}}, \bibinfo {author}
  {\bibfnamefont {Q.}~\bibnamefont {Duan}}, \bibinfo {author} {\bibfnamefont
  {H.}~\bibnamefont {Wang}},\ and\ \bibinfo {author} {\bibfnamefont
  {Z.}~\bibnamefont {Ma}},\ }\bibfield  {title} {\bibinfo {title} {Adiabatic
  quantum algorithm for factorization with growing minimum energy gap},\ }\href
  {https://doi.org/10.1002/que2.59} {\bibfield  {journal} {\bibinfo  {journal}
  {Quan. Eng.}\ }\textbf {\bibinfo {volume} {3}},\ \bibinfo {pages} {e59}
  (\bibinfo {year} {2021})}\BibitemShut {NoStop}%
\bibitem [{\citenamefont {Anschuetz}\ \emph {et~al.}(2019)\citenamefont
  {Anschuetz}, \citenamefont {Olson}, \citenamefont {Aspuru-Guzik},\ and\
  \citenamefont {Cao}}]{anschuetz2019variational}%
  \BibitemOpen
  \bibfield  {author} {\bibinfo {author} {\bibfnamefont {E.}~\bibnamefont
  {Anschuetz}}, \bibinfo {author} {\bibfnamefont {J.}~\bibnamefont {Olson}},
  \bibinfo {author} {\bibfnamefont {A.}~\bibnamefont {Aspuru-Guzik}},\ and\
  \bibinfo {author} {\bibfnamefont {Y.}~\bibnamefont {Cao}},\ }\bibfield
  {title} {\bibinfo {title} {Variational quantum factoring},\ }in\ \href
  {https://link.springer.com/chapter/10.1007/978-3-030-14082-3_7} {\emph
  {\bibinfo {booktitle} {Int. Worksh. on Quantum Technology and Optimization
  Problems}}}\ (\bibinfo {organization} {Springer},\ \bibinfo {year} {2019})\
  pp.\ \bibinfo {pages} {74--85}\BibitemShut {NoStop}%
\bibitem [{\citenamefont {Xu}\ \emph {et~al.}(2017)\citenamefont {Xu},
  \citenamefont {Xie}, \citenamefont {Li}, \citenamefont {Xu}, \citenamefont
  {Wang}, \citenamefont {Ye}, \citenamefont {Kong}, \citenamefont {Geng},
  \citenamefont {Duan}, \citenamefont {Shi} \emph
  {et~al.}}]{xu2017experimental}%
  \BibitemOpen
  \bibfield  {author} {\bibinfo {author} {\bibfnamefont {K.}~\bibnamefont
  {Xu}}, \bibinfo {author} {\bibfnamefont {T.}~\bibnamefont {Xie}}, \bibinfo
  {author} {\bibfnamefont {Z.}~\bibnamefont {Li}}, \bibinfo {author}
  {\bibfnamefont {X.}~\bibnamefont {Xu}}, \bibinfo {author} {\bibfnamefont
  {M.}~\bibnamefont {Wang}}, \bibinfo {author} {\bibfnamefont {X.}~\bibnamefont
  {Ye}}, \bibinfo {author} {\bibfnamefont {F.}~\bibnamefont {Kong}}, \bibinfo
  {author} {\bibfnamefont {J.}~\bibnamefont {Geng}}, \bibinfo {author}
  {\bibfnamefont {C.}~\bibnamefont {Duan}}, \bibinfo {author} {\bibfnamefont
  {F.}~\bibnamefont {Shi}}, \emph {et~al.},\ }\bibfield  {title} {\bibinfo
  {title} {Experimental adiabatic quantum factorization under ambient
  conditions based on a solid-state single spin system},\ }\href
  {https://doi.org/10.1103/PhysRevLett.118.130504} {\bibfield  {journal}
  {\bibinfo  {journal} {Phys. Rev. Lett.}\ }\textbf {\bibinfo {volume} {118}},\
  \bibinfo {pages} {130504} (\bibinfo {year} {2017})}\BibitemShut {NoStop}%
\bibitem [{\citenamefont {Jiang}\ \emph {et~al.}(2018)\citenamefont {Jiang},
  \citenamefont {Britt}, \citenamefont {McCaskey}, \citenamefont {Humble},\
  and\ \citenamefont {Kais}}]{jiang2018quantum}%
  \BibitemOpen
  \bibfield  {author} {\bibinfo {author} {\bibfnamefont {S.}~\bibnamefont
  {Jiang}}, \bibinfo {author} {\bibfnamefont {K.~A.}\ \bibnamefont {Britt}},
  \bibinfo {author} {\bibfnamefont {A.~J.}\ \bibnamefont {McCaskey}}, \bibinfo
  {author} {\bibfnamefont {T.~S.}\ \bibnamefont {Humble}},\ and\ \bibinfo
  {author} {\bibfnamefont {S.}~\bibnamefont {Kais}},\ }\bibfield  {title}
  {\bibinfo {title} {Quantum annealing for prime factorization},\ }\href
  {https://doi.org/10.1038/s41598-018-36058-z} {\bibfield  {journal} {\bibinfo
  {journal} {Sci. Rep.}\ }\textbf {\bibinfo {volume} {8}},\ \bibinfo {pages}
  {1} (\bibinfo {year} {2018})}\BibitemShut {NoStop}%
\bibitem [{\citenamefont {Li}\ \emph {et~al.}(2017)\citenamefont {Li},
  \citenamefont {Dattani}, \citenamefont {Chen}, \citenamefont {Liu},
  \citenamefont {Wang}, \citenamefont {Tanburn}, \citenamefont {Chen},
  \citenamefont {Peng},\ and\ \citenamefont {Du}}]{li2017high}%
  \BibitemOpen
  \bibfield  {author} {\bibinfo {author} {\bibfnamefont {Z.}~\bibnamefont
  {Li}}, \bibinfo {author} {\bibfnamefont {N.~S.}\ \bibnamefont {Dattani}},
  \bibinfo {author} {\bibfnamefont {X.}~\bibnamefont {Chen}}, \bibinfo {author}
  {\bibfnamefont {X.}~\bibnamefont {Liu}}, \bibinfo {author} {\bibfnamefont
  {H.}~\bibnamefont {Wang}}, \bibinfo {author} {\bibfnamefont {R.}~\bibnamefont
  {Tanburn}}, \bibinfo {author} {\bibfnamefont {H.}~\bibnamefont {Chen}},
  \bibinfo {author} {\bibfnamefont {X.}~\bibnamefont {Peng}},\ and\ \bibinfo
  {author} {\bibfnamefont {J.}~\bibnamefont {Du}},\ }\bibfield  {title}
  {\bibinfo {title} {High-fidelity adiabatic quantum computation using the
  intrinsic hamiltonian of a spin system: Application to the experimental
  factorization of 291311},\ }\href {https://arxiv.org/pdf/1706.08061.pdf}
  {\bibfield  {journal} {\bibinfo  {journal} {arXiv:1706.08061}\ } (\bibinfo
  {year} {2017})}\BibitemShut {NoStop}%
\bibitem [{\citenamefont {Karamlou}\ \emph {et~al.}(2021)\citenamefont
  {Karamlou}, \citenamefont {Simon}, \citenamefont {Katabarwa}, \citenamefont
  {Scholten}, \citenamefont {Peropadre},\ and\ \citenamefont
  {Cao}}]{karamlou2021analyzing}%
  \BibitemOpen
  \bibfield  {author} {\bibinfo {author} {\bibfnamefont {A.~H.}\ \bibnamefont
  {Karamlou}}, \bibinfo {author} {\bibfnamefont {W.~A.}\ \bibnamefont {Simon}},
  \bibinfo {author} {\bibfnamefont {A.}~\bibnamefont {Katabarwa}}, \bibinfo
  {author} {\bibfnamefont {T.~L.}\ \bibnamefont {Scholten}}, \bibinfo {author}
  {\bibfnamefont {B.}~\bibnamefont {Peropadre}},\ and\ \bibinfo {author}
  {\bibfnamefont {Y.}~\bibnamefont {Cao}},\ }\bibfield  {title} {\bibinfo
  {title} {Analyzing the performance of variational quantum factoring on a
  superconducting quantum processor},\ }\href
  {https://doi.org/10.1038/s41534-021-00478-z} {\bibfield  {journal} {\bibinfo
  {journal} {npj Quantum Inf.}\ }\textbf {\bibinfo {volume} {7}},\ \bibinfo
  {pages} {1} (\bibinfo {year} {2021})}\BibitemShut {NoStop}%
\bibitem [{\citenamefont {Mosca}\ and\ \citenamefont
  {Verschoor}(2022)}]{mosca2022factoring}%
  \BibitemOpen
  \bibfield  {author} {\bibinfo {author} {\bibfnamefont {M.}~\bibnamefont
  {Mosca}}\ and\ \bibinfo {author} {\bibfnamefont {S.~R.}\ \bibnamefont
  {Verschoor}},\ }\bibfield  {title} {\bibinfo {title} {Factoring semi-primes
  with (quantum) {SAT}-solvers},\ }\href
  {https://doi.org/10.1038/s41598-022-11687-7} {\bibfield  {journal} {\bibinfo
  {journal} {Sci. Rep.}\ }\textbf {\bibinfo {volume} {12}},\ \bibinfo {pages}
  {1} (\bibinfo {year} {2022})}\BibitemShut {NoStop}%
\bibitem [{\citenamefont {Schnorr}(2013)}]{schnorr2013factoring}%
  \BibitemOpen
  \bibfield  {author} {\bibinfo {author} {\bibfnamefont {C.~P.}\ \bibnamefont
  {Schnorr}},\ }\bibfield  {title} {\bibinfo {title} {Factoring integers by
  {CVP} algorithms},\ }in\ \href {https://doi.org/10.1007/978-3-642-42001-6_6}
  {\emph {\bibinfo {booktitle} {Number Theory and Cryptography}}}\ (\bibinfo
  {publisher} {Springer},\ \bibinfo {year} {2013})\ pp.\ \bibinfo {pages}
  {73--93}\BibitemShut {NoStop}%
\bibitem [{\citenamefont {Schnorr}(2021)}]{schnorr2021fast}%
  \BibitemOpen
  \bibfield  {author} {\bibinfo {author} {\bibfnamefont {C.~P.}\ \bibnamefont
  {Schnorr}},\ }\bibfield  {title} {\bibinfo {title} {Fast factoring integers
  by {SVP} algorithms, corrected},\ }\href {https://eprint.iacr.org/2021/933}
  {\bibfield  {journal} {\bibinfo  {journal} {Cryptology ePrint Archive}\ }
  (\bibinfo {year} {2021})}\BibitemShut {NoStop}%
\bibitem [{se()}]{se}%
  \BibitemOpen
  \bibinfo {title} {See supplementary materials}\BibitemShut {NoStop}%
\bibitem [{\citenamefont {Babai}(1986)}]{babai1986lovasz}%
  \BibitemOpen
\bibfield  {title} {  }\bibfield  {author} {\bibinfo {author} {\bibfnamefont
  {L.}~\bibnamefont {Babai}},\ }\bibfield  {title} {\bibinfo {title} {On
  lov{\'a}sz'lattice reduction and the nearest lattice point problem},\ }\href
  {https://doi.org/10.1007/BF02579403} {\bibfield  {journal} {\bibinfo
  {journal} {Combinatorica}\ }\textbf {\bibinfo {volume} {6}},\ \bibinfo
  {pages} {1} (\bibinfo {year} {1986})}\BibitemShut {NoStop}%
\bibitem [{\citenamefont {Micciancio}(2001)}]{micciancio2001hardness}%
  \BibitemOpen
  \bibfield  {author} {\bibinfo {author} {\bibfnamefont {D.}~\bibnamefont
  {Micciancio}},\ }\bibfield  {title} {\bibinfo {title} {The hardness of the
  closest vector problem with preprocessing},\ }\href
  {https://ieeexplore.ieee.org/document/915688/} {\bibfield  {journal}
  {\bibinfo  {journal} {IEEE Trans. Inf. Theory}\ }\textbf {\bibinfo {volume}
  {47}},\ \bibinfo {pages} {1212} (\bibinfo {year} {2001})}\BibitemShut
  {NoStop}%
\bibitem [{\citenamefont {Zhang}\ \emph {et~al.}(2022)\citenamefont {Zhang},
  \citenamefont {Jiang}, \citenamefont {Deng}, \citenamefont {Wang},
  \citenamefont {Chen}, \citenamefont {Zhang}, \citenamefont {Ren},
  \citenamefont {Dong}, \citenamefont {Xu}, \citenamefont {Gao} \emph
  {et~al.}}]{zhang2022digital}%
  \BibitemOpen
  \bibfield  {author} {\bibinfo {author} {\bibfnamefont {X.}~\bibnamefont
  {Zhang}}, \bibinfo {author} {\bibfnamefont {W.}~\bibnamefont {Jiang}},
  \bibinfo {author} {\bibfnamefont {J.}~\bibnamefont {Deng}}, \bibinfo {author}
  {\bibfnamefont {K.}~\bibnamefont {Wang}}, \bibinfo {author} {\bibfnamefont
  {J.}~\bibnamefont {Chen}}, \bibinfo {author} {\bibfnamefont {P.}~\bibnamefont
  {Zhang}}, \bibinfo {author} {\bibfnamefont {W.}~\bibnamefont {Ren}}, \bibinfo
  {author} {\bibfnamefont {H.}~\bibnamefont {Dong}}, \bibinfo {author}
  {\bibfnamefont {S.}~\bibnamefont {Xu}}, \bibinfo {author} {\bibfnamefont
  {Y.}~\bibnamefont {Gao}}, \emph {et~al.},\ }\bibfield  {title} {\bibinfo
  {title} {Digital quantum simulation of {Floquet} symmetry-protected
  topological phases},\ }\href {https://doi.org/10.1038/s41586-022-04854-3}
  {\bibfield  {journal} {\bibinfo  {journal} {Nature}\ }\textbf {\bibinfo
  {volume} {607}},\ \bibinfo {pages} {468} (\bibinfo {year}
  {2022})}\BibitemShut {NoStop}%
\bibitem [{\citenamefont {Lenstra}\ \emph {et~al.}(1982)\citenamefont
  {Lenstra}, \citenamefont {Lenstra},\ and\ \citenamefont
  {Lov{\'a}sz}}]{lenstra1982factoring}%
  \BibitemOpen
  \bibfield  {author} {\bibinfo {author} {\bibfnamefont {A.~K.}\ \bibnamefont
  {Lenstra}}, \bibinfo {author} {\bibfnamefont {H.~W.}\ \bibnamefont
  {Lenstra}},\ and\ \bibinfo {author} {\bibnamefont {Lov{\'a}sz}},\ }\bibfield
  {title} {\bibinfo {title} {Factoring polynomials with rational
  coefficients},\ }\href {https://doi.org/10.1007/BF01457454} {\bibfield
  {journal} {\bibinfo  {journal} {Math. Ann}\ }\textbf {\bibinfo {volume}
  {261}},\ \bibinfo {pages} {515} (\bibinfo {year} {1982})}\BibitemShut
  {NoStop}%
\bibitem [{\citenamefont {Ajtai}\ \emph {et~al.}(2001)\citenamefont {Ajtai},
  \citenamefont {Kumar},\ and\ \citenamefont {Sivakumar}}]{ajtai2001sieve}%
  \BibitemOpen
  \bibfield  {author} {\bibinfo {author} {\bibfnamefont {M.}~\bibnamefont
  {Ajtai}}, \bibinfo {author} {\bibfnamefont {R.}~\bibnamefont {Kumar}},\ and\
  \bibinfo {author} {\bibfnamefont {D.}~\bibnamefont {Sivakumar}},\ }\bibfield
  {title} {\bibinfo {title} {A sieve algorithm for the shortest lattice vector
  problem},\ }in\ \href {https://doi.org/10.1145/380752.380857} {\emph
  {\bibinfo {booktitle} {STOC '01}}}\ (\bibinfo {year} {2001})\ pp.\ \bibinfo
  {pages} {601--610}\BibitemShut {NoStop}%
\bibitem [{\citenamefont {Schnorr}\ and\ \citenamefont
  {Euchner}(1994)}]{schnorr1994lattice}%
  \BibitemOpen
  \bibfield  {author} {\bibinfo {author} {\bibfnamefont {C.-P.}\ \bibnamefont
  {Schnorr}}\ and\ \bibinfo {author} {\bibfnamefont {M.}~\bibnamefont
  {Euchner}},\ }\bibfield  {title} {\bibinfo {title} {Lattice basis reduction:
  Improved practical algorithms and solving subset sum problems},\ }\href
  {https://doi.org/10.1007/BF01581144} {\bibfield  {journal} {\bibinfo
  {journal} {Math Program}\ }\textbf {\bibinfo {volume} {66}},\ \bibinfo
  {pages} {181} (\bibinfo {year} {1994})}\BibitemShut {NoStop}%
\bibitem [{\citenamefont {Fincke}\ and\ \citenamefont
  {Pohst}(1985)}]{fincke1985improved}%
  \BibitemOpen
  \bibfield  {author} {\bibinfo {author} {\bibfnamefont {U.}~\bibnamefont
  {Fincke}}\ and\ \bibinfo {author} {\bibfnamefont {M.}~\bibnamefont {Pohst}},\
  }\bibfield  {title} {\bibinfo {title} {Improved methods for calculating
  vectors of short length in a lattice, including a complexity analysis},\
  }\href {https://doi.org/10.1090/S0025-5718-1985-0777278-8} {\bibfield
  {journal} {\bibinfo  {journal} {Math. Comp}\ }\textbf {\bibinfo {volume}
  {44}},\ \bibinfo {pages} {463} (\bibinfo {year} {1985})}\BibitemShut
  {NoStop}%
\bibitem [{\citenamefont {Schnorr}\ and\ \citenamefont
  {H{\"o}rner}(1995)}]{schnorr1995attacking}%
  \BibitemOpen
  \bibfield  {author} {\bibinfo {author} {\bibfnamefont {C.-P.}\ \bibnamefont
  {Schnorr}}\ and\ \bibinfo {author} {\bibfnamefont {H.~H.}\ \bibnamefont
  {H{\"o}rner}},\ }\bibfield  {title} {\bibinfo {title} {Attacking the
  {C}hor-{R}ivest cryptosystem by improved lattice reduction},\ }in\ \href
  {https://doi.org/10.1007/3-540-49264-X_1} {\emph {\bibinfo {booktitle} {Proc.
  EUROCRYPT '95}}}\ (\bibinfo {organization} {Springer},\ \bibinfo {year}
  {1995})\ pp.\ \bibinfo {pages} {1--12}\BibitemShut {NoStop}%
\bibitem [{\citenamefont {Gama}\ \emph {et~al.}(2010)\citenamefont {Gama},
  \citenamefont {Nguyen},\ and\ \citenamefont {Regev}}]{gama2010lattice}%
  \BibitemOpen
  \bibfield  {author} {\bibinfo {author} {\bibfnamefont {N.}~\bibnamefont
  {Gama}}, \bibinfo {author} {\bibfnamefont {P.~Q.}\ \bibnamefont {Nguyen}},\
  and\ \bibinfo {author} {\bibfnamefont {O.}~\bibnamefont {Regev}},\ }\bibfield
   {title} {\bibinfo {title} {Lattice enumeration using extreme pruning},\ }in\
  \href {https://doi.org/10.1007/978-3-642-13190-5_13} {\emph {\bibinfo
  {booktitle} {Proc. EUROCRYPT '10}}}\ (\bibinfo {organization} {Springer},\
  \bibinfo {year} {2010})\ pp.\ \bibinfo {pages} {257--278}\BibitemShut
  {NoStop}%
\bibitem [{\citenamefont {Schnorr}(1991)}]{schnorr91factoring}%
  \BibitemOpen
  \bibfield  {author} {\bibinfo {author} {\bibfnamefont {C.}~\bibnamefont
  {Schnorr}},\ }\bibfield  {title} {\bibinfo {title} {Factoring integers and
  computing discrete logarithms via diophantine approximation},\ }in\ \href
  {https://doi.org/10.1007/3-540-46416-6_24} {\emph {\bibinfo {booktitle}
  {Proc. EUROCRYPT '91}}}\ (\bibinfo {year} {1991})\ pp.\ \bibinfo {pages}
  {281--293}\BibitemShut {NoStop}%
\bibitem [{\citenamefont {Cassels}(2012)}]{cassels2012introduction}%
  \BibitemOpen
  \bibfield  {author} {\bibinfo {author} {\bibfnamefont {J.~W.~S.}\
  \bibnamefont {Cassels}},\ }\href
  {https://sc.panda321.com/extdomains/books.google.com} {\emph {\bibinfo
  {title} {An introduction to the geometry of numbers}}}\ (\bibinfo
  {publisher} {Springer Science \& Business Media},\ \bibinfo {year}
  {2012})\BibitemShut {NoStop}%
\bibitem [{\citenamefont {Kabatiansky}\ and\ \citenamefont
  {Levenshtein}(1978)}]{kabatiansky1978bounds}%
  \BibitemOpen
  \bibfield  {author} {\bibinfo {author} {\bibfnamefont {G.~A.}\ \bibnamefont
  {Kabatiansky}}\ and\ \bibinfo {author} {\bibfnamefont {V.~I.}\ \bibnamefont
  {Levenshtein}},\ }\bibfield  {title} {\bibinfo {title} {On bounds for
  packings on a sphere and in space},\ }\href {http://mi.mathnet.ru/ppi1518}
  {\bibfield  {journal} {\bibinfo  {journal} {Probl. Peredachi Inf.}\ }\textbf
  {\bibinfo {volume} {14}},\ \bibinfo {pages} {3} (\bibinfo {year}
  {1978})}\BibitemShut {NoStop}%
\bibitem [{\citenamefont {Xu}\ \emph {et~al.}(2022)\citenamefont {Xu},
  \citenamefont {Sun}, \citenamefont {Wang}, \citenamefont {Xiang},
  \citenamefont {Bao}, \citenamefont {Zhu}, \citenamefont {Shen}, \citenamefont
  {Song}, \citenamefont {Zhang}, \citenamefont {Ren} \emph
  {et~al.}}]{xu2022digital}%
  \BibitemOpen
  \bibfield  {author} {\bibinfo {author} {\bibfnamefont {S.}~\bibnamefont
  {Xu}}, \bibinfo {author} {\bibfnamefont {Z.-Z.}\ \bibnamefont {Sun}},
  \bibinfo {author} {\bibfnamefont {K.}~\bibnamefont {Wang}}, \bibinfo {author}
  {\bibfnamefont {L.}~\bibnamefont {Xiang}}, \bibinfo {author} {\bibfnamefont
  {Z.}~\bibnamefont {Bao}}, \bibinfo {author} {\bibfnamefont {Z.}~\bibnamefont
  {Zhu}}, \bibinfo {author} {\bibfnamefont {F.}~\bibnamefont {Shen}}, \bibinfo
  {author} {\bibfnamefont {Z.}~\bibnamefont {Song}}, \bibinfo {author}
  {\bibfnamefont {P.}~\bibnamefont {Zhang}}, \bibinfo {author} {\bibfnamefont
  {W.}~\bibnamefont {Ren}}, \emph {et~al.},\ }\bibfield  {title} {\bibinfo
  {title} {Digital simulation of non-{Abelian} anyons with 68 programmable
  superconducting qubits},\ }\href {http://arxiv.org/abs/2211.09802} {\bibfield
   {journal} {\bibinfo  {journal} {arXiv:2211.09802}\ } (\bibinfo {year}
  {2022})}\BibitemShut {NoStop}%
\bibitem [{\citenamefont {Wang}\ \emph {et~al.}(2021)\citenamefont {Wang},
  \citenamefont {Chen}, \citenamefont {Song}, \citenamefont {Qin},
  \citenamefont {Li}, \citenamefont {Guo}, \citenamefont {Wang}, \citenamefont
  {Song},\ and\ \citenamefont {Li}}]{wang2021scalable}%
  \BibitemOpen
  \bibfield  {author} {\bibinfo {author} {\bibfnamefont {Z.}~\bibnamefont
  {Wang}}, \bibinfo {author} {\bibfnamefont {Y.}~\bibnamefont {Chen}}, \bibinfo
  {author} {\bibfnamefont {Z.}~\bibnamefont {Song}}, \bibinfo {author}
  {\bibfnamefont {D.}~\bibnamefont {Qin}}, \bibinfo {author} {\bibfnamefont
  {H.}~\bibnamefont {Li}}, \bibinfo {author} {\bibfnamefont {Q.}~\bibnamefont
  {Guo}}, \bibinfo {author} {\bibfnamefont {H.}~\bibnamefont {Wang}}, \bibinfo
  {author} {\bibfnamefont {C.}~\bibnamefont {Song}},\ and\ \bibinfo {author}
  {\bibfnamefont {Y.}~\bibnamefont {Li}},\ }\bibfield  {title} {\bibinfo
  {title} {Scalable evaluation of quantum-circuit error loss using clifford
  sampling},\ }\href {https://doi.org/10.1103/PhysRevLett.126.080501}
  {\bibfield  {journal} {\bibinfo  {journal} {Phys. Rev. Lett.}\ }\textbf
  {\bibinfo {volume} {126}},\ \bibinfo {pages} {080501} (\bibinfo {year}
  {2021})}\BibitemShut {NoStop}%
\bibitem [{\citenamefont {McKay}\ \emph {et~al.}(2017)\citenamefont {McKay},
  \citenamefont {Wood}, \citenamefont {Sheldon}, \citenamefont {Chow},\ and\
  \citenamefont {Gambetta}}]{Mckay2017VZ}%
  \BibitemOpen
  \bibfield  {author} {\bibinfo {author} {\bibfnamefont {D.~C.}\ \bibnamefont
  {McKay}}, \bibinfo {author} {\bibfnamefont {C.~J.}\ \bibnamefont {Wood}},
  \bibinfo {author} {\bibfnamefont {S.}~\bibnamefont {Sheldon}}, \bibinfo
  {author} {\bibfnamefont {J.~M.}\ \bibnamefont {Chow}},\ and\ \bibinfo
  {author} {\bibfnamefont {J.~M.}\ \bibnamefont {Gambetta}},\ }\bibfield
  {title} {\bibinfo {title} {Efficient $z$ gates for quantum computing},\
  }\href {https://doi.org/10.1103/PhysRevA.96.022330} {\bibfield  {journal}
  {\bibinfo  {journal} {Phys. Rev. A}\ }\textbf {\bibinfo {volume} {96}},\
  \bibinfo {pages} {022330} (\bibinfo {year} {2017})}\BibitemShut {NoStop}%
\bibitem [{\citenamefont {Ren}\ \emph {et~al.}(2022)\citenamefont {Ren},
  \citenamefont {Li}, \citenamefont {Xu}, \citenamefont {Wang}, \citenamefont
  {Jiang}, \citenamefont {Jin}, \citenamefont {Zhu}, \citenamefont {Chen},
  \citenamefont {Zhang}, \citenamefont {Dong} \emph
  {et~al.}}]{ren2022experimental}%
  \BibitemOpen
  \bibfield  {author} {\bibinfo {author} {\bibfnamefont {W.}~\bibnamefont
  {Ren}}, \bibinfo {author} {\bibfnamefont {W.}~\bibnamefont {Li}}, \bibinfo
  {author} {\bibfnamefont {S.}~\bibnamefont {Xu}}, \bibinfo {author}
  {\bibfnamefont {K.}~\bibnamefont {Wang}}, \bibinfo {author} {\bibfnamefont
  {W.}~\bibnamefont {Jiang}}, \bibinfo {author} {\bibfnamefont
  {F.}~\bibnamefont {Jin}}, \bibinfo {author} {\bibfnamefont {X.}~\bibnamefont
  {Zhu}}, \bibinfo {author} {\bibfnamefont {J.}~\bibnamefont {Chen}}, \bibinfo
  {author} {\bibfnamefont {P.}~\bibnamefont {Zhang}}, \bibinfo {author}
  {\bibfnamefont {H.}~\bibnamefont {Dong}}, \emph {et~al.},\ }\bibfield
  {title} {\bibinfo {title} {Experimental quantum adversarial learning with
  programmable superconducting qubits},\ }\href
  {https://arxiv.org/abs/2204.01738} {\bibfield  {journal} {\bibinfo  {journal}
  {arXiv:2204.01738}\ } (\bibinfo {year} {2022})}\BibitemShut {NoStop}%
\bibitem [{\citenamefont {Sung}\ \emph {et~al.}(2020)\citenamefont {Sung},
  \citenamefont {Yao}, \citenamefont {Harrigan}, \citenamefont {Rubin},
  \citenamefont {Jiang}, \citenamefont {Lin}, \citenamefont {Babbush},\ and\
  \citenamefont {McClean}}]{sung2020using}%
  \BibitemOpen
  \bibfield  {author} {\bibinfo {author} {\bibfnamefont {K.~J.}\ \bibnamefont
  {Sung}}, \bibinfo {author} {\bibfnamefont {J.}~\bibnamefont {Yao}}, \bibinfo
  {author} {\bibfnamefont {M.~P.}\ \bibnamefont {Harrigan}}, \bibinfo {author}
  {\bibfnamefont {N.~C.}\ \bibnamefont {Rubin}}, \bibinfo {author}
  {\bibfnamefont {Z.}~\bibnamefont {Jiang}}, \bibinfo {author} {\bibfnamefont
  {L.}~\bibnamefont {Lin}}, \bibinfo {author} {\bibfnamefont {R.}~\bibnamefont
  {Babbush}},\ and\ \bibinfo {author} {\bibfnamefont {J.~R.}\ \bibnamefont
  {McClean}},\ }\bibfield  {title} {\bibinfo {title} {Using models to improve
  optimizers for variational quantum algorithms},\ }\href
  {https://doi.org/10.1088/2058-9565/abb6d9} {\bibfield  {journal} {\bibinfo
  {journal} {Quantum Sci. Technol.}\ }\textbf {\bibinfo {volume} {5}},\
  \bibinfo {pages} {044008} (\bibinfo {year} {2020})}\BibitemShut {NoStop}%
\bibitem [{\citenamefont {Lagarias}\ \emph {et~al.}(1998)\citenamefont
  {Lagarias}, \citenamefont {Reeds}, \citenamefont {Wright},\ and\
  \citenamefont {Wright}}]{lagarias1998convergence}%
  \BibitemOpen
  \bibfield  {author} {\bibinfo {author} {\bibfnamefont {J.~C.}\ \bibnamefont
  {Lagarias}}, \bibinfo {author} {\bibfnamefont {J.~A.}\ \bibnamefont {Reeds}},
  \bibinfo {author} {\bibfnamefont {M.~H.}\ \bibnamefont {Wright}},\ and\
  \bibinfo {author} {\bibfnamefont {P.~E.}\ \bibnamefont {Wright}},\ }\bibfield
   {title} {\bibinfo {title} {Convergence properties of the {Nelder}--{Mead}
  simplex method in low dimensions},\ }\href
  {https://doi.org/10.1137/S1052623496303470} {\bibfield  {journal} {\bibinfo
  {journal} {SIAM J. Optim.}\ }\textbf {\bibinfo {volume} {9}},\ \bibinfo
  {pages} {112} (\bibinfo {year} {1998})}\BibitemShut {NoStop}%
\bibitem [{\citenamefont {Broyden}(1970)}]{broyden1970convergence}%
  \BibitemOpen
  \bibfield  {author} {\bibinfo {author} {\bibfnamefont {C.~G.}\ \bibnamefont
  {Broyden}},\ }\bibfield  {title} {\bibinfo {title} {The convergence of a
  class of double-rank minimization algorithms 1. general considerations},\
  }\href {https://doi.org/10.1093/imamat/6.1.76} {\bibfield  {journal}
  {\bibinfo  {journal} {IMA J Appl Math}\ }\textbf {\bibinfo {volume} {6}},\
  \bibinfo {pages} {76} (\bibinfo {year} {1970})}\BibitemShut {NoStop}%
\bibitem [{\citenamefont {Liu}\ and\ \citenamefont
  {Nocedal}(1989)}]{liu1989limited}%
  \BibitemOpen
  \bibfield  {author} {\bibinfo {author} {\bibfnamefont {D.~C.}\ \bibnamefont
  {Liu}}\ and\ \bibinfo {author} {\bibfnamefont {J.}~\bibnamefont {Nocedal}},\
  }\bibfield  {title} {\bibinfo {title} {On the limited memory {BFGS} method
  for large scale optimization},\ }\href {https://doi.org/10.1007/BF01589116}
  {\bibfield  {journal} {\bibinfo  {journal} {Math Program}\ }\textbf {\bibinfo
  {volume} {45}},\ \bibinfo {pages} {503} (\bibinfo {year} {1989})}\BibitemShut
  {NoStop}%
\bibitem [{\citenamefont {Pagano}\ \emph {et~al.}(2020)\citenamefont {Pagano},
  \citenamefont {Bapat}, \citenamefont {Becker}, \citenamefont {Collins},
  \citenamefont {De}, \citenamefont {Hess}, \citenamefont {Kaplan},
  \citenamefont {Kyprianidis}, \citenamefont {Tan}, \citenamefont {Baldwin}
  \emph {et~al.}}]{pagano2020quantum}%
  \BibitemOpen
  \bibfield  {author} {\bibinfo {author} {\bibfnamefont {G.}~\bibnamefont
  {Pagano}}, \bibinfo {author} {\bibfnamefont {A.}~\bibnamefont {Bapat}},
  \bibinfo {author} {\bibfnamefont {P.}~\bibnamefont {Becker}}, \bibinfo
  {author} {\bibfnamefont {K.~S.}\ \bibnamefont {Collins}}, \bibinfo {author}
  {\bibfnamefont {A.}~\bibnamefont {De}}, \bibinfo {author} {\bibfnamefont
  {P.~W.}\ \bibnamefont {Hess}}, \bibinfo {author} {\bibfnamefont {H.~B.}\
  \bibnamefont {Kaplan}}, \bibinfo {author} {\bibfnamefont {A.}~\bibnamefont
  {Kyprianidis}}, \bibinfo {author} {\bibfnamefont {W.~L.}\ \bibnamefont
  {Tan}}, \bibinfo {author} {\bibfnamefont {C.}~\bibnamefont {Baldwin}}, \emph
  {et~al.},\ }\bibfield  {title} {\bibinfo {title} {Quantum approximate
  optimization of the long-range {Ising} model with a trapped-ion quantum
  simulator},\ }\href {https://doi.org/10.1073/pnas.2006373117} {\bibfield
  {journal} {\bibinfo  {journal} {PNAS}\ }\textbf {\bibinfo {volume} {117}},\
  \bibinfo {pages} {25396} (\bibinfo {year} {2020})}\BibitemShut {NoStop}%
\bibitem [{\citenamefont {Takahashi}\ \emph {et~al.}(2007)\citenamefont
  {Takahashi}, \citenamefont {Kunihiro},\ and\ \citenamefont
  {Ohta}}]{takahashi2007quantum}%
  \BibitemOpen
  \bibfield  {author} {\bibinfo {author} {\bibfnamefont {Y.}~\bibnamefont
  {Takahashi}}, \bibinfo {author} {\bibfnamefont {N.}~\bibnamefont
  {Kunihiro}},\ and\ \bibinfo {author} {\bibfnamefont {K.}~\bibnamefont
  {Ohta}},\ }\bibfield  {title} {\bibinfo {title} {The quantum fourier
  transform on a linear nearest neighbor architecture},\ }\href
  {https://doi.org/10.5555/2011725.2011732} {\bibfield  {journal} {\bibinfo
  {journal} {Quantum Info. Comput.}\ }\textbf {\bibinfo {volume} {7}},\
  \bibinfo {pages} {383} (\bibinfo {year} {2007})}\BibitemShut {NoStop}%
\bibitem [{\citenamefont {Kutin}(2006)}]{kutin2006shor}%
  \BibitemOpen
  \bibfield  {author} {\bibinfo {author} {\bibfnamefont {S.~A.}\ \bibnamefont
  {Kutin}},\ }\bibfield  {title} {\bibinfo {title} {Shor's algorithm on a
  nearest-neighbor machine},\ }\href
  {https://doi.org/10.48550/arXiv.quant-ph/0609001} {\bibfield  {journal}
  {\bibinfo  {journal} {arXiv:quant-ph/0609001}\ } (\bibinfo {year}
  {2006})}\BibitemShut {NoStop}%
\bibitem [{\citenamefont {Cheung}\ \emph {et~al.}(2007)\citenamefont {Cheung},
  \citenamefont {Maslov},\ and\ \citenamefont
  {Severini}}]{cheung2007translation}%
  \BibitemOpen
  \bibfield  {author} {\bibinfo {author} {\bibfnamefont {D.}~\bibnamefont
  {Cheung}}, \bibinfo {author} {\bibfnamefont {D.}~\bibnamefont {Maslov}},\
  and\ \bibinfo {author} {\bibfnamefont {S.}~\bibnamefont {Severini}},\
  }\bibfield  {title} {\bibinfo {title} {Translation techniques between quantum
  circuit architectures},\ }in\ \href@noop {} {\emph {\bibinfo {booktitle}
  {Workshop on Quant. Inf. Proc.}}}\ (\bibinfo {organization} {Citeseer},\
  \bibinfo {year} {2007})\BibitemShut {NoStop}%
\bibitem [{\citenamefont {Hirata}\ \emph {et~al.}(2009)\citenamefont {Hirata},
  \citenamefont {Nakanishi}, \citenamefont {Yamashita},\ and\ \citenamefont
  {Nakashima}}]{hirata2009efficient}%
  \BibitemOpen
  \bibfield  {author} {\bibinfo {author} {\bibfnamefont {Y.}~\bibnamefont
  {Hirata}}, \bibinfo {author} {\bibfnamefont {M.}~\bibnamefont {Nakanishi}},
  \bibinfo {author} {\bibfnamefont {S.}~\bibnamefont {Yamashita}},\ and\
  \bibinfo {author} {\bibfnamefont {Y.}~\bibnamefont {Nakashima}},\ }\bibfield
  {title} {\bibinfo {title} {An efficient method to convert arbitrary quantum
  circuits to ones on a linear nearest neighbor architecture},\ }in\ \href
  {https://doi.org/10.1109/ICQNM.2009.25} {\emph {\bibinfo {booktitle} {ICQNM
  '09}}}\ (\bibinfo {organization} {IEEE},\ \bibinfo {year} {2009})\ pp.\
  \bibinfo {pages} {26--33}\BibitemShut {NoStop}%
\bibitem [{\citenamefont {Saeedi}\ \emph {et~al.}(2011)\citenamefont {Saeedi},
  \citenamefont {Wille},\ and\ \citenamefont
  {Drechsler}}]{saeedi2011synthesis}%
  \BibitemOpen
  \bibfield  {author} {\bibinfo {author} {\bibfnamefont {M.}~\bibnamefont
  {Saeedi}}, \bibinfo {author} {\bibfnamefont {R.}~\bibnamefont {Wille}},\ and\
  \bibinfo {author} {\bibfnamefont {R.}~\bibnamefont {Drechsler}},\ }\bibfield
  {title} {\bibinfo {title} {Synthesis of quantum circuits for linear nearest
  neighbor architectures},\ }\href {https://doi.org/10.1007/s11128-010-0201-2}
  {\bibfield  {journal} {\bibinfo  {journal} {Quantum Inf Process}\ }\textbf
  {\bibinfo {volume} {10}},\ \bibinfo {pages} {355} (\bibinfo {year}
  {2011})}\BibitemShut {NoStop}%
\bibitem [{\citenamefont {Wille}\ \emph {et~al.}(2016)\citenamefont {Wille},
  \citenamefont {Keszocze}, \citenamefont {Walter}, \citenamefont {Rohrs},
  \citenamefont {Chattopadhyay},\ and\ \citenamefont
  {Drechsler}}]{wille2016look}%
  \BibitemOpen
  \bibfield  {author} {\bibinfo {author} {\bibfnamefont {R.}~\bibnamefont
  {Wille}}, \bibinfo {author} {\bibfnamefont {O.}~\bibnamefont {Keszocze}},
  \bibinfo {author} {\bibfnamefont {M.}~\bibnamefont {Walter}}, \bibinfo
  {author} {\bibfnamefont {P.}~\bibnamefont {Rohrs}}, \bibinfo {author}
  {\bibfnamefont {A.}~\bibnamefont {Chattopadhyay}},\ and\ \bibinfo {author}
  {\bibfnamefont {R.}~\bibnamefont {Drechsler}},\ }\bibfield  {title} {\bibinfo
  {title} {Look-ahead schemes for nearest neighbor optimization of {1D} and
  {2D} quantum circuits},\ }in\ \href
  {https://doi.org/10.1109/ASPDAC.2016.7428026} {\emph {\bibinfo {booktitle}
  {ASP-DAC '16}}}\ (\bibinfo {organization} {IEEE},\ \bibinfo {year} {2016})\
  pp.\ \bibinfo {pages} {292--297}\BibitemShut {NoStop}%
\bibitem [{\citenamefont {Farghadan}\ and\ \citenamefont
  {Mohammadzadeh}(2017)}]{farghadan2017quantum}%
  \BibitemOpen
  \bibfield  {author} {\bibinfo {author} {\bibfnamefont {A.}~\bibnamefont
  {Farghadan}}\ and\ \bibinfo {author} {\bibfnamefont {N.}~\bibnamefont
  {Mohammadzadeh}},\ }\bibfield  {title} {\bibinfo {title} {Quantum circuit
  physical design flow for {2D} nearest-neighbor architectures},\ }\href
  {https://doi.org/10.1002/cta.2335} {\bibfield  {journal} {\bibinfo  {journal}
  {Int. J. Circ. Theor. Appl.}\ }\textbf {\bibinfo {volume} {45}},\ \bibinfo
  {pages} {989} (\bibinfo {year} {2017})}\BibitemShut {NoStop}%
\end{thebibliography}%


%

\vspace{.5cm}
\noindent\textbf{Acknowledgements:} 
We thank H.Fan, K.Xu and C.Chen for helpful discussions. The device was fabricated at the Micro-Nano Fabrication Center of Zhejiang University. The experiment was performed on the quantum computing platform at Zhejiang University. 

\noindent\textbf{Funding:} This research was supported by the National Natural Science Foundation of China (Grant Nos. U20A2076, 12274367, 12174342, 12005015, 61972413, 61901525, 11974205, 11774197), the Zhejiang Province Key Research and Development Program (Grant No. 2020C01019), the Fundamental Research Funds for the Central Universities (Grant No. 2022QZJH03), the National Key Research and  Development Program of China (2017YFA0303700), the Key Research and  Development Program of Guangdong province (2018B030325002). 

\noindent\textbf{Author contributions:} B.Y. proposed the SQIF algorithm and designed the experiment scheme. Z.T. and C.Z carried out the experiments and collected results under the supervision of Z.W.. J.C., X.Z. and F.J. designed the device, and H.L. fabricated the device supervised by H.W.. S.-J.W., H.W., Q.D. contributed to the theory and experiment design. H.J., W.W., L.L., W.S., Y.H. performed numerical simulations. Y.L., Y.F., X.M., Z.S. contributed to the depth analysis. Z.M. and G.-L.L. initiated and supervised this project. All authors contributed to the writing of the manuscript.

\noindent\textbf{Competing interests:}  All authors declare no competing interests.

\noindent\textbf{Data and materials availability:}
The data presented in the figures and that support the other findings of this study
will be publically available upon its publication.

\clearpage

\setcounter{secnumdepth}{3}

\makeatletter
\setcounter{figure}{0}
\setcounter{equation}{0}
\renewcommand{\thefigure}{S\@arabic\c@figure}
\renewcommand \theequation{S\@arabic\c@equation}
\renewcommand \thetable{S\@arabic\c@table}

\begin{center} 
	{\large \bf Supplementary material for  ``Factoring integers with sublinear resources on a superconducting quantum processor"}
\end{center}

\maketitle
\tableofcontents

\section{Background knowledge about lattice}\label{sec1}

In recent years,  lattices are used as algorithmic tools to solve a wide variety of problems in computer science, mathematics and cryptography, especially in quantum-resistant cryptography protocols. The following introduces some basic concepts and well-known algorithms in lattices that are closely related to our work.
\subsection{Basic concepts}

Let $\Vert \cdot \Vert$be the Euclidean norm of the vectors in $\mathbb{R}^{m}$. Vectors will be written in bold and we use row-representation for matrices. For a matrix $M$, we usually denote its coefficients by $m_{i,j}$. We also use superscript 'T' to represent the transpose of matrices or vectors. 
\begin{itemize}
\item{ \textbf{Lattice:}}
Let $\mathbf{b}_1,...,\mathbf{b}_n\in\mathbb{R}^{m}$ be a group of linearly independent column vectors, then we call the set  generated by the linear combination of its integer coefficients a lattice, denoted as
\begin{equation}
\begin{aligned}
\Lambda(B)&=\{B\mathbf{x}\mid\;\mathbf x \in \mathbb{Z}^n\}
\\&=\{\mathbf{b}=x_1\mathbf{b_1}+...+x_n\mathbf{ b_n}\mid\;x_1,...,x_n \in \mathbb{Z}\},
\end{aligned}
\end{equation}
  where ${B}=[\mathbf{b}_1,...,\mathbf{b}_n]\in\mathbb{R}^{m\times n}$ is called a basis matrix, which could also be used to represent a lattice for simplicity. $\{\mathbf{b}_1,...,\mathbf{b}_n\}$is a group of basis of lattice $\Lambda(B)$. The dimension of lattice $\Lambda$ is $n$. The determinant of  $\Lambda$ is  $\det \Lambda=(\det{{B^{\it T} B}})^{1/2}$, here ${B}^T$ is the transpose of ${B}$. For a square matrix ${B}$, it is directly $\det \Lambda=\det{ B}$. The determinant also represents the volume of the lattice in geometry perspective, denoted as $\text{vol}(\Lambda)$. The length of the lattice point $\mathbf{b}\in \mathbb{R}^m$ is defined as $\lVert\mathbf{b}\rVert=(\mathbf{b^{\it T} b})^{1/2}$. 

\item{\textbf{Successive minima:}}
The successive minima of an $n$-dimensional lattice $ \Lambda$ are the positive quantities $\lambda_1(\Lambda)\le\lambda_2(\Lambda)\le...\le\lambda_n(\Lambda)$, where $\lambda_k(\Lambda)$ is the smallest radius of a zero-centered ball containing $k$ linearly independent vectors of $\Lambda$.  Denote $\lambda_1=\lambda_1(\Lambda)$ as the length of the shortest nonzero vector of $ \Lambda$. 
\item{\textbf{Hermite's constant:}}
 The Hermite invariant of the lattice $\Lambda$ is defined by
\begin{equation}
\gamma(\Lambda)=\lambda_1^2(\Lambda)/\text{vol}(\Lambda)^{2/n}
=\lambda_1^2(\Lambda)/\text{det}(\Lambda)^{2/n}.
\end{equation}

Hermite's constant $\gamma_n$ is the maximal value $\gamma( \Lambda)$ over all $n$-dimensional lattices, or the minimal constant $\gamma$ which enables $\lambda_1( \Lambda)^2\le \gamma(\det \Lambda)^{2/n}$ satisfied for all $n$-dimensional lattices equivalently.
\item{ \textbf{${QR}$-decomposition:}}
The lattice basis matrix ${B}$ has the unique decomposition ${B}={QR}\in\mathbb{R}^{m\times n},{R}=[r_{i,j}]_{1\le i,j\le n}\in\mathbb{R}^{n\times n}$,here ${Q}\in \mathbb{R}^{m\times n}$ is isometric (with pairwise orthogonal column vectors of length 1) and  ${R}\in\mathbb{R}^{n\times n}$  is an upper-triangular matrix with positive diagonal entries  $r_{i,i}$ . The Gram-Schmidt (GS) coefficients $\mu_{j,i}=r_{i,j}/r_{i,i}$ can be obtained easily by the ${QR}$-decomposition. For an integer matrix ${B}$ , the GS coefficients are usually rational.
\item{\textbf{Shortest Vector Problem (SVP):}}{
Given a group of basis ${B}$ of a lattice $\Lambda$,
\subitem{Shortest Vector Problem (SVP):} Find a vector $\mathbf v \in\Lambda$, such that $\lVert\mathbf v\rVert=\lambda_1(\Lambda)$.
\subitem{Approximate Shortest Vector Problem (${\alpha}$-SVP):} Find a nonzero vector $\mathbf v \in\Lambda$, such that $\lVert\mathbf v\rVert\le\alpha\cdot\lambda_1(\Lambda)$.

\subitem{Hermite Shortest Vector Problem ($r$-Hermite SVP):} Find a nonzero vector $\mathbf v \in\Lambda$, such that $\lVert\mathbf v\rVert\le r\cdot\det(\Lambda)^{1/n}$.
}

The parameter $\alpha\ge1$ in ${\alpha}$-SVP is called the approximation factor. Usually, the problem becomes easier when $\alpha$ gets bigger. When $\alpha =1$, ${\alpha}$-SVP and SVP are the same problem.  The real value of $\lambda_1$ in  $\alpha $-SVP is hard to obtain because of the hardness of SVP. Thus the solution of ${\alpha}$-SVP is hard to check in some cases. The problem $r$-Hermite SVP is defined by  a computable (ralatively easy to compute) value $\text{det}(\Lambda)^{1/n}$ instead of  $\lambda_1$ to qualify the solution. As a result, we can check the solution easily but lack a comparison with the shortest vector.

\item{ \textbf{Closest Vector Problem (CVP):}}{
Given a group of basis ${B}$ of a lattice $\Lambda$, and a target vector $\mathbf t\in \text{span}(B)$, 
\subitem{Closest Vector Problem (CVP): Find a vector $\mathbf v \in\Lambda$, such that the distance $\lVert\mathbf {v-t}\rVert$ could be minimized, namely $\lVert\mathbf {v-t}\rVert=\text{dist}(\Lambda,\mathbf t)$.}
\subitem{$\alpha$-Approximate Closest Vector Problem ($\alpha$-CVP): Find a vector $\mathbf v \in\Lambda$, such that the distance $\lVert\mathbf {v-t}\rVert\le \alpha\cdot \text{dist}(\Lambda,\mathbf t)$.}
\subitem{$r$-Approximate Closest Vector Problem ($r$-$Abs$CVP): Find a vector $\mathbf v \in\Lambda$, such that the distance $\lVert\mathbf {v-t}\rVert\le r$.}}

Here the problem definitions are similar to those in SVP, the role of parameter $\alpha\ge1$ in $\alpha$-CVP is the same as $\alpha$-SVP. In $r$-${Abs}$CVP, the parameter $r$ can be any reasonable value which is comparable to $\text{dist}(\Lambda,\mathbf t)$, such like $\det(\Lambda)^{1/n}$ in $r$-Hermite SVP.
\end{itemize}
\subsection{ LLL algorithm}
The LLL algorithm is one of the most famous algorithms in the field of lattice reduction, proposed by A. K. Lenstra, H. W. Lenstra, Jr., and L. Lovasz in 1982~\cite{lenstra1982factoring}.  For an $n$-dimensional lattice, the algorithm can be used to solve the  $\it{\alpha}$-SVP with $\alpha=(\frac{2}{\sqrt 3})^{n} $ in polynomial time. The related concepts and algorithms are as follows.
\begin{itemize}
\item{ \textbf{LLL basis:}} A basis ${B=QR}$ is called LLL-reduced or a LLL basis, given LLL-reduction parameter $\delta\in(\frac{1}{4},1]$, if it satisfies:

  i. $\mid r_{i,j}\mid /r_{i,i}\le \frac{1}{2}$, for all $j>i$;     

  ii. $\delta r_{i,i}^2\le r_{i,i+1}^2+r_{i+1,i+1}^2$ , for $i=1,..,n-1$.

  Obviously, LLL basis also satisfies $r_{i,i}^2\le\alpha r_{i+1,i+1}^2$, for $\alpha=1/(\delta-\frac{1}{4})$. 

  The parameters considered in the original literature of the LLL algorithm are $\delta=3/4, \; \alpha=2$. A well-known result about LLL basis shows that for any $\delta<1$, LLL basis can be obtained in polynomial time and that they nicely approximate the successive minima : 

  iii. $\alpha^{-i+1}\le \lVert \mathbf{b}_i\rVert^2\lambda_i^{-2}\le\alpha^{n-1}$,  for $i=1,...,n$;

  iv. $\lVert\mathbf{b}_1\rVert^2\le\alpha^{\frac{n-1}{2}}(\det \Lambda)^{2/n}$.
\end{itemize}
\begin{itemize}
\item{\textbf{LLL algorithm:}}
 Given a group of basis ${B}=[\mathbf{b}_1,...,\mathbf{b}_n]\in\mathbb{Z}^{m\times n}$, the algorithm can make it LLL-reduced or convert it into a LLL basis. The algorithm consists of three main steps: Gram-Schmidt orthogonalization, reduction, and swap. The specific steps can be found in Algorithm \ref{LLL}.
\end{itemize}

\begin{algorithm} 
\SetAlgoNoLine
\caption{LLL-reduction algorithm}\label{LLL}
\KwIn{lattice basis $\mathbf{b}_1,...,\mathbf{b}_n\in\mathbb{Z}^{m}$, parameter $\delta$ }
\KwOut{$\delta$-LLL-reduced basis}

1.Gram-Schmidt orthogonalization\\
Imply the Gram-Schmidt orthogonalization to basis $\mathbf{b}_1,...,\mathbf{b}_n$, denote the results as: $\mathbf{\tilde b}_1,...,\mathbf{\tilde b}_n\in\mathbb{R}^{m}$.\\
2.Reduction step\\
    \For{i from 2 to $n$ } {          
        \For{j from i-1 to 1}{
        
                $\mathbf b_i\gets \mathbf b_i-c_{i,j}\mathbf b_j$, where $c_{i,j}=\lceil \langle \mathbf b_i,\mathbf{\tilde b}_j\rangle\langle \mathbf{\tilde b}_j,\mathbf{\tilde b}_j\rangle\rfloor.$ 
        }
    }
3.Swap step\\
      \If {$\exists \;i$  s. t.  $\delta\lVert\mathbf{\tilde b}_i\rVert^2>\lVert\mu_{i+1,i}\mathbf{\tilde b}_i+\mathbf{\tilde b}_{i+1}\rVert^2$ }{
      
            $\mathbf{b}_i\leftrightarrow \mathbf{b}_{i+1}$,\\
            
             go to 1.
      }
4.Output $\mathbf{b}_1,...,\mathbf{b}_n$.

\end{algorithm} 

\subsection{ Babai's nearest plane algorithm}

Babai's nearest plane algorithm~\cite{babai1986lovasz} (Babai's algorithm for short) can be used to solve CVP.  For an $n$-dimensional lattice, the algorithm can obtain an approximation factor of $\alpha=2(\frac{2}{\sqrt{3}})^n$ for $\it{\alpha}$-CVP. The algorithm  consists of two steps, the first is to reduce the input lattice basis with the LLL algorithm. The second is a size reduction procedure, which mainly calculates the linear combination of integer coefficients closest to the target vector $\mathbf t$ under the LLL basis. This step is essentially the same as the second step in LLL reduction. The specific steps of the algorithm can be found in Algorithm \ref{Babai's}.

\begin{algorithm}  
\SetAlgoNoLine
\caption{Babai's  algorithm} \label{Babai's} 
 
\KwIn{ lattice basis $\mathbf{b}_1,...,\mathbf{b}_n\in\mathbb{Z}^{m}$, parameter $\delta=3/4$ and target  $\mathbf{t}\in \mathbb{Z}^m$}
\KwOut{ a vector $\mathbf{x}\in \Lambda(B)$, such that $\lVert \mathbf{x-t}\rVert\le 2^{\frac{n}{2}}\text{dist}(\mathbf{t},\Lambda(B))$\\}
1. LLL reduction\\
Apply the LLL reduction on basis ${B}$ with parameter $\delta$. Denote the results as $\mathbf{\tilde b}_1,...,\mathbf{\tilde b}_n\in\mathbb{R}^{m}$.\\

2.Size reduction\\
$\mathbf{b}\gets \mathbf{t}$\\
    \For{j from $n$ to 1 }{
    
        $\mathbf{b}\gets \mathbf{b}-c_j\mathbf{b}_j$, where $c_j=\lceil\langle \mathbf{b,\tilde b}_j\rangle/\langle\mathbf{\tilde b}_j,\mathbf{\tilde b}_j\rangle\rfloor$.
    }
3.Output $\mathbf{t-b}$.

\end{algorithm}  

\section{Schnorr's integer factoring algorithm}
\subsection{Schnorr's sieve method}
Consider a general integer factoring situation in which the integer to be factored into two non-trivial factors, namely given $N$, finding the factors $p, q\;(p<q)$ such that $N=p\times q$. The sieve method to factor an integer firstly needs to define the smooth relation pair. Let $p_{i},i=1,...,n$ be the first $n$ primes together with $p_0$ which satisfy $-1=p_0<1<p_1<...<p_n<p$. The set $P=\{p_i\}_{i=0,...,n}$ is called a prime basis. The $p_0=-1$  is not a prime, nevertheless, it is included to characterize the sign of an integer.  An integer is called $p_n$-smooth if all of its prime factors are less than $p_n$, here $p_n$ is also called the smooth bound. The integer pair $(u_j,v_j)$ is called $p_n$-smooth pair, if both  $u_j$ and $v_j$ are $p_n$-smooth. Further more, a pair of integers $(u_j,v_j)$ is called $p_n$-smooth relation pair (abbreviate as sr-pair), if:
\begin{equation}\label{relation}
u_j=\prod_{i=1}^{n}{p_i^{e_{i,j}}},\;u_j-v_jN=\prod_{i=0}^{n}{p_i^{e_{i,j}'}},
\end{equation}
where $e_{i,j},e_{i,j}'\in\mathbb{N}$, then we have
\begin{equation}
(u_j-v_jN)/u_j\equiv\prod_{i=0}^n{p_i^{e_{i,j}'-e_{i,j}}}\equiv1\quad \text{mod}N.
\end{equation}

It should be noted  that the smooth pair is different with sr-pair in which the sr-pair not only need to be smooth, but also to meet more severe conditions in Eq.~\ref{relation}.
Let $S=\{(u_j,v_j)\}_{j=1,...,n+1}$  be a set with $n+1$ sr-pairs. If there exists a group of coefficients $t_1,...,t_{n+1}\in\{0,1\}$, such that
\begin{equation}
\sum_{j=1}^{n+1}{t_j({e_{i,j}'-e_{i,j}})}\equiv0\quad\text{mod}\;2,i=0,1,...,n.
\end{equation}
Denote $X=\prod_{i=0}^n{p_i^{\frac{1}{2}\sum_{j=1}^{n+1}{t_j}({e_{i,j}'-e_{i,j}})}}$, then we have
\begin{equation}
X^2-1=(X+1)(X-1)\equiv0\quad \text{mod}N.
\end{equation}
If $X\not\equiv\pm1\quad\text{mod}N$, then we'll obtain a nontrivial factor of $N$ by $\text{gcd}(X\pm1,N)$.

Since the dimension of the linear equation system is $O(n)$, and it can be solved within $O(n^3)$ operations. We neglect this minor part of the workload for factoring $N$. Hence the factoring problem is reduced to the sr-pair problem. This problem will be transformed into the closest vector problem on a lattice in the following part.

\subsection{The construction of the lattice and target vector} 
The sr-pairs will be obtained from the approximate solution of CVP in Schnorr's algorithm. We first introduce the construction of the prime lattice $\Lambda({B}_{n,c})$ and the target vector $\mathbf{t}\in\mathbb{R}^{n+1}$, here $c>0$ is an adjustable parameter. The matrix form of the lattice ${B}_{n,c}=[\mathbf{b}_1,...,\mathbf{b}_n]\in\mathbb{R}^{{(n+1)}\times n}$ can be constructed as
\begin{equation}\label{construction}
{B}_{n,c}=\begin{pmatrix}
 f(1)& 0 &...&0 \\
0 &f(2)&...&0\\
\vdots &\vdots &\ddots &\vdots\\
0 & 0 & ...&f(n)\\
N^c\text{ln}p_1 &N^c\text{ln}p_2 &...&N^c\text{ln}p_n
\end{pmatrix},
\quad \mathbf{t}=\begin{pmatrix}0\\\vdots\\0\\N^c\text{ln}N
\end{pmatrix},
\end{equation}
where the functions $f(i)$ for $i=1,...,n$ are the random permutations of diagonal elements $(\sqrt{\text{ln}p_1}, \sqrt{\text{ln}p_2},...,\sqrt{\text{ln}p_n})$.

A lattice point or vector can be represented by the integer combination of the lattice basis as $\mathbf{b}=\sum_{i=1}^n{e_i\mathbf{b}_i\in\Lambda({B}_{n,c})}$, here $e_i\in \mathbb Z$ for $i=1,...,n$.  In the following, we'll assume $(u,v)$ is $p_n$-smooth and $\text{gcd}(u,v)=1$. Then $u,v$ can be represented by the product of primes on the prime basis, namely:
\begin{equation}
u=\prod_{e_i>0}{p_i}^{e_i},\quad v=\prod_{e_i<0}p_i^{-e_i}.
\end{equation}
Under this representation, the smooth pair $(u,v)$ corresponds to the vector $\mathbf{b}=(e_1,...,e_n)$ in the lattice one-to-one, denoted as $\mathbf{b}\sim(u,v)$. Therefore, a vector on a lattice encodes a smooth pair.  

The closest vector problem (CVP) is to find a vector $\mathbf{b}_0\in \Lambda({B}_{n,c})$ which is closest to the target vector $\mathbf t$, mathematically expressed as
\begin{equation}
\mathbf b_0=\text{arg}\;\mathop{\min_{\mathbf b\in\Lambda}}\lVert\mathbf b-\mathbf t\rVert.
\end{equation}
According to the above definition, the following relationship  is established
\begin{equation}\label{CVP}
\|\mathbf b-\mathbf t\|^2\ge\text{ln}(uv)+N^{2c}\mid\text{ln}\frac{u}{vN}\mid^2.
\end{equation}
The equation is established if and only if $e_i\in\{-1,0,1\}$, that is, $u\;, v$ do not contain square factors. The constant $N^{2c}$ acts as a "weight" which is controlled by adjusting  the parameter $c$. When $N^{2c}>>\text{ln}(uv)$, the body of the equation is $N^{2c}\mid\text{ln}\frac{u}{vN}\mid^2$. Hence the quality $\mid\text{ln}\frac{u}{vN}\mid^2$ , or further on, $\mid u-vN \mid$ can be effected by parameter $c$, which is also called precision parameter. According to the inequality~\ref{CVP}, we can find that the shorter the length of distance vector $\mathbf b-\mathbf t$ , the smaller $\mid u-vN \mid$ could be, hence the higher probability for $(u,v)$ being an sr-pair. Further discussion about this relationship can be found in the next part of this Material. 

\subsection{Solving the CVP}
There are mainly two well-studied approaches to solve CVP or approximate CVP. One is based on the sieve method which is firstly proposed by Ajtai et al. in 2001~\cite{ajtai2001sieve}. The other is based on Babai's algorithm, in which a lattice reduction method such as LLL algorithm is firstly implemented to obtain a group of relatively short basis, then apply the size-reduction procedure to get the approximate closest vector solution. Schnorr adopted the latter approach to solve CVP. In fact, some superior lattice reduction methods such as BKZ~\cite{schnorr1994lattice}, HKZ, ENUM~\cite{fincke1985improved,schnorr1994lattice,schnorr1995attacking,gama2010lattice} and so on, are involved to get a better efficiency of the algorithm. However, these methods are too complicated and need more professional knowledge which is out of the scope of this paper. We adopt the LLL lattice reduction algorithm when we mention Babai's algorithm in the following part (and in the main text), which is simple and relatively easy to understand. Besides the principle of quantum enhancement of Babai's algorithm is general for any of the lattice reduction algorithm.

\section{The sublinear scheme about lattice dimension}
\subsection{The history results}
In this section, we discuss the dimension selection of lattices in Schnorr's algorithm. The dimension $n$ of the lattice depends on the size of the prime basis, meantime has an important influence on the efficiency of the algorithm. On the one hand, the number of smooth relation pairs on the prime basis will increase greatly when $n$ is large, which is more conducive to obtaining smooth relation pairs. On the other hand, $n$ cannot be too large, because the time complexity of the lattice reduction process and the linear equations solving procedure is positively correlated with $n$. Choosing an appropriate $n$ requires a balance between the two facts. This issue is not clearly explained by Schnorr in the original text~\cite{schnorr91factoring,schnorr2013factoring,schnorr2021fast}, and there are different descriptions or applications in different places. In Schnorr's near edition in 2021~\cite{schnorr2021fast}, when analyzing specific examples, a sub-linear magnitude of lattice dimension is used, but the author does not explain the choice of the lattice dimension scheme. For example, when discussing the factoring of a 400-bit integer, the lattice dimension is 48, which is close to the sublinear scheme $400/\text{log}_2400\sim 46$. In many other works, however, the lattice dimension $n$ is usually assumed to be polynomial order of the binary length $m$ of a large integer $N$. The specific description is given based on the restriction of the smooth bound $p_n$. In Schnorr's sieve method, it is usually assumed that the smooth bound $p_n$ satisfies
\begin{equation}\label{bound}
p_n\approx (\text{log}N)^{\alpha}=m^{\alpha},\;\alpha>0.
\end{equation}
According to the prime number theorem, we have
\begin{equation}
n\approx \frac{(\text{log}N)^{\alpha}}{\alpha\text{loglog}N}=m^{\alpha}/\alpha\text{log}m.
\end{equation}
When taking $\alpha=1$,  the dimension is
\begin{equation}
n=m/\text{log}m,
\end{equation}
which is a sublinear scale of the bit length of $N$. When $\alpha>1$, $n$ is typically polynomial scale of $m$. Therefore, the specific value of  $\alpha$ determines the dimension of the lattice.

The value of $\alpha$ is mainly determined by the mathematical relationship between the short vector and the smooth relation pair. Regarding what conditions short vectors satisfy to obtain smooth relation pairs, Schnorr gives the following lemma:

\begin{lemma}\label{lama1}
 If $\Vert\mathbf b-\mathbf t\Vert^2=O(\text{log}N)$ and $v\le N^{c-1}p_n(n/\text{log}N)^{1/2}$, then most likely $\mid u-vN\mid=O(p_n)$.
\end{lemma}

Here $c$ is the precision parameter. The lemma answers that when the square norm of a short vector is  $O(\text{log}N)$, then most likely the sr-pairs can be obtained. Here we set the short vector length $O(\text{log}N)$  as a theoretical bound.

The next important question is whether short vectors  satisfying  this condition exist, or whether there are enough of them.  Schnorr proved that there will be a large number of short vectors that satisfy the theoretical bound when $\alpha>2$. Specifically, the size of $\alpha$ is proportional to the size of the smooth bound according to the Eq.~\ref{bound}. In the sieve method, the larger the smooth bound $p_n$ is, the easier it is to obtain smooth relation pairs. However, the number of smooth relation pairs required as whole increases accordingly. Schnorr pointed out that there will be a large number of short vectors that can generate smooth relation pairs according to the density polynomial of smooth numbers when $\alpha>(2c-1)/(c-1)>2$~\cite{schnorr91factoring,schnorr2013factoring,schnorr2021fast}, which leads to a polynomial dimension scheme. 

We discuss the relationship between the short vector and the smooth relation pair based on the former. That is, to discuss the condition that $\alpha$ or the dimension $n$ of the lattice needs to satisfy from the perspective of the existence of the short vector. We first give a linear scheme of the lattice dimension $n$ under Minkowski's first theorem~\cite{cassels2012introduction}. Under the density assumption in Schnorr's algorithm~\cite{schnorr2021fast}, a sublinear dimension scheme is given.

\subsection{Linear scheme}

The existence problem refers to whether there is a vector $\mathbf b\in\Lambda(B_{n,c}),$ such that $\Vert\mathbf b-\mathbf t\Vert^2=O(\text{log}N)$ holds. Here, we estimate the distance from the target vector $\mathbf t$ to the lattice $\Lambda$ by considering the length $\lambda_1$ of the shortest vector on the extended lattice $\bar B_{n,c}=[B_{n,c},\mathbf t]$. Further, since the determinant of the extended lattice $\bar B_{n,c}$ can be obtained, the upper bound of $\lambda_1$ can be estimated according to Minkowski's first theorem, which is described as follows.
\begin{lemma}
(Minkowski's first theorem) For any full rank lattice $\Lambda$ with dimension $n$,
\begin{equation}
\lambda_1(\Lambda)^2\le n (\text{det}\Lambda)^{2/n}.
\end{equation}
\end{lemma}

Minkowski's first theorem considers the upper bound of the shortest nonzero vector, i.e., the first successive minimum $\lambda_1$. With this bound, we have the following results.
\begin{proposition}
If the dimension $n$ of the lattice $ B_{n,c}$ satisfies $n= \text{log}N$, then there exists a vector $\mathbf b\in  \Lambda(\bar B_{n,c}),$ such that
\begin{equation}
\Vert\mathbf {b-t}\Vert^2=O(\text{log}N).
\end{equation}
\end{proposition}
\begin{proof}
Let the length of the shortest vector on the extended lattice $ \bar B_{n,c}$ be $\lambda_1$.  Here we use the scale of $\lambda_1$ to estimate the $\text{dist}(B_{n,c},\mathbf t)$ between the lattice and the target vector, that is, assuming $\text{dist}(B_{n,c},\mathbf t)=O(\lambda_1)$. Then according to Minkowski's first theorem, we have
\begin{equation}
\lambda_1^2 \le (n+1)(\text{det}{\bar B_{n,c}})^{2/{n+1}}.
\end{equation}
According to the construction of the lattice, we have
\begin{equation}\label{lattice}
(\text{det}\bar B_{n,c})^{2/{n+1}}=(\prod_{i=1}^{n}f(i))^{2/{n+1}}(N^c\text{log}N)^{2/{n+1}}.
\end{equation}
Here we  set the diagonal elements  belong to the set $\{1,2\}$. And, we choose the diagonal elements as $ 2$ in a proportion  of  $(n+1)/3n$,  to ensure  the number of different arrangements is large enough to generate random lattices. Then, we have
\begin{equation}\label{fi}
(\prod_{i=1}^{n}f(i))^{2/(n+1)}=(2^{(n+1)/3})^{2/(n+1)}=2^{2/3}=O(1).
\end{equation}
Then substitute Eq. \ref{fi} and $n= \text{log}N$ into Eq. \ref{lattice} , we have
\begin{equation}
(\text{det}\bar B_{n,c})^{2/(n+1)}=O(N^{2c/(n+1)})=O(2^{\frac{2cn}{n+1}})=O(1).
\end{equation}
Hence we have
\begin{equation}
\lambda_1^2 \le nO(1)=O(\text{log}N).
\end{equation}
This completes the proof.
\end{proof}

It should be point out that the construction of the lattice is modified in that the diagonal elements are generated from the set $\{1, 2\}$, and the number of 2s is about $(n+1)/3n$. This condition can be further generalized on the condition of

\begin{equation}\label{ Herm}
\prod_{i=1}^{n}f(i))^{2/{n+1}}\sim O(1).
\end{equation}
In Minkowski's first theorem, a tighter upper bound can be obtained when we introduce Hermitian constants. Consider the following relationship
\begin{equation}\label{ Hermitian}
\gamma=\frac{\lambda_1^2(\Lambda)}{(\text{det}\Lambda)^{2/n}}.
\end{equation}

\begin{definition}
Denote $\gamma_n$ as the maximum value (upper bound) that satisfies Eq. \ref{ Hermitian} in all $n$ dimensional lattices, then  $\gamma_n$  is called the Hermitian constant of dimension $n$.
\end{definition}
In fact, $\gamma_n$ is also a supremum, that is, for any $n>1$, there is an $n$ dimensional lattice $\Lambda$ such that $\gamma_n={\lambda_1^2(\Lambda)}/{(\text{det}\Lambda)^{2/n}}$ holds. Such lattices are also commonly referred to as being critical. But calculating the exact $\gamma_n$ is usually difficult, which is also the central problem in the study of Minkowski's geometric numbers~\cite{cassels2012introduction}. Currently, we only know the results when $1\le n\le8$ and $n=24$. Asymptotically, the tightest bound~\cite{kabatiansky1978bounds} known is
\begin{equation}\label{Hermit}
\lambda_1^2 \le \gamma_n(\text{det}{\Lambda})^{2/n}\le\frac{1.744n}{2e\pi}(\text{det}\Lambda)^{2/n}.
\end{equation}
By using Eq.~\ref{Hermit}  to estimate $\lambda_1$, the same conclusion as Proposition 1 can be obtained.

\subsection{Sublinear scheme}
Since Minkowski's first theorem gives an upper bound on the value of the shortest vector, for many random lattices, the real shortest vector is quite different from this upper bound. This gap can be measured by the relative density $rd(\Lambda)$ of the lattice. The relative density $rd(\Lambda)$ of the lattice refers to the ratio between the actual length of the shortest vector $\lambda_1$ and the upper bound of the shortest vector estimated by the Hermitian constant. According to Eq. \ref{Hermit}, it is obvious that $0<rd( \Lambda)\le1$, and we specifically defined as
\begin{equation}
rd(\Lambda)=\frac{\lambda_1}{\sqrt{\gamma_n}(\text{det}\Lambda)^{1/n}}.
\end{equation}
When the relative density is close to 1, it indicates that the optimal lattice basis vectors are of the same size, and the lattice points are dense.

Schnorr has made the following assumption about the relative density of the lattices used for finding smooth relation pairs when discussing the efficiency of the algorithm.
\begin{assumption}\label{assum}
 The random lattice $\Lambda$ with basis $B=[\mathbf b_1,...,\mathbf b_n]$ has relative density which satisfies
\begin{equation}
rd(\Lambda)\le(\sqrt{\frac{e\pi}{2n}}\frac{\lambda_1}{\Vert\mathbf b_1\Vert})^{1/2}.
\end{equation}
\end{assumption}
That is, both $\mathbf b_1$ and  $rd(\Lambda)$ are relatively small. Since $\lambda_1/{\Vert\mathbf b_1\Vert}\le1$, according to this assumption, we have
\begin{equation}\label{rd}
rd(\Lambda)=\frac{\lambda_1}{\sqrt{\gamma_n}(\text{det}\Lambda)^{1/n}}\le(\frac{e\pi}{2n})^{1/4}.
\end{equation}
Hence we have the following results.
\begin{proposition}
If the dimension $n$ of the lattice $ B_{n,c}$ satisfies $n= 2c\text{log}N/\text{loglog}N$, and the relative density of the lattice satisfies Assumption 1, then there exists a vector $\mathbf b\in  \Lambda( B_{n,c}),$ such that:
\begin{equation}
\Vert\mathbf {b-t}\Vert^2=O(\log N).
\end{equation}
\end{proposition}
\begin{proof}
According to Eq. \ref{rd}, we have
\begin{equation}
{\lambda_1}^2\le(\frac{e\pi}{2n})^{1/2}{\gamma_n(\text{det}\Lambda)^{2/n}}.
\end{equation}
Substituting Eq. \ref{lattice} and  Eq. \ref{fi} into the above equation, we have
\begin{equation}
{\lambda_1}^2\le(\frac{e\pi}{2n})^{1/2}{\gamma_n(N)^{2c/n}}.
\end{equation}
At this time, if we choose $n=2c\text{log}N/\text{loglog}N$, then we have
\begin{equation}
{\lambda_1}^2\le(\frac{e\pi}{2})^{1/2}\frac{1.744}{2e\pi}\sqrt{2c\text{log}N/\text{loglog}N}{\text{log}N}=O(\text{log}N).
\end{equation}

Here, since $\sqrt{2c\text{log}N/\text{loglog}N}$ is a lower order quantity compared to $\text{log}N$, it is ignored in the final expression.

This completes the proof.
\end{proof}

 It is reasonable to ignore this lower order quantity mentioned in the proof. Choosing $c=1$, for $N\approx 2^{1024}$ as an example, we have
\begin{equation}
(\frac{e\pi}{2})^{1/2}\frac{1.744}{2e\pi}\sqrt{2c\text{ln}N/\text{loglog}N}\approx3.0960\sim O(1).
\end{equation}
Or for $N\approx 2^{2048}$ as another example, we have
\begin{equation}
(\frac{e\pi}{2})^{1/2}\frac{1.744}{2e\pi}\sqrt{2c\text{log}N/\text{loglog}N}\approx4.1641\sim O(1).
\end{equation}
This indicates that under the density assumption in Assumption~\ref{assum}, taking the dimension of the lattice $n$ as $2c\text{log}N/\text{loglog}N$ is reasonable, and the length (square norm) of the shortest vector in the lattice can be guaranteed to be $O(\text{log}N)$. That is, a smooth relation pair can be obtained from the closest vector of the lattice with a high probability, as described in Lemma \ref{lama1}.

\section{Preprocessing: the details about the factoring cases}\label{pre}
\subsection{The construction of the lattice and target vector}
We'll take the factorization  of $N=48567227$ in 5 qubits as an example to introduce the computational steps before the quantum part, which include the construction of lattice and target vector, LLL-reduction and the solution process of Babai's nearest plane algorithm. The 3-qubit case and 10-qubit case will be  shown directly.  Here we adopt the sublinear lattice dimension scheme. The lattice dimension required to factorize the integer $N=48567227$ is $logN/loglogN= 26/5\approx 5$. The prime basis consists of the first five prime numbers, which is $\{-1,2,3,5,7,11\}$. 

In order to generate enough random integer lattices, we roughly adjust the lattice construction in Eq. \ref{construction}. Firstly, using $\lceil i/2\rfloor$ to replace the original diagonal $\sqrt{\text{ln}p_i}$, where $\lceil \;\rfloor$ is the nearest rounding function. Secondly, in order to get distinct fac-relations, a random permutation function $f$ is used to perform random permutation on the diagonal elements of the lattice. In addition, using "$10^{c}$" to replace the 'weight' item "$N^c$" in the original lattice. In this way, if $c$ is an integer, it will be easy to convert the lattice to an integer lattice  and the parameter $c$ will  directly represent the precision. The specific lattice structure is presented in~\ref{bnc} and ~\ref{tn1}:
\begin{equation}
\label{bnc}
{B}_{n,c}=\begin{pmatrix}
 f(1)& 0 &...&0  \\
 0 & f(2)&...&0 \\
\vdots &\vdots &\ddots &\vdots \\
 0 & 0 & ...&f(n) \\
 \lceil{10^{c}\text{ln}2\rfloor} &\lceil{10^{c}\text{ln}3 \rfloor}  & ...&\lceil{10^{c}\text{ln}11\rfloor}
\end{pmatrix}
\end{equation}

\begin{equation}
\label{tn1}
\mathbf{t}_n=\begin{pmatrix}
0  \\ \vdots \\ 0 \\  \lceil{10^c\text{ln}N\rfloor}
\end{pmatrix}.
\end{equation}
Here  ${B}_{n,c}$ is the matrix form of the lattice with every column as a basis vector. The subscript represents the dimension $n$ of the lattice and the precision parameter $c$.  In the 5-qubit case, the dimension is $5$ and the precision parameter is $4$. The $f(i)$ elements on the diagonal are random permutations of elements in $\{\lceil{1/2}\rfloor,...,\lceil{5/2}\rfloor\}=\{1,1,2,2,3\}$.   Thus, the exact lattice and the target vector corresponding to the 
sr-pair are presented in ~\ref{b54} and ~\ref{tn2}:
\begin{equation}
\label{b54}
{B}_{5,4}=\begin{pmatrix}  2  & 0& 0& 0  &0  \\ 0 & 1& 0&0  &0 \\ 0  & 0& 3& 0  &0 \\ 0 & 0& 0&2  &0 \\ 0 & 0& 0&0  &1 \\ 
6931 &10986  &16094  &19459  &23979
\end{pmatrix}
\end{equation}

\begin{equation}
\label{tn2}
 \quad \mathbf{t}_5=\begin{pmatrix} 0 \\0 \\0 \\0 \\0 \\176985 \end{pmatrix}.
\end{equation}
 Similarly, in the 3-qubit case, the dimension satisfies $n=3$ and the precision parameter satisfies  $c=1.5$. The exact lattice and the target vector corresponding to the sr-pair are
\begin{equation}
{B}_{3,1.5}=\begin{pmatrix}  1  & 0& 0 \\ 0 & 1& 0 \\ 0  & 0& 2\\ 22  &35 & 51   \end{pmatrix},
\quad	\mathbf{t}_3=\begin{pmatrix} 0 \\0 \\0 \\240 \end{pmatrix}.
\end{equation}

 In the 10-qubit case, the dimension satisfies  $n=10$ and the precision parameter satisfies  $c=4$. The exact lattice and the target vector corresponding to the sr-pair are presented  in ~\ref{b104} and  ~\ref{t104}:
\begin{figure*}
\begin{equation}
\label{b104}
{B}_{10,4}=
\begin{pmatrix}  
3  & 0  & 0  & 0  & 0  & 0  & 0  & 0  & 0  & 0\\
0  & 2  & 0  & 0  & 0  & 0  & 0  & 0  & 0  & 0\\
0  & 0  & 3  & 0  & 0  & 0  & 0  & 0  & 0  & 0\\
0  & 0  & 0  & 1  & 0  & 0  & 0  & 0  & 0  & 0\\
0  & 0  & 0  & 0  & 1  & 0  & 0  & 0  & 0  & 0\\
0  & 0  & 0  & 0  & 0  & 3  & 0  & 0  & 0  & 0\\
0  & 0  & 0  & 0  & 0  & 0  & 1  & 0  & 0  & 0\\
0  & 0  & 0  & 0  & 0  & 0  & 0  & 1  & 0  & 0\\
0  & 0  & 0  & 0  & 0  & 0  & 0  & 0  & 2  & 0\\
0  & 0  & 0  & 0  & 0  & 0  & 0  & 0  & 0  & 2\\
6931 & 10986 & 16094 & 19459 & 23979 & 25649 & 28332 & 29444 & 31355 & 33673
\end{pmatrix},
\end{equation}
\end{figure*}

\begin{equation}
\label{t104}
\mathbf{t}_{10}=(0 \quad  0 \quad  0 \quad  0 \quad  0 \quad  0 \quad  0 \quad  0 \quad  0\quad   0 \quad 331993 )^T.
\end{equation}
\subsection{Solving the CVP using Babai's algorithm} 
The smooth relation pair can be obtained by solving the CVP on the above lattice. Before using the quantum method, an approximate optimal solution of the CVP can be obtained by  the classical lattice reduction algorithm~(the Babai's algorithm). Firstly, a LLL-reduction with parameter $\delta=3/4$ is performed on the lattice basis. The LLL-reduced basis is ${D}_{3,1.5}$ (\ref{d315}), ${D}_{5,4}$ (\ref{d54}) and ${D}_{10,4}$ (\ref{d104}) for the three factoring cases respectively.
\begin{equation}
\label{d315}
{D}_{3,1.5}=\begin{pmatrix}  
1  &-4  &-3  \\
-2 &1   &2   \\
2  &2   &0   \\
3  &-2  &4   \\
 \end{pmatrix}
 \end{equation}

\begin{equation}
\label{d54}
{D}_{5,4}=\begin{pmatrix} 
6  &-8  &2  &-4  &-4 \\
-4 &-3  &11 &-5  &-3 \\
6  &6   &3  &0   &-3 \\
4  &-2  &0  &12  &4  \\
-2 &2   &-6 &-2  &1  \\
-3 &5   &-3 &4   &-17\\
\end{pmatrix},
\end{equation}

\begin{equation}
\label{d104}
 {D}_{10,4}=\begin{pmatrix} 
0  & 0  & 3  & 0  & 0  & 0  & 3  & 0  & -3  & -3\\
0  & 0  & 2  & 0  & 4  & -4  & 0  & 4  & -2  & 4\\
-3  & 0  & 0  & 0  & 0  & 0  & -3  & 0  & 0  & 0\\
1  & 2  & 1  & 4  & -4  & -2  & -2  & 0  & -1  & 0\\
2  & 0  & 2  & -2  & 0  & 0  & 1  & -1  & 0  & 4\\
0  & 0  & -3  & -3  & 0  & 0  & 0  & 0  & -3  & 3\\
-3  & 3  & -1  & 0  & 1  & 2  & 1  & 2  & -2  & -1\\
0  & -2  & 0  & 1  & 2  & -1  & 1  & -3  & 3  & -3\\
0  & -2  & -2  & 0  & -2  & 0  & 0  & 0  & 2  & 2\\
2  & -2  & 0  & -2  & 0  & 2  & -2  & 2  & 0  & 0\\
0  & -2  & -2  & 0  & 1  & 3  & 1  & -2  & -2  & -1
\end{pmatrix}.
\end{equation}

Secondly, perform the size reduction procedure. This process takes the largest basis vector in the LLL basis (the rightmost column of the matrix ${D}_{5,4}$) as the starting point, subtracts it from the target vector $\mathbf{t}_5$ in turn according to the round function of GS-coefficient values $\mu_i,i=1,...,5$,  until the shortest basis vector in the LLL basis (the leftmost column) is subtracted. The distance vector $\mathbf{\bar{t}_5}$ and the approximate nearest vector $\mathbf b_{op}$ are obtained at the end of this procedure. Since the length of the distance vector represents the quality of the CVP solution, we also referred it the short vector in CVP. Here, the classical optimal solutions obtained by the Babai's algorithm are presented in ~\ref{b_op} to ~\ref{t_10}.

\begin{figure*}
\begin{equation}
    \label{b_op}
 \mathbf{b}_{op}=(2 \quad4\quad9 \quad8 \quad0 \quad176993 )^T,\quad
\mathbf{\bar{t}}_5=\mathbf{b}_{op}-\mathbf{t}_{5}=\begin{pmatrix}
 2 &4 &9 &8 &0 &8
 \end{pmatrix}^T.
\end{equation}
The corresponding results for the 3-qubit case are:
\begin{equation}
\mathbf{b}_{op}=(0 \quad4\quad4 \quad242 )^T,
\quad \mathbf{\bar{t}}_3=\mathbf{b}_{op}-\mathbf{t}_{3}=\begin{pmatrix} 0 &4 &4 &2 \end{pmatrix}^T.
\end{equation}
The corresponding results for the 10-qubit case are:
\begin{gather}
\mathbf{b}_{op}=(3 \quad4 \quad0 \quad1 \quad2 \quad3 \quad2 \quad3 \quad2 \quad2 \quad331993)^T,
\\ 
\label{t_10}
\mathbf{\bar{t}}_{10}=\mathbf{b}_{op}-\mathbf{t}_{10}=(3 \quad4 \quad0 \quad1 \quad2 \quad3 \quad2 \quad3 \quad2 \quad2 \quad 0)^T.
\end{gather}
\end{figure*}
The approximate closest  vector is relatively far from the target vector $\mathbf t_5$, which is $\lVert\mathbf{\bar{t}_{5}}\rVert^2=229$. In the three factoring cases, a vector that is closer (or shorter) than that of Babai's algorithm can be obtained by the quantum optimization.

\subsection{The problem Hamiltonian} 
In the main text, we have introduced the construction of the problem Hamiltonian by mapping the binary variables $x_i,\{i=1,...,n\}$ to the Pauli-Z items, namely
\begin{equation}
Hc=\|\mathbf t- \sum_{i=1}^{n}{\hat x_i\mathbf d_i-\mathbf b_{op}}\|^2
=\sum_{j=1}^{n+1}{\mid t_j-\sum_{i=1}^{n}{\hat x_id_{i,j}}-b_{op}^j\mid^2},
\end{equation}

We use a single qubit to encode the floating variables $x_i\in\{-1,0,1\},i=1,2,\dots,n$, according to the intermediate calculation results of Babai's algorithm. The quantum operator $\hat x_i$ is  mapped to the Pauli-Z basis according to the following rules:
\begin{equation}
\hat x_i=\begin{cases}
\frac{I-\sigma_z^i}{2}, &\mbox{if}\;c_i\le\mu_i\\
\frac{\sigma_z^i-I}{2}, &\mbox{if}\;c_i>\mu_i
\end{cases}
\end{equation}
As shown above, if the function is rounding down to the nearest integer, namely $c_i\le \mu_i$, the coefficient value $c_i$ will be floated up by 1 or unchanged. In this case, the floating value $x_i\in\{0,1\}$ corresponds to the eigenvalues of quantum operator $\frac{I-\sigma_z^i}{2}$, and vice versa. Therefore, we can use the rounding information of the coefficient $c_i$ in Babai's algorithm to determine the encoding of the floating value $x_i$. It is easy to see that the lower energy state of the Hamiltonian system will result in an approximate close vector solution in lattice $\Lambda$ according to the correspondence of the problem Hamiltonian and loss function. 

The problem Hamiltonian for the 5-qubit case can be construct as $H_{c5}=\sum_{j=1}^6{\hat{h}_j}$,
where 
\begin{equation}
\begin{cases}\hat{h}_1=(6\hat{x}_1-8\hat{x}_2+2\hat{x}_3-4\hat{x}_4-4\hat{x}_5+2)^2\\
 \hat{h}_2=(-4\hat{x}_1-3\hat{x}_2+11\hat{x}_3-5\hat{x}_4-3\hat{x}_5+4)^2\\ 
\hat{h}_3=(6\hat{x}_1+6\hat{x}_2+3\hat{x}_3-0\hat{x}_4-3\hat{x}_5+9)^2\\
\hat{h}_4=(4\hat{x}_1-2\hat{x}_2+0\hat{x}_3+12\hat{x}_4+4\hat{x}_5+8)^2\\ 
\hat{h}_5=(-2\hat{x}_1+2\hat{x}_2-6\hat{x}_3-2\hat{x}_4+\hat{x}_5)^2\\
 \hat{h}_6=(-3\hat{x}_1+5\hat{x}_2-3\hat{x}_3+4\hat{x}_4-17\hat{x}_5+8)^2\\
\end{cases}
\end{equation}
We can determine the specific encoding process of each variable $x_i,i=1,...,5$ according to the intermediate calculation results of Babai's algorithm, which is shown in Table~\ref{5coding}. 

\begin{table}[ht]
\begin{center}
\caption{Qubits encoding information for the 5-qubit case. The  subscript  "$j$" decreases sequentially from left to right.}\label{5coding}
\begin{tabular}{@{}cccccc@{}}
\toprule
steps & 1 ($x_5$) &2 ($x_4$) & 3 ($x_3$) & 4 ($x_2$) & 5 ($x_1$) \\
\hline
 $\mu_j$        & -8731.5607         & 3882.5019              & -1837.4760             & -354.467               & -3092.4957             \\
 $c_j$        & -8732              & 3883                   & -1837                  & -354                   & -3092                  \\
 $\mu_j-c_j$     & 0.4393             & -0.4981                & -0.4760                & -0.4669                & -0.4957               \\
 coding    & (0,1)              & (0,-1)                 & (0,-1)                 & (0,-1)                 &      (0,-1)                  \\
\botrule
\end{tabular}
\end{center}
\end{table}

\begin{figure*}
\begin{equation}
\label{5_q_Hc}
\begin{aligned}
H_{c5}=&781I - 142\sigma_z^1  - 64\sigma_z^2 - 81\sigma_z^3 - 213\sigma_z^4  - 4.5\sigma_z^5- 13.5\sigma_z^1\sigma_z^2 + 3.5\sigma_z^1\sigma_z^3 + 18\sigma_z^1\sigma_z^4 + 17.5\sigma_z^1\sigma_z^5 \\&-29\sigma_z^2\sigma_z^3 + 19.5\sigma_z^2\sigma_z^4 - 34\sigma_z^2\sigma_z^5- 31.5\sigma_z^3\sigma_z^4 - 2.5\sigma_z^3\sigma_z^5 + 4.5\sigma_z^4\sigma_z^5 .
\end{aligned}
\end{equation}
\end{figure*}

Thus, the 5-qubit Hamiltonian is reduced to Eq.~\ref{5_q_Hc}. The qubits encoding information and the problem Hamiltonian corresponding to the 3-qubit case are presented in Table~\ref{3coding} and  Eq.~\ref{3_q_Hc}. The qubits encoding information and the problem Hamiltonian corresponding to the 10-qubit case are presented in Table~\ref{10coding} and Eq.~\ref{10_q_Hc}.

\begin{table}[ht]
\begin{center}
\caption{Qubits encoding information for the 3-qubit case. The  subscript  '$j$' decreases sequentially from left to right.}\label{3coding}
\setlength{\tabcolsep}{5mm}{
\begin{tabular}{@{}cccc@{}}
\toprule
steps  & 1 ($x_3$) & 2 ($x_2$) & 3 ($x_1$) \\
\hline
 $\mu_j$      & 33.5812 &-20.4974 & 21.6667 \\
      $c_j$         &34       & -20     & 22      \\
      $\mu_j-c_j$  &-0.4188  & -0.4974 &-0.3333  \\
      coding     &(0,-1)   &(0,-1)   &(0,-1)   \\
\botrule
\end{tabular}
}
\end{center}
\end{table}

\begin{figure*}
\begin{equation}
\label{3_q_Hc}
\begin{aligned}
H_{c3}=& 43.5I- 4\sigma_z^1\sigma_z^2 + 2.5\sigma_z^1\sigma_z^3 -1.5\sigma_z^1 + 3\sigma_z^2\sigma_z^3 -3.5\sigma_z^2 -4\sigma_z^3.
\end{aligned}
\end{equation}
\end{figure*}

\begin{table*}[ht]
\begin{center}
\caption{Qubits encoding information for the 10-qubit case. The  subscript '$j$' decreases sequentially from left to right.}\label{10coding}
\setlength{\tabcolsep}{0.8 mm}{
\begin{tabular}{@{}ccccccccccc@{}}
\toprule
steps  & 1 ($x_{10}$) & 2 ($x_9$) & 3 ($x_8$) & 4 ($x_7$) & 5 ($x_6$) & 6 ($x_5$)& 7 ($x_4$) & 8 ($x_3$) & 9 ($x_2$) & 10 ($x_1$) \\
\hline
 $\mu_j$      & 21514.149 &-45688.541 & -29225.45  & -5953.325 &29891.446 & 23868.721  & 42395.337 &-18221.276  & -29823.805  &5952.889 \\
      $c_j$     &21514    & -45689    & -29225     &-5953      &29891     & 23869      &42395      & -18221     & -29824 &5953\\
      $\mu_j-c_j$  &0.149 & 0.459     &-0.45       &-0.325     &0.446     &-0.279      &0.337      &-0.276      &  0.195  &-0.111        \\
      coding     &(0,1)   &(0,1)      &(0,-1)      &(0,-1)     &(0,1)     &(0, -1)     &(0,1)      &(0,-1)   &(0,1)  &(0,-1)\\
\botrule
\end{tabular}}
\end{center}
\end{table*}

\begin{figure*}
\begin{equation}
\label{10_q_Hc}
\begin{aligned}
H_{c10}&= (708I+22\sigma_z^1\sigma_z^2 +16\sigma_z^1\sigma_z^3+8\sigma_z^1\sigma_z^4-14\sigma_z^1\sigma_z^5+8\sigma_z^1\sigma_z^6 + 4\sigma_z^1\sigma_z^7- 8\sigma_z^1\sigma_z^8 - 10\sigma_z^1\sigma_z^9 - 22\sigma_z^1\sigma_z^{10}- 46\sigma_z^1  - 14\sigma_z^2\sigma_z^3 \\
&+ 20\sigma_z^2\sigma_z^4 + 14\sigma_z^2\sigma_z^5- 12\sigma_z^2\sigma_z^6 + 2\sigma_z^2\sigma_z^7 -24\sigma_z^2\sigma_z^8 - 28\sigma_z^2\sigma_z^9 + 2\sigma_z^2\sigma_z^{10} - 16\sigma_z^2  - 18\sigma_z^3\sigma_z^4 + 10\sigma_z^3\sigma_z^5 + 36\sigma_z^3\sigma_z^6 + 12\sigma_z^3\sigma_z^7\\
&+ 16\sigma_z^3\sigma_z^8 + 6\sigma_z^3\sigma_z^9 - 30\sigma_z^3\sigma_z^{10} - 78\sigma_z^3  + 28\sigma_z^4\sigma_z^5 - 26\sigma_z^4\sigma_z^6 + 10\sigma_z^4\sigma_z^7 + 10\sigma_z^4\sigma_z^8 + 16\sigma_z^4\sigma_z^9 - 4\sigma_z^4\sigma_z^{10} - 72\sigma_z^4  + 10\sigma_z^5\sigma_z^6 \\
&+ 24\sigma_z^5\sigma_z^7 + 20\sigma_z^5\sigma_z^8 + 12\sigma_z^5\sigma_z^9 - 8\sigma_z^5\sigma_z^{10} - 116\sigma_z^5  - 8\sigma_z^6\sigma_z^7 + 22\sigma_z^6\sigma_z^8 - 6\sigma_z^6\sigma_z^9 - 36\sigma_z^6\sigma_z^{10} - 12\sigma_z^6  - 16\sigma_z^7\sigma_z^8 + 16\sigma_z^7\sigma_z^9 \\
&+ 20\sigma_z^7\sigma_z^{10} - 84\sigma_z^7 + 34\sigma_z^8\sigma_z^9 - 42\sigma_z^8\sigma_z^{10} - 36\sigma_z^8  + 18\sigma_z^9\sigma_z^{10} - 74\sigma_z^9 - 24\sigma_z^{10})/4.
\end{aligned}
\end{equation}
\end{figure*}

\subsection{The energy spectrum and the target state}

We numerically traverse the energy spectrum of the problem Hamiltonian. Here we only show the lowest ten energy levels and the corresponding quantum states, meantime, the corresponding sr-pairs, if there are. It should be noted that the smooth bound $B_2$ about $\mid u-vN\mid$ is different from the smooth bound $B_1$ of $(u,v)$. In the 5-qubit case, we choose $B_2=p_{50}=229$, and the dimension of the corresponding linear equation system (denote by eq-dim) is 51. The details for the 3- and 10-qubit case can be found in Table~\ref{dim1}. The dimension of the corresponding linear equation system is $\sim 2n^2$, which is the polynomial scale of $n$. Hence it is reasonable to relax the $B_2$ bound here.
\begin{table}[ht]
\begin{center}
\caption{Two smooth bounds for the three factoring cases.}\label{dim1}
\setlength{\tabcolsep}{2mm}{
\begin{tabular}{@{}cccccc@{}}
\toprule
case          &B1-dim & B1 & B2-dim & B2  &eq-dim\\
\hline
 3-qubit      & 3     &5   & 15     &47   & 16\\
 5-qubit      &5      &11  & 50     &229  & 51 \\
10-qubit      &10     &29  & 200    &1223 & 201   \\
     
\botrule
\end{tabular}
}
\end{center}
\end{table}
On the one hand, the dimension of the prime basis (lattice dimension) is low in the Schnorr's sieve method, solving the system of linear equations  consumes  relatively less computational resources. On the other hand, the algorithm has higher quality requirements for short vectors, which increases the overall computational complexity of the algorithm drastically. Therefore, we can reduce the quality requirements of the algorithm for short vectors by appropriately relaxing the smooth bound $B_2$, which will increase the  amount of the linear equation system. However, with the balance between the amount of calculation  the short vector and solving the linear equation system, the efficiency of the whole algorithm is improved.
\begin{table}[ht]
\begin{center}
\caption{ The first ten lowest energy levels and the corresponding quantum states. The fourth excited state  generates a smooth relation pair, and its corresponding value $\mid u-vN\mid=12097706=2*41*43*47*73$ is smooth on the $B_2$ bound, which made it a target state  required  for the 5-qubit case.}\label{5energy}
\begin{tabular}{@{}ccccccc@{}}
\toprule
level & energy &state & u & v & $\mid u-vN\mid$ & smooth\\
\hline

 0    & 186  & 0 0 1 1 0   & 21435888100 & 441   & 89*199337     & no   \\
 1    & 189  & 0 1 1 1 0   & 340139712   & 7       & 53*3191       & no   \\
 2    & 193  & 1 1 1 0 0   & 1215290846  & 25     & 3*370057      & no   \\
 3    & 198  & 1 0 0 0 1   & 776562633   & 16      & 512999        & no   \\
 4    & 215  & 0 0 1 1 1   & 11789738455 & 243   & 2*41*43*47*73 & yes  \\
 5    & 218  & 1 1 0 0 0   & 243045684   & 5       & 209549        & no   \\
 6    & 222  & 1 1 1 1 0   & 4.16714E+11 & 8575   & 249693139     & no   \\
 7    & 229  & 0 0 0 0 0   & 48620250    & 1        & 17*3119       & no   \\
 8    & 230  & 1 0 0 0 0   & 194500845   & 4       & 41*5657       & no   \\
 9    & 232  & 1 0 1 1 1   & 2.85312E+11 & 5880   & 37*7124977    & no   \\
\botrule

\end{tabular}
\end{center}
\end{table}

In Table~\ref{5energy}, the first column represents the first ten lowest energy levels of the Hamiltonian in Eq. \ref{5_q_Hc}, and the second column represents the  energys corresponding to the energy level. It also represents the square norm value of the short vector. Columns $3$ represent the eigenstate of the Hamiltonian. The first row represents the ground state, and the corresponding energy value is $186$.  Here we find that the length of the solution vector corresponding to the ground state is the shortest, but no sr-pair is obtained. This is because the relation between the short vector and the sr-pair is probabilistic. The energy corresponding to the fourth excited state is $215$, and the corresponding $\mid u-vN\mid=12097706=2*41*43*47*73$ is smooth on the $B_2$ bound. A set of sr-pair is obtained from the state $(00111)$.
Meanwhile, the square norm of the solution vector corresponding to the seventh excited state $(00000)$ is 229,  which is the optimal solution obtained by the Babai algorithm. Hence we can see that the quantum method leads to shorter vectors and obtains an sr-pair for factoring. Therefore, the quantum state $(00111)$ is the target state  required in the following.

Making a relatively low energy  state as a target state is a good idea since QAOA is often challenging to iterate to the lowest energy state. When the energy of the quantum state prepared by the QAOA circuit is low enough, the quantum states of low energy levels will prevail. Therefore, even if the target state is not the ground state, there will be a considerable probability of being measurable.  This is verified  by the experiments results in the next part.

\begin{table}[ht]
\begin{center}
\caption{ The first four lowest energy levels and the corresponding quantum states for the 3-qubit case.}\label{4energy}
\setlength{\tabcolsep}{2mm}{
\begin{tabular}{@{}cccccccc@{}}
\toprule
 levels & energy & state            & u    & v    & $\mid u-vN\mid$  & smooth \\
\hline
 0        & 33     & 0 0 1          & 1800 & 1         & 7*23    & yes  \\     
 1        & 35     & 1 1 0          & 1944 & 1        & 17      & yes    \\   
 2        & 36     & 0 0 0          & 2025 & 1        &$ 2^6$     & yes      \\ 
 3        & 42     & 1 0 0          & 3645 & 2        & 277     & no        \\

\botrule
\end{tabular}
}
\end{center}
\end{table}

Likewise, we give the first four low-energy eigenstates for the 3-qubit case. In Table~\ref{4energy}, we find that sr-pairs can be obtained from the first three low-energy states including the ground state. Here the second excited state (000) corresponds to the optimal solution obtained by the Babai's algorithm. Although the solution itself can obtain an sr-pair, after quantum optimization, shorter vectors are obtained with the square norms of 33 and 35 respectively. And new sr-pairs can also be obtained. We set the target state in the 3-qubit case  as the ground state (001) which will be prepared in the experiments. For the 10-qubit case, the details are shown in Table~\ref{10energy2}. The ground state (0100010010) would lead to an sr-pair, which makes it a target state in the 10-qubit case.

\begin{table*}[ht]
\begin{center}

\caption{ The first ten lowest energy levels and the corresponding quantum states for the 10-qubit case. The ground state (0100010010) generates a smooth relation pair, and its corresponding value $\mid u-vN\mid=2*31*97*109*163*433$ is smooth on the $B_2$ bound, which made it a target state  required  for 10-qubit case.}\label{10energy2}
\setlength{\tabcolsep}{2.5 mm}{
\begin{tabular}{@{}ccccccc@{}}
\toprule
level & energy & state & u & v & $\mid u-vN\mid$ & smooth\\
\hline
0 & 51 & 0 1 0 0 0 1 0 0 1 0 & 785989264048241 & 3 & $2*31*97*109*163*433$   & yes \\
1 & 57 & 0 1 0 0 0 0 0 0 1 0 & 261933899831373 & 1  & $2^3*29*203014633$   & no\\
2 & 60 & 0 0 0 0 0 0 0 0 0 0 & 262049748526566 & 1  & $47*139*10523389$   & no\\
3 & 60 & 0 0 0 1 0 1 0 0 0 0 & 262123789565918 & 1  & $3803*37546763$   & no \\
4 & 61 & 0 1 0 0 0 0 1 1 0 0 & 262027921960805 & 1  & $2^4*457*2243*2861$  & no\\
5 & 65 & 0 1 0 0 0 0 0 1 0 0 & 7599879238585630 & 29 & $3^2*211*1531*835897$  & no \\
6 & 66 & 0 0 0 0 0 0 1 1 0 0 & 4455399847833940 & 17 & $2*3*24407*11764801$   & no \\
7 & 68 & 0 0 0 0 0 0 0 0 1 0 & 261988302332823 & 1  & $2*7*22717*22963$  & no\\
8 & 70 & 0 0 0 0 0 0 1 0 0 0 & 262012871275155 & 1  & $2*1693*9412891$   & no\\
9 & 70 & 0 0 0 0 1 0 0 0 0 0 & 262002304109546 & 1  & $21304883317$  & no\\
\botrule

\end{tabular}}
\end{center}
\end{table*}

\section{Experimental details}

\subsection{Device parameters}
We perform our experiment on a flip-chip superconducting quantum processor~\cite{xu2022digital} with 10 qubits ($Q_1$ $\sim$ $Q_{10}$) and 9 couplers ($C_1$ $\sim$ $C_9$) alternately arranged in a chain topology (Fig.~2\textbf{A}). All qubits and couplers are of transmon type, and their frequencies can be tuned independently by applying slow flux pulses (up to hundreds of microseconds in length) or fast Z flux pulses (up to tens of microseconds in length) on their corresponding control lines. The maximum frequencies of the qubits (couplers) are around 4.7 GHz (9.0 GHz) with nonlinearities around -210 MHz (-150 MHz). The coupling strength between each pair of neighboring qubits can be tuned from nearly off to -10 MHz by modulating the frequency of the coupler between them. Control lines for all qubits can also be used to apply microwave pulses to implement single-qubit gates. The energy relaxation times, dephasing times, gate fidelities and other relevant parameters are summarized in Table ~\ref{tab:dev_para}.

\subsection{Benchmarking the experimental gates}
 We initialize all qubits in the ground states via a reset procedure. More specifically, we first tune one of the qubit's adjacent couplers into resonance with the qubit for several nanoseconds, which swaps the excitation into the coupler. We then tune this coupler to its maximum frequency, where it has a very short $T_1$, so that the excitation decays quickly. We repeat the above steps several times until the qubit is thoroughly initialized to the ground state. Then we implement the experimental circuits (Fig.~2\textbf{D}). Note that for some long idle positions, we insert double $R_x(\pi)$ gates to protect the qubit from dephasing. After measuring the raw probabilities of all bitstrings, a readout correction is performed to eliminate the readout errors~\cite{wang2021scalable}.

The experimental circuit consists of $R_x$($\theta$), $R_y$($\theta$), $R_z$($\theta$) (rotations around $x$-, $y$- and $z$-axis by $\theta$), Hadamard and CZ gates. $R_x$($\theta$), $R_y$($\theta$) gates are implemented by 30-ns-long microwave pulses with controlled phases and amplitudes. $R_z$($\theta$) gates are implemented by virtual Z gates that could be considered ideal gates that take zero time simply by changing the phases of all subsequent microwave pulses~\cite{Mckay2017VZ}. Hadamard gates are implemented by the composition of $R_z$($\pi$) gates followed by$R_y$($\pi/2$) gates, written as $H=R_{y}(\pi/2)R_{z}(\pi)$. CZ gates are realized by applying a well-designed flux pulses on the neighboring qubits and their shared coupler, and detailed procedures for the individual implementation of them on neighboring qubits were described in our previous study~\cite{ren2022experimental, zhang2022digital}. If possible, consecutive single-qubit gates are combined to one gate in order to reduce the total circuit depth.

Here we discuss the case of optimizing the gate fidelities when executing the CZ gates in parallel, which is similar to the individual implementation except for optimizing the pulse parameters for the couplers. As requested by the experimental circuit, we divide the nine couplers into two groups, \{$C_1, C_3, C_5, C_7, C_9$\} for group A and \{$C_2, C_4, C_6, C_8$\} for group B. We apply sine-decorated rectangular flux pulses of the form $A(t)=z_{C_j}\cdot\left [ 1-r_{C_j}+r_{C_j}\sin \left ( \pi\frac{t}{t_\text{gate}}\right )\right ]$ to all couplers in group A or B simultaneously. Here $z_{C_j}$ is the maximum flux amplitude applied on $C_j$, $r_{C_j}$ is a modulated parameter fixed around 0.1 and $t_\text{gate}$ is 50 ns. Note that 
5-ns spacings are applied before and after this flux pulse. To optimize $z_{C_j}$, we prepare $Q_j$, $Q_{j+1}$ in the state of $\ket{11}$ and apply m cycles of CZ pulses with $m\in \{1,3,5\}$. Then we directly measure the $\ket{2}$-state leakage of the qubit that has a higher resonant frequency in a pair, and minimize the leakage populations for all pairs simultaneously by fine-tuning all $z_{C_i}$s.

\begin{table*}
\begin{center}
\caption{Device parameters. $\omega_j^0$ is the idle frequency of $Q_j$ where the qubit is initialized. $\eta_j$ is the nonlinearity of $Q_j$. $T_{1,j}$ and $T_{2,j}$ are the energy relaxation time and Ramsey dephasing time of $Q_j$ at idle frequency, respectively. $F_{0,j}$ and $F_{1,j}$ are the measure fidelities of $Q_j$ prepared in $\ket{0}$ and $\ket{1}$ respectively. $e_j^S$ is the simultaneous single-qubit gate Pauli error of $Q_j$. $e_{j,A (B)}^\text{CZ}$ is the simultaneous CZ-gate Pauli error of $Q_j$ and $Q_{j+1}$ in group A (B). $(\omega_j^{A (B)}, \omega_{j+1}^{A (B)})$ are the estimated qubit frequencies of $Q_j$ and $Q_{j+1}$ in group A (B) when we perform CZ gates.}
\label{tab:dev_para}
    \setlength{\tabcolsep}{2mm}{
    \begin{tabular}{c|c|c|c|c|c|c|c|c|c|c|c}
    \toprule
    Qubit & $Q_1$ & $Q_2$ & $Q_3$ & $Q_4$ & $Q_5$ & $Q_6$ & $Q_7$ & $Q_8$ & $Q_9$ & $Q_{10}$ & mean\\\hline
    $\omega_j^0/2\pi$ (GHz)& 4.420 & 4.500 & 4.553 & 4.460 & 4.370 & 4.600 & 4.430 & 4.515 & 4.445 & 4.570 & 4.486\\\hline
    $\eta_j/2\pi$ (MHz) & -213 & -209 & -208 & 209 & -211 & -211 & -209 & -210 & -211 & -210 & -210\\\hline
    $T_{1,j}$ ($\mu$s) & 91.4 & 83.3 & 113.6 & 131.1 & 111.6 & 99.5 & 116.0 & 108.2 & 123.3 & 115.1 & 109.3\\\hline
    $T_{2,j}$ ($\mu$s) & 5.3 & 7.0 & 5.7 & 7.2 & 4.3 & 6.2 & 4.9 & 5.8 & 8.2 & 6.9 & 6.2\\\hline
    $F_{0,j}$ ($\mu$s) & 0.982 & 0.984 & 0.976 & 0.991 & 0.972 & 0.990 & 0.981 & 0.978 & 0.974 & 0.978 & 0.981\\\hline
    $F_{1,j}$ ($\mu$s) & 0.949 & 0.967 & 0.942 & 0.957 & 0.951 & 0.958 & 0.960 & 0.958 & 0.927 & 0.923 & 0.949\\\hline
    $e_j^S$ (\%) & 0.11 & 0.07 & 0.09 & 0.09 & 0.15 & 0.12 & 0.06 & 0.07 & 0.08 & 0.09 & 0.09\\\hline
    $(\omega_j^A, \omega_{j+1}^A)/2\pi$ (GHz) & \multicolumn{2}{c|}{ 4.315, 4.520 } & \multicolumn{2}{c|}{ 4.666, 4.460 } & \multicolumn{2}{c|}{ 4.600, 4.392 } & \multicolumn{2}{c|}{ 4.335, 4.540 } & \multicolumn{2}{c|}{ 4.570, 4.364 } & -\\\hline
    $e_{j,A}^\text{CZ}$ (\%) & \multicolumn{2}{c|}{ 0.65 } & \multicolumn{2}{c|}{ 0.72 } & \multicolumn{2}{c|}{ 0.65 } & \multicolumn{2}{c|}{ 0.67 } & \multicolumn{2}{c|}{ 0.76 } & 0.69\\\hline
    $(\omega_j^B, \omega_{j+1}^B)/2\pi$ (GHz) & - & \multicolumn{2}{c|}{ 4.348, 4.553 } & \multicolumn{2}{c|}{ 4.510, 4.304 } & \multicolumn{2}{c|}{ 4.609, 4.400 } & \multicolumn{2}{c|}{ 4.532, 4.325 } & - & -\\\hline
    $e_{j,B}^\text{CZ}$ (\%) & - & \multicolumn{2}{c|}{ 0.54 } & \multicolumn{2}{c|}{ 0.70 } & \multicolumn{2}{c|}{ 0.58 } & \multicolumn{2}{c|}{ 0.57 } & - & 0.60\\\hline
    \botrule
    \end{tabular}}
\end{center}
\end{table*}

\begin{figure*}
\centering
\includegraphics[width=1.0\linewidth]{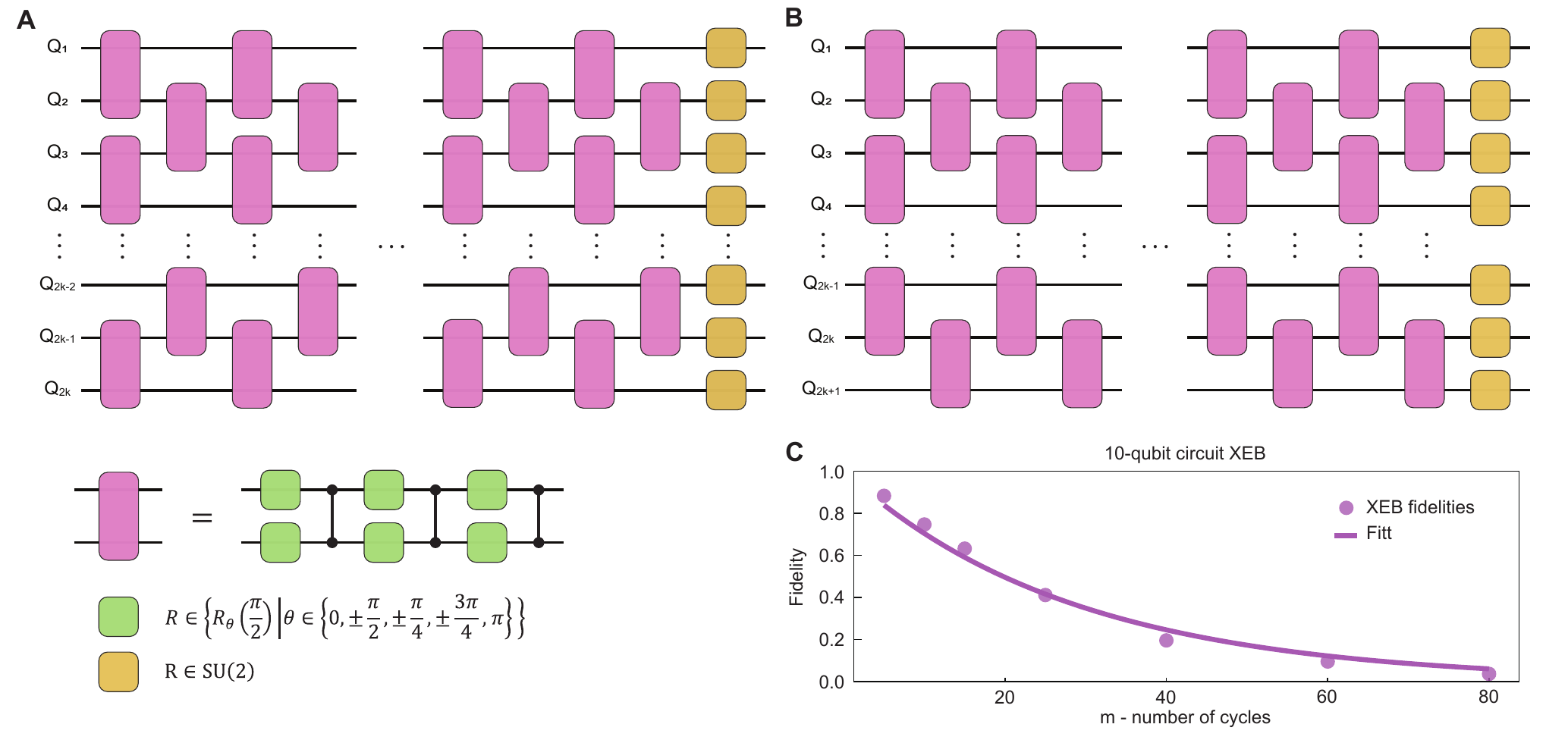}
\caption{\textbf{Alternative quantum circuits used to benchmark the CZ gates}, for even (\textbf{A}) and odd (\textbf{B}) number of qubits. Green squares represent randomly chosen half-$\pi$ rotations around 8 axes, where $\theta$ is the angle between the rotation axis and $x$-axis. Yellow squares represent gates randomly chosen from SU(2). \textbf{C}, XEB fidelity as a function of numbers of gate cycles for the 10-qubit case. The cycle error is fitted to be around 3.45\%, where each cycle contains a layer of 11 single-qubit gates followed by a layer of 4.5 CZ gates on average.}
\label{aaabbb}
\end{figure*}

We adopt the cross entropy benchmarking (XEB)~\cite{arute2019quantum} to evaluate the performance of our quantum gates. The Pauli errors for simultaneous single-qubit gates average to 0.09$\%$. The Pauli errors for CZ gates in group A and B average to 0.69$\%$ and 0.60$\%$, respectively. Note that the structure of the experimental circuit is different from the standard benchmarking circuits. As such, we take alternative quantum circuits to further verify the performance of our CZ gates, which have structures similar to the circuits for our algorithm (Fig.~\ref{aaabbb}). Figure~\ref{aaabbb}\textbf{C} depicts the XEB fidelity as a function of circuit depth (number of cycles), showing a cycle error around 3.45$\%$, based on which we estimate an average CZ gate error of 0.55$\%$.

\subsection{QAOA procedure and the convergence}

\noindent 
QAOA can find the approximate ground state of the Hamiltonian system by updating the parameters. For the $p$-layer QAOA, $2p$ variational parameters $\mathbf{\gamma}=(\gamma_1,...,\gamma_p),\mathbf\beta=(\beta_1,...,\beta_p)$ are involved. The main job for the quantum processor is to repeatedly prepare the following parameterized wave function
\begin{equation}
\ket{\mathbf{{\gamma}},\mathbf{\beta}}=e^{-i\beta_pH_b}e^{-i\gamma_pH_c}...e^{-i\beta_1H_b}e^{-i\gamma_1H_c}\ket{+}^n,
\end{equation}
where $H_b=\sum _{j =1}^{n}{\sigma_x^j}$ is the mixing Hamiltonian. This state can be prepared by applying the unitaries $U(H_c,\gamma)=e^{-i\gamma H_c}$ and $U(H_b,\beta)=e^{-i\beta H_b}$ alternately with different parameters in the uniform superposition state $\ket{+}^n$.
A classical optimizer is used to find the optimal parameters $(\mathbf{\gamma^*},\mathbf{\beta^*})$ that minimize the expected energy value of the problem Hamiltonian:
\begin{equation}
E(\mathbf{\gamma},\mathbf{\beta})=\bra{\mathbf{\gamma},\mathbf{\beta}}H_c\ket{\mathbf{\gamma},\mathbf{\beta}}.
\end{equation}

This energy function can be calculated by repeatedly preparing the wave function $\ket{\mathbf\gamma,\mathbf\beta}$ in the quantum register and measuring it on the computational basis. Finally, the quantum state $\ket{\mathbf{\gamma}^*,\mathbf{\beta}^*}$ corresponding to the approximate solution is obtained. The algorithm is shown graphically in Fig.~2\textbf{C}. 

Here we briefly introduce the classical optimizer adopt in QAOA during the parameter optimization procedure. The optimizer is called model gradient descent(MGD) method~\cite{sung2020using}. In fact, there are a lot of other classical optimization algorithms can be considered for the optimizer, like Nelder-Mead simplex method~\cite{lagarias1998convergence}, quasi-Newton method~\cite{broyden1970convergence,liu1989limited}. The performance of these methods often varies depending on the problem. Model gradient descent has been shown both numerically and experimentally perform well on some variational quantum ansatz~\cite{sung2020using,harrigan2021quantum}. The core idea of MGD is using model to estimate the gradient of the objective function, which is a continuous surface or hypersurface. To estimate the gradient of a given point in the surface, several points in the vicinity are randomly chosen and their objective function values need to be evaluated.  Then a quadratic model is fit to the surface of these points in the vicinity using least-squares regression. The gradient of this quadratic model is then used as a surrogate for the true gradient, and the algorithm descends in the corresponding direction. The pseudocode is given in Algorithm ~\ref{Gradient}.

\begin{algorithm}  
\SetAlgoNoLine
\LinesNumbered
    \caption{Model Gradient Descent}  \label{Gradient}
    
        \KwIn{ Initial point $x_0$, learning rate $\gamma$, sample radius $\delta$, sample number $k$, rate decay exponent $\alpha$, stability constant $A$, sample radius decay exponent $\xi$, tolerance $\epsilon$, maximum evaluations $n$}     
        Initialize a list $L$.\\
        Let $x\gets x_0$.\\
        Let $m\gets 0$.\\
            
        \While {(\#function evaluations so far)+ $k$ does not exceed $n$ } {       
            Add the tuple $(x,f(x))$ to the list $L$.\\           
            Let $\delta'\gets \delta/(m+1)^{\xi}$.\\         
            Sample $k$ points uniformly at random from the $\delta'$-neighborhood of $x$. Call the resulting set $S$.\\
            \For {each $x'$ in $S$} {           
                Add$(x',f(x'))$ to $L$.
            }
            Initialize a list $L'$.\\
            \For {each tuple $(x',y')$ in $L$}{              
                \If {$\mid x'-x\mid<\delta'$}{                
                    Add $(x',y')$ to $L'$.
                }
            }
            Fit a quadratic model to the points in $L'$ using least squares linear regression with polynomial features.\\
            Let $g$ be the gradient of the quadratic model evaluated at $x$.\\
            Let $\gamma'=\gamma/(m+1+A)^{\alpha}$. \\           
            \If {$\gamma'\cdot\mid g\mid<\epsilon$} {           
                \Return $x$
            }  
            Let $x\gets x-\gamma'\cdot g$.\\            
            Let $m\gets m+1$.
        }
        \Return $x$.
\end{algorithm}  

\begin{figure}
\centering
\includegraphics[width=0.48\textwidth]{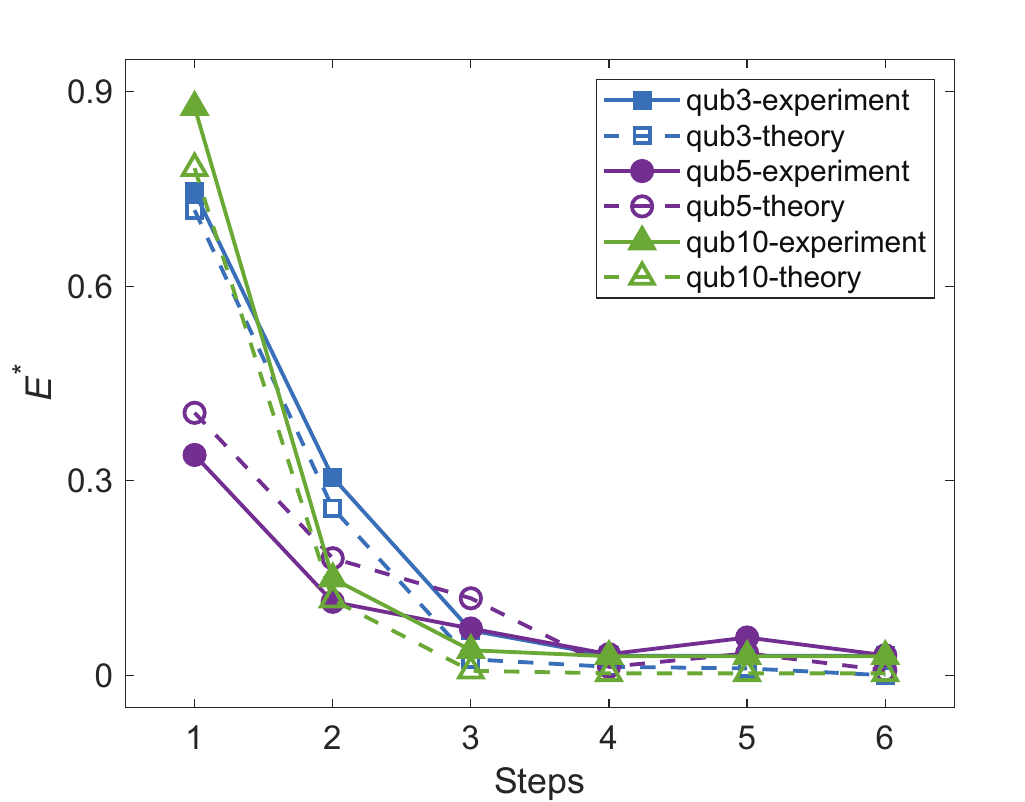}
\caption{Details about the convergence paths for the three factoring cases. The squares, circles and triangles represent for the 3, 5 and 10 qubits cases respectively. The solid (hollow) symbols represent the experiment (theory) results. The vertical coordinates represent the normalized energy function value $E^*$ while the horizontal coordinates represent the computational iteration steps.}\label{path1}
\end{figure}

In our experiment, the MGD method performs well for the three factoring cases. It can converge to the local or global optimum within 10 steps from randomly chosen initial points. Meantime, the experimental convergence results are comparable to the theoretical results at the current scale. The details about the convergence traces can be found in Fig.~\ref{path1}. 

\subsection{10-qubit case up to $p=3$}

\begin{figure*}
\centering
\includegraphics[width=1\textwidth]{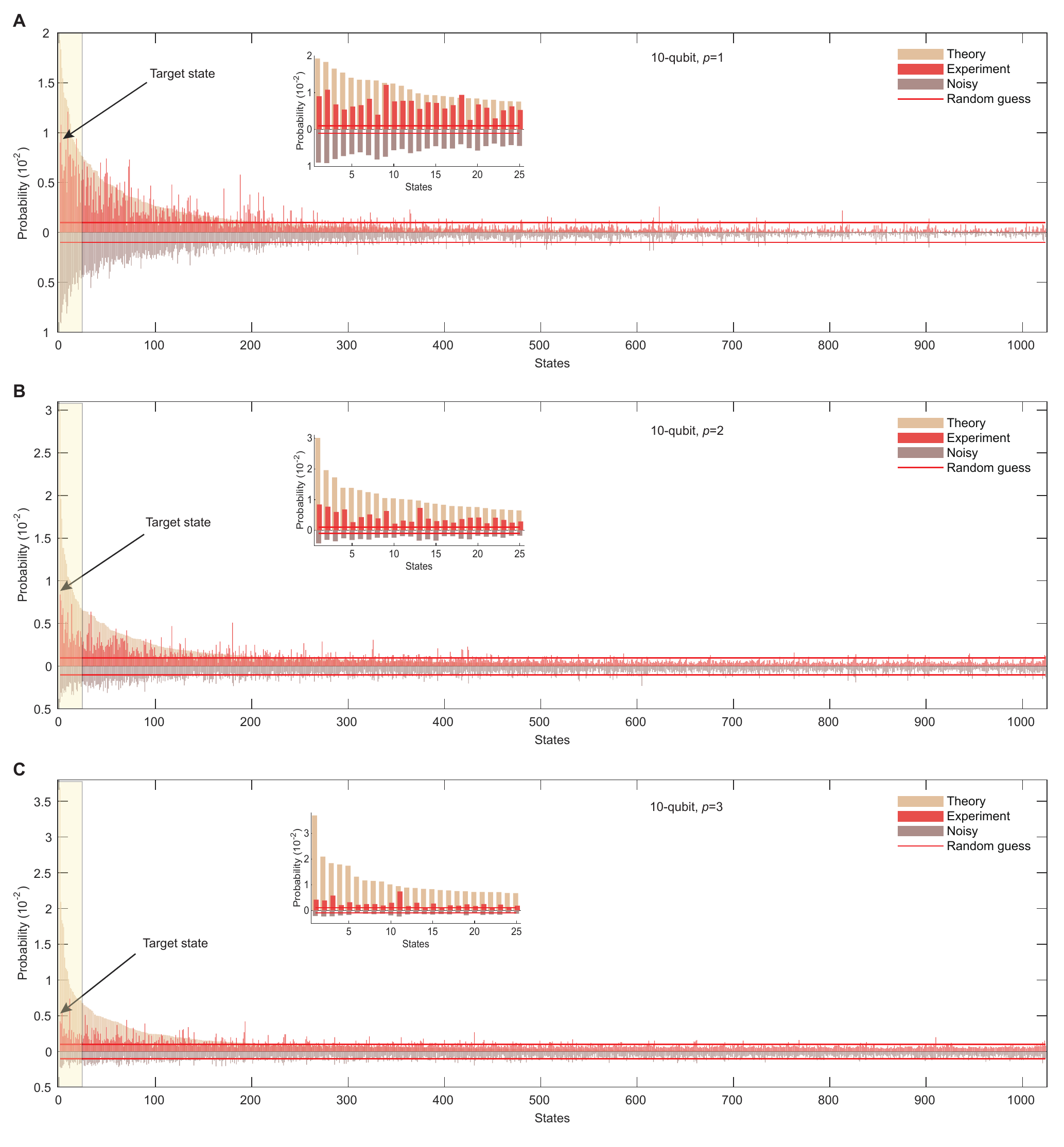}
\caption{Whole statistical histogram for the 1024 states in the  experiment of 10-qubit factoring case, \textbf{A} for $p=1$ case, \textbf{B, C} for $p=2$ and $p=3$ cases respectively. The states are sorted by the probability of the noiseless simulation results and  the target state can be found in the far left of the histogram. The horizontal red lines represent the results of random guess. The inner plot gives the amplified details of the highlight zone.}\label{p3his}
\end{figure*}

Here we present the whole statistical histogram for the 1024 states in the  experiment of 10-qubit factoring case. We also present the noiseless simulation results and 0.01-noise simulation results for comparison. For the noisy simulation, we randomly implement a group of single qubit
Pauli gates in $[X,Y,Z,I]$, and two-qubit Pauli gates in $[X,Y,Z,I]^{\otimes 2}$ after every operation in
the quantum circuit with error rates of $0.2\%$ and $1\%$, respectively. As shown in Fig.~\ref{p3his}\textbf{A}, the histogram results are obtained by excuting the QAOA circuit 30000 times repeatedly. The states are sorted by the probability of the noiseless simulation results which can be take as theory results.  The target state is pointed out with an arrow which can be found in the far left of the histogram. The experimental results for $p=2$ and $p=3$ are also given, see Fig.~\ref{p3his}\textbf{B, C}. 

The performance of QAOA will be improved by increasing the depth of hyperparameter $p$ in theory. However, the errors are accumulated during the increasing of circuit depth and the bonus of the computation can be counteracted. The best performance of QAOA should make balance of the computation bonus and the effects of noise. As can be found in Fig.~\ref{p3his}, the relative ratio of the target state is more significant than the other states as $p$ grows. However, the absolute ratio is reduced. Meanwhile, it is still significantly higher than the results of random guess when $p=3$.   

\section{Postprocessing: the smooth relation pairs and linear equations}
Attached here are other  smooth relation pairs obtained in the examples of factoring integers 1961, 48567227 and 261980999226229, as shown in the following lists. The first column is the sequence number of the sr-pair, and $ \mid u-vN\mid$ is presented in terms of  the corresponded prime basis. In the 3-qubit case, we give 20 independent smooth relation pairs, and the corresponding Boolean matrix is combined by 20 vectors of 16 dimension. Hence there must be a group of linearly dependent vectors, that is, the linear equation system has at least one group of solution. 

\subsection{The 3-qubit case}
\begin{figure*}
\begin{lstlisting}[mathescape]

The smooth relation pairs for the 3 qubits factoring case:

        sn    u                     v           $\mid u-vN \mid$
        1     2^3 * 3^5             1           17                            
        2     2^4 * 5^3             1           3 * 13                        
        3     2^7 * 3 * 5           1           41                            
        4     3^4 * 5^2             1           2^6                           
        5     3 * 5^4               1           2 * 43                        
        6     2^3 * 3^2 * 5^2       1           7 * 23                        
        7     2 * 3^2 * 5^3         1           17^2                          
        8     2^2 * 3^4 * 5         1           11 * 31                       
        9     2^6 * 5^2             1           19^2                          
        10    2^2 * 5^4             1           7^2 * 11                      
        11    2 * 3^3 * 5^2         1           13 * 47                       
        12    2^4 * 3^4             1           5 * 7 * 19                    
        13    3^2 * 5^3             1           2^2 *11*19                 
        14    2^3 * 5^3             1           31^2                          
        15    2^2 * 3^5             1           23 * 43                       
        16    2^5 * 5^2             1           3^3 * 43                      
        17    2^4 * 3^6             5           11 * 13^2                     
        18    2^4 * 3^5             1           41 * 47                       
        19    3^5 * 5^2             1           2 *11^2*17                 
        20    3 * 5^5               2           7 *19*41 
\end{lstlisting}
\end{figure*}

\begin{table}[ht]
\begin{center}
\caption{ Boolean exponential vectors corresponding to smooth relation pairs.  The first column represents the sequence number of the smooth relation pair $(u, v)$. The second column is the sign basis, which represents the positive or negative of $u/(u-vN)$. Columns 3 to 17 represent the Boolean exponents on the first 15 prime basis, respectively.}\label{exponential vector}
\begin{tabular}{@{}ccccccccccccccccc@{}}

\toprule
   sn & sign & $p_1$ & $p_2$ & $p_3$ & $p_4$ & $p_5$ & $p_6$ & $p_7$ & $p_8$ & $p_9$ & $p_{10}$ & $p_{11}$ & $p_{12}$ & $p_{13}$ & $p_{14}$ & $p_{15}$ \\
\hline

 1     & 1    & 1     & 1     & 0     & 0     & 0     & 0     & 1     & 0     & 0     & 0      & 0      & 0      & 0      & 0      & 0      \\
 2     & 0    & 0     & 1     & 0     & 0     & 0     & 1     & 0     & 0     & 0     & 0      & 0      & 0      & 0      & 0      & 0      \\
 3     & 1    & 1     & 1     & 0     & 0     & 0     & 0     & 0     & 0     & 0     & 0      & 0      & 0      & 1      & 0      & 0      \\
 4     & 0    & 0     & 0     & 0     & 0     & 0     & 0     & 0     & 0     & 0     & 0      & 0      & 0      & 0      & 0      & 0      \\
 5     & 1    & 1     & 1     & 0     & 0     & 0     & 0     & 0     & 0     & 0     & 0      & 0      & 0      & 0      & 1      & 0      \\
 6     & 1    & 1     & 0     & 0     & 1     & 0     & 0     & 0     & 0     & 1     & 0      & 0      & 0      & 0      & 0      & 0      \\
 7     & 0    & 1     & 0     & 0     & 0     & 0     & 0     & 0     & 0     & 0     & 0      & 0      & 0      & 0      & 0      & 0      \\
 8     & 1    & 0     & 0     & 0     & 0     & 1     & 0     & 0     & 0     & 0     & 0      & 1      & 0      & 0      & 0      & 0      \\
 9     & 1    & 0     & 0     & 0     & 0     & 0     & 0     & 0     & 0     & 0     & 0      & 0      & 0      & 0      & 0      & 0      \\
 10    & 0    & 0     & 0     & 0     & 0     & 1     & 0     & 0     & 0     & 0     & 0      & 0      & 0      & 0      & 0      & 0      \\
 11    & 1    & 1     & 1     & 0     & 0     & 0     & 1     & 0     & 0     & 0     & 0      & 0      & 0      & 0      & 0      & 1      \\
 12    & 1    & 0     & 0     & 1     & 1     & 0     & 0     & 0     & 1     & 0     & 0      & 0      & 0      & 0      & 0      & 0      \\
 13    & 1    & 0     & 0     & 0     & 0     & 1     & 0     & 0     & 1     & 0     & 0      & 0      & 0      & 0      & 0      & 0      \\
 14    & 1    & 1     & 0     & 0     & 0     & 0     & 0     & 0     & 0     & 0     & 0      & 0      & 0      & 0      & 0      & 0      \\
 15    & 1    & 0     & 1     & 0     & 0     & 0     & 0     & 0     & 0     & 1     & 0      & 0      & 0      & 0      & 1      & 0      \\
 16    & 1    & 1     & 1     & 0     & 0     & 0     & 0     & 0     & 0     & 0     & 0      & 0      & 0      & 0      & 1      & 0      \\
 17    & 0    & 0     & 0     & 0     & 0     & 1     & 0     & 0     & 0     & 0     & 0      & 0      & 0      & 0      & 0      & 0      \\
 18    & 0    & 0     & 1     & 0     & 0     & 0     & 0     & 0     & 0     & 0     & 0      & 0      & 0      & 1      & 0      & 1      \\
 19    & 0    & 1     & 1     & 0     & 0     & 0     & 0     & 1     & 0     & 0     & 0      & 0      & 0      & 0      & 0      & 0      \\
 20    & 0    & 0     & 1     & 0     & 1     & 0     & 0     & 0     & 1     & 0     & 0      & 0      & 0      & 1      & 0      & 0      \\
\botrule
\end{tabular}
\end{center}
\end{table}

Table~\ref{exponential vector} shows a list of Boolean vectors corresponding to the exponents of the prime basis consist of $u/(u-vN)$.
We can see that the fourth vector is an all-zero vector, which is itself a linear correlation vector. The 10-th vector  and  the 17-th vector, the 5-th vector and the 16-th vector, are two groups of linearly dependent vectors.  Other linearly dependent vectors need to be  obtained by solving linear equations. Here each group of linear correlations will correspond to a quadratic congruence equation of the form $X^2\equiv Y^2 \; \text{mod}\;N$. There is a high probability that the factorization of the integer $N$ will be obtained by this equation. Below we will give the details about factorization of $N=1961$ in combination with the specific smooth relation pairs.

According to the above discussion, from the above smooth relation pairs   we can find the solutions of linear equations, such as
\begin{itemize}
\item{Eg.1:} The 4th pair, where $u=3^4 * 5^2, v=1, \mid u-vN\mid=2^6$, which made a quadratic congruence: $(9*5)^2-8^2=N$. Then we have: $p=\text{gcd}(45+8,N)=53, q=\text{gcd}(45-8,N)=37$.

\item{Eg.2:} The 9th pair, where $u=2^6 * 5^2,v=1,\mid u-vN\mid=(19)^2,(8*5)^2+19^2=N$, then according the factoring method of Gauss, it'll lead to a pair of factors. Let
\begin{equation}\label{pq}
 p=x^2+y^2,\quad q=a^2+b^2.
\end{equation} 
Then, we have
  \begin{equation}
  \begin{cases} 
  \mid ax-by\mid=40, \\
  \mid bx+ay\mid=19, 
  \end{cases} 
  ~ or~
  \begin{cases} 
  \mid ax-by\mid=19, \\
  \mid bx+ay\mid=40. 
  \end{cases}
  \end{equation}
  Solving the above equations, we have $a=1,b=6,x=2,y=7$. Substitute the result into Eq.~\ref{pq}, we have $p=53,q=37$. Or $a=2,b=7,x=6,y=-1$, we have $p=37,q=53$.

\item{Eg.3:} The combination of the 10-th and 17-th pair, we have
  \begin{equation}
  (2*5^2*2^2*3^3)^2\equiv(7*11*13)^2\quad\text{mod}\;1961.
  \end{equation}
  Then we have 

\begin{align}
  &p=\text{gcd}(5400+1001,1961)=37,\\ &q=\text{gcd}(5400-1001,1961)=53.
  \end{align}

\item{Eg.4:} The combination of the 5-th and 16-th pair, we have 
\begin{equation}
2^5 * 5^2*3 * 5^4 \equiv 2 * 43*3^3 * 43 \quad \text{mod}\;1961,
\end{equation}
namely
\begin{equation}
  (2^2 * 5^3)^2 \equiv (43*3)^2  \quad \text{mod}\;1961.
\end{equation}
Hence we have
\begin{align}
&p=\text{gcd}(500+129,1961)=37,\\&q=\text{gcd}(500-129,1961)=53.
\end{align}
\end{itemize}
In addition, prime factors can also be obtained from the solution of linear equations of other relationships, which will not be listed here.

\subsection{The 5-qubit case}
In the 5-qubit case, we present 55 independent smooth relation pairs in the following list. The corresponding Boolean matrix containing 55 vectors of 51 dimension (50 dimension for the prime basis plus 1 dimension sign basis). Similarly, there must be a group of linearly dependent vectors.

\begin{figure*}
\centering
\begin{lstlisting}[mathescape]
The smooth relation pairs for the 5 qubits factoring case:

sn    u                              v        |u-vN|
1     5^6 * 7 * 11^3                 3        2^2 * 17 * 23 * 79            
2     5^11                           1        2 * 3 * 11 * 59 * 67          
3     3^3 * 5 * 7^2 * 11^4           2        31 * 53 * 173                 
4     5 * 11^7                       2        3^4 * 61^2                    
5     3^2 * 5^6 * 7^3                1        2^2 * 13 * 37 * 173           
6     2 * 3 * 5^3 * 7^2 * 11^3       1        37 * 83 * 113                 
7     3^6 * 5^3 * 7^2 * 11           1        2^2 * 23 * 47 * 127           
8     2 * 3 * 7^9                    5        13 * 17 * 53 * 61             
9     3^3 * 11^6                     1        2^3 * 5 * 17 * 23 * 47        
10    3^2 * 5^9 * 11                 4        53 * 131^2                    
11    2^3 * 3^4 * 5 * 11^4           1        37 * 137 * 223                
12    2^4 * 5^2 * 7^6                1        11 * 23 * 59 * 101            
13    3^2 * 5^3 * 7^3 * 11^2         1        2^7 * 107 * 137               
14    2^4 * 3^7 * 11^3               1        5^3 * 107 * 149               
15    2^7 * 3 * 5 * 7^4 * 11         1        13 * 37 * 61 * 73             
16    2^8 * 5^7 * 7                  3        17^2 * 109 * 181              
17    2^7 * 3^7 * 7 * 11^2           5        17 * 19 * 113 * 157           
18    2 * 3 * 5^4 * 11^4             1        23 * 43^2 * 149               
19    7^3 * 11^5                     1        2 * 3^6 * 23 * 199            
20    2^7 * 5 * 7^2 * 11^3           1        3^2 * 13 * 23 * 43 * 59       
21    5^7 * 11^3                     2        3 * 13 * 37 * 47 * 101        
22    2^2 * 5^3 * 7^6                1        3^5 * 13 * 17 * 191           
23    2^6 * 3^5 * 7^4                1        5^4 * 11 * 23 * 71            
24    3^7 * 5^2 * 7^3 * 11           4        17 * 61 * 67 * 173            
25    5 * 11^9                       243      2 * 41 * 43 * 47 * 73         
26    5 * 7 * 11^6                   1        2^5 * 3 * 19 * 53 * 139       
27    2^2 * 3^8 * 7^4                1        11^2 * 19 * 61 * 103          
28    2^2 * 3 * 5^8 * 7              1        31 * 43 * 53 * 223            
29    2^6 * 5^2 * 7 * 11^4           3        13^2 * 37^2 * 79              
30    2^4 * 11^6                     1        7^3 * 19 * 29 * 107           
31    3^4 * 5 * 7^5 * 11             2        29 * 41 * 97 * 193            
32    2^2 * 5^7 * 7 * 11             1        13 * 71 * 139 * 191           
33    2^2 * 5^8 * 7 * 11             3        29 * 67 * 73 * 179            
34    3^2 * 5^2 * 7 * 11^4           1        2^2 * 19 * 37 * 47 * 193      
35    2^7 * 3^6 * 5 * 7^2            1        37 * 43 * 107 * 151           
36    3^3 * 7^7                      1        2 * 43 * 53^2 * 109           
37    2^2 * 5^2 * 7^5 * 11           1        3 * 19 * 41 * 61 * 211        
38    2^5 * 3^3 * 5^2 * 7 * 11^2     1        29 * 31 * 151 * 223           
39    2 * 5^6 * 7^2 * 11             1        29 * 61 * 79 * 227            
40    2^2 * 5^2 * 11^5               1        3^5 * 19 * 79 * 89            
41    3^2 * 5^4 * 11^4               1        2 * 7 * 97 * 139 * 179        
42    5^4 * 7^4 * 11^2               3        2^3 * 17 * 31 * 67 * 127      
43    2^3 * 3^2 * 5^6 * 7 * 11       1        13 * 29^2 * 59^2              
44    3^2 * 5^4 * 7^3 * 11^2         4        19^3 * 29 * 197               
45    7^4 * 11^5                     9        2^3 * 13 * 17 * 19^2 * 79     
46    2 * 5^4 * 11^5                 3        19 * 103 * 157 * 181          
47    3 * 5^2 * 11^6                 4        37 * 73 * 127 * 179           
48    2^2 * 3^4 * 5^5 * 11^2         1        29 * 109 * 149 * 157          
49    2^2 * 5^3 * 7^2 * 11^4         9        13 * 29^2 * 71 * 101          
50    3^7 * 5^2 * 7^4                1        2^3 * 17 * 23 * 137 * 193     
51    2 * 5 * 7^3 * 11^4             3        23 * 29 * 37 * 53 * 73        
52    2^2 * 3 * 7^7 * 11             5        47 * 107 * 149 * 179          
53    3 * 11^8                       10       7 * 29 * 67 * 71 * 163        
54    2 * 5^5 * 7^6                  11       3 * 43 * 83 * 89 * 211        
55    3 * 5^5 * 7^2 * 11^3           8        23 * 41 * 47^2 * 107     

\end{lstlisting}
\end{figure*}

\begin{figure*}
\begin{equation}
\label{xy_solution}
\begin{split}
X=~&75695763106501556705305764502754936587819598184067351577433032507856332859\\&
5679423010251901850133526999585453943501610097294132617819824833831101214591\\&
4436530877754829641771815233949422940291369528064063220752769498277701462410\\&
8310326460216549269998237692558639726079463565070924375206289830057722589435\\&
9044218428126425544707265515259392507522823290552782790244143095915540083328\\&
6647355489098701203665266097543096430161981191012698547957753245615400642156\\&
6009521484375000000000000000000000000000000000000000000,\\
Y=~&89703025676439146909634318953050859996463498582497692905363926386871531215\\&
1470596839682183291916847167330486410267725009210425053361066685989493430651\\&
6222448335298694730188139419712140836303477738769684013638969809410021181628\\&
1484584566446454557099758814179896192053526153001893986178123643916393821728\\&
8509975066081055665376292175821267217313751458334859802980440111341258224039\\&
13885046671262199158753471668113044162973340975659170623801.
\end{split}
\end{equation}
\end{figure*}

Due to the large dimension of the vectors corresponding to the 5-qubit case,  we just present a set of solution for the linear equation system:

\begin{equation}
\begin{split}
x=&(0, 0, 0, 0, 0, 0, 0, 0, 0, 1, 0, 0,
1, 1, 0, 0, 0, 1, 0, 0,  \\& 1, 1,0, 0, 0, 1, 0, 0, 0, 0, 0, 1, 0, 1, 0, 0, 
0, 0, 0, 0, \\&0, 0, 0, 0, 0,  0,0, 0, 1, 1, 0, 0, 0, 0, 0).
\end{split}
\end{equation}
The corresponding solution for the quadratic conjugation is

\begin{equation}
\begin{split}
&X=639232456435359657331994419097900390625,\\
&Y=12136572734325633629343926054845304.
\end{split}
\end{equation}

It is easy to verify that the solution satisfies the equation:
\begin{equation}
X^2\equiv Y^2\quad \text{mod}\;N
\end{equation}
Furthermore, we have
\begin{equation}
\begin{split}
p&=\text{gcd}(X+Y,N)
\\&=(639232456435359657331994419097900390625+\\& 12136572734325633629343926054845304,48567227)
\\&=7919,
\\
q&=\text{gcd}(X-Y,N)
\\&=(639232456435359657331994419097900390625-\\& 12136572734325633629343926054845304,48567227)
\\&=6133.
\end{split}
\end{equation}

Finally, we obtain the factorization result: $N=48567227=7919\times 6133$.

\subsection{The 10-qubit case}
In the 10-qubit case, we present 221 independent smooth relation pairs in the Supplementary Data. The corresponding Boolean matrix containing 221 vectors of 201 dimension (200 dimension for the prime basis plus 1 dimension sign basis). We present a set of solution for the linear equation system:

\begin{equation}
\begin{split}
x=&(1, 0, 0, 0, 0, 0, 0, 0, 0, 0, 0, 0, 0, 0, 0, 0, 0, 0, 0, 0,\\&
0, 0, 0, 0, 0, 0, 0, 0, 0, 0, 0, 0, 0, 0, 0, 0, 0, 0, 0, 0, 1, \\&
1, 1, 0, 1, 0, 0, 0, 0, 0, 1, 0, 0, 0, 0, 0, 1, 0, 1, 0, 0, 0, \\&
0, 0, 0, 0, 1, 1, 0, 0, 0, 1, 0, 1, 1, 1, 1, 1, 0, 0, 1, 0, 1, \\&
1, 1, 1, 0, 1, 0, 0, 1, 0, 1, 0, 1, 1, 1, 0, 0, 1, 1, 0, 1, 0, \\&
0, 1, 1, 1, 0, 0, 1, 0, 1, 0, 0, 0, 0, 1, 0, 0, 1, 1, 0, 1, 1, \\&
1, 0, 0, 0, 1, 1, 0, 0, 0, 0, 0, 1, 0, 0, 1, 1, 0, 0, 1, 1, 1, \\&
0, 0, 0, 1, 0, 0, 0, 0, 0, 0, 1, 0, 1, 1, 1, 1, 0, 1, 0, 0, 1, \\&
0, 0, 0, 0, 0, 0, 0, 0, 1, 1, 1, 0, 0, 1, 0, 0, 0, 0, 0, 0, 0, \\&
0, 1, 0, 0, 0, 0, 1, 1, 1, 0, 0, 0, 1, 0, 0, 0, 1, 1, 1, 1, 1, \\&
1, 0, 0, 0, 0, 0, 1, 0, 0, 1, 0, 0).
\end{split}
\end{equation}
The corresponding solution for the quadratic conjugation is presented in Eq. \ref{xy_solution}.

It is easy to verify that the solution satisfies the equation:
\begin{equation}
X^2\equiv Y^2\quad \text{mod}\;N
\end{equation}
Furthermore, we have
\begin{equation}
\begin{split}
p=\text{gcd}(X+Y,N)=15538213,
\\
q=\text{gcd}(X-Y,N)=16860433.
\end{split}
\end{equation}

Finally, we obtain the factorization result: 
\begin{equation}
N=261980999226229=15538213\times 16860433.
\end{equation}

\begin{figure}
\centering
\includegraphics[width=0.48\textwidth]{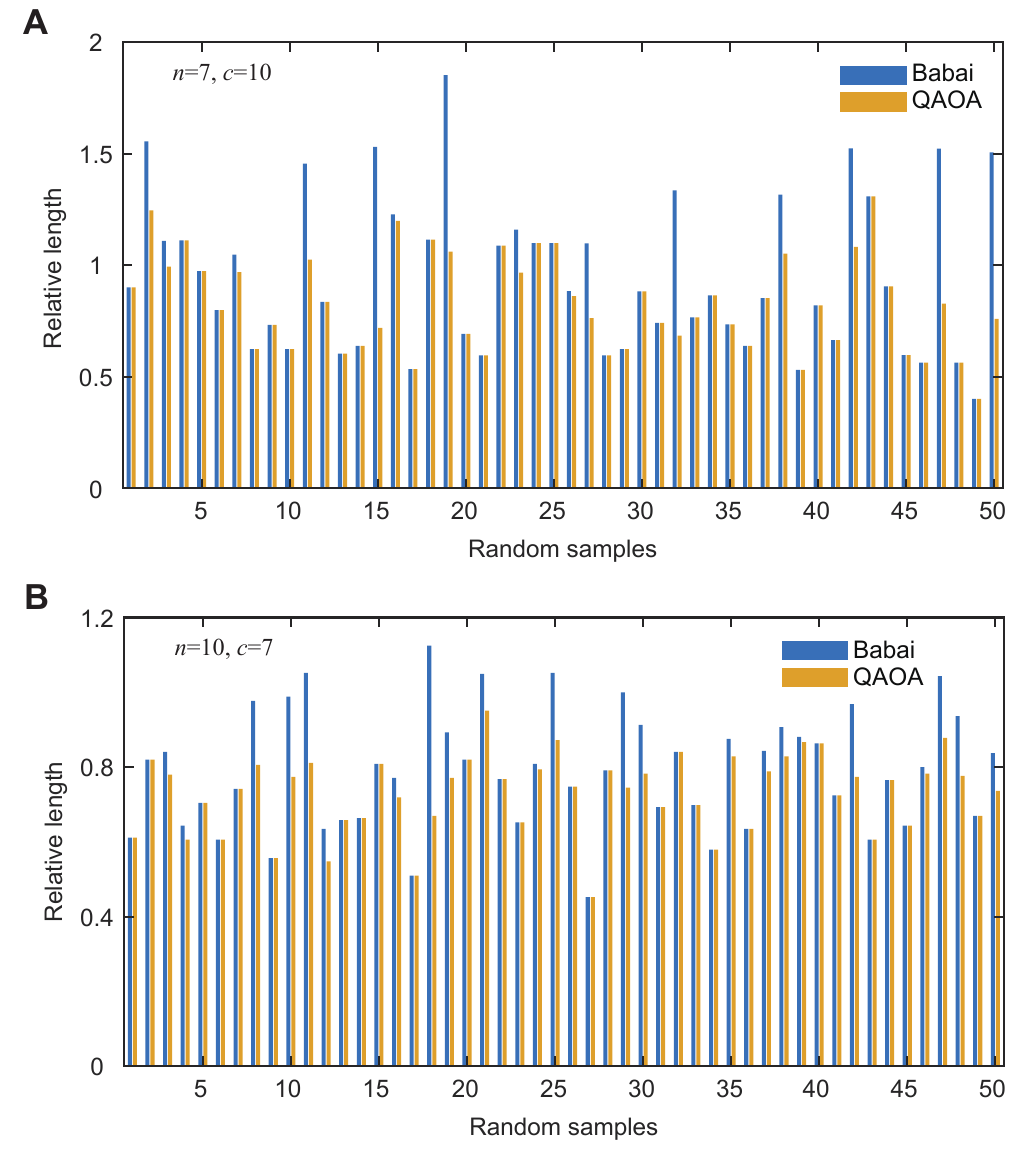}
\caption{Performance on random samples for the quantum optimizer (QAOA) over Babai's algorithm. \textbf{A (B)} for the results of 50 random CVP samples under the condition of $n=7, c=10~(n=10, c=7)$. The yellow (blue) bars represent the results of the quantum optimizer (classical Babai's algorithm). We can observe that the quantum optimization results are better than the classical results in many cases.}\label{rand1}
\end{figure}

\section{The exploration of quantum advantage}
In this part, we explore the advantage of the quantum optimizer compared to Babai's algorithm numerically. The  measurable criteria considered is the quality of the short vectors for CVP. The quality of the short vector is positively related to the efficiency of obtaining smooth relation pairs in Schnorr's sieve method. The higher the quality of the short vector, the more efficient the factoring method. Since it is an open question  to estimate the analytical complexity of the QAOA algorithm at present, in the discussion here, it is assumed that the QAOA procedure can give the optimal solution to the optimization problem in a limited time.  Here we use the relative distance parameter $r$ to measure the length of the vector instead of  the Euclid norm or square norm. The parameter is specifically defined by
\begin{equation}
r=\lVert\mathbf b-\mathbf t\rVert^2/\text{det}(\mathbf B'_{n,c})^{\frac{2}{n}},
\end{equation}
where $\mathbf B'_{n,c}=[\mathbf B_{n,c},\mathbf N_c]$. This parameter uses the $2/{n}$-power of the determinant of the extended lattice $\mathbf B'_{n,c}$  to measure the relative length of the short vector $\mathbf b-\mathbf t$, which can reduce the effects of different determinant of lattices to a certain extent on short vector quality.
\subsection{The random sample results}
We first study the performance  of the quantum optimizer and the classical Babai's algorithm on random samples of CVP. Here, we generate 50 random CVP (lattice and target vector) samples under the condition of lattice dimension $n=7$,  precision $c=10$ and $n=10 $, $c=7$ respectively. For each random sample,  the lattice determinant and the target vector are the same, only the main diagonal elements of the lattice are randomly permuted. The results can be found in Fig.~\ref{rand1}.  Here the horizontal axis represents random samples, and the vertical axis represents the relative quality  $r$ of the result vector.  The blue (yellow) bars represent the results of the classical Babai's algorithm (quantum optimization). As shown in the picture, the quantum-optimized results are not worse than the classical results. And in many cases, the quantum optimization results are significantly better than the classical results, i.e., shorter vectors are obtained. 

\begin{figure}[t]
\centering
\includegraphics[width=0.48\textwidth]{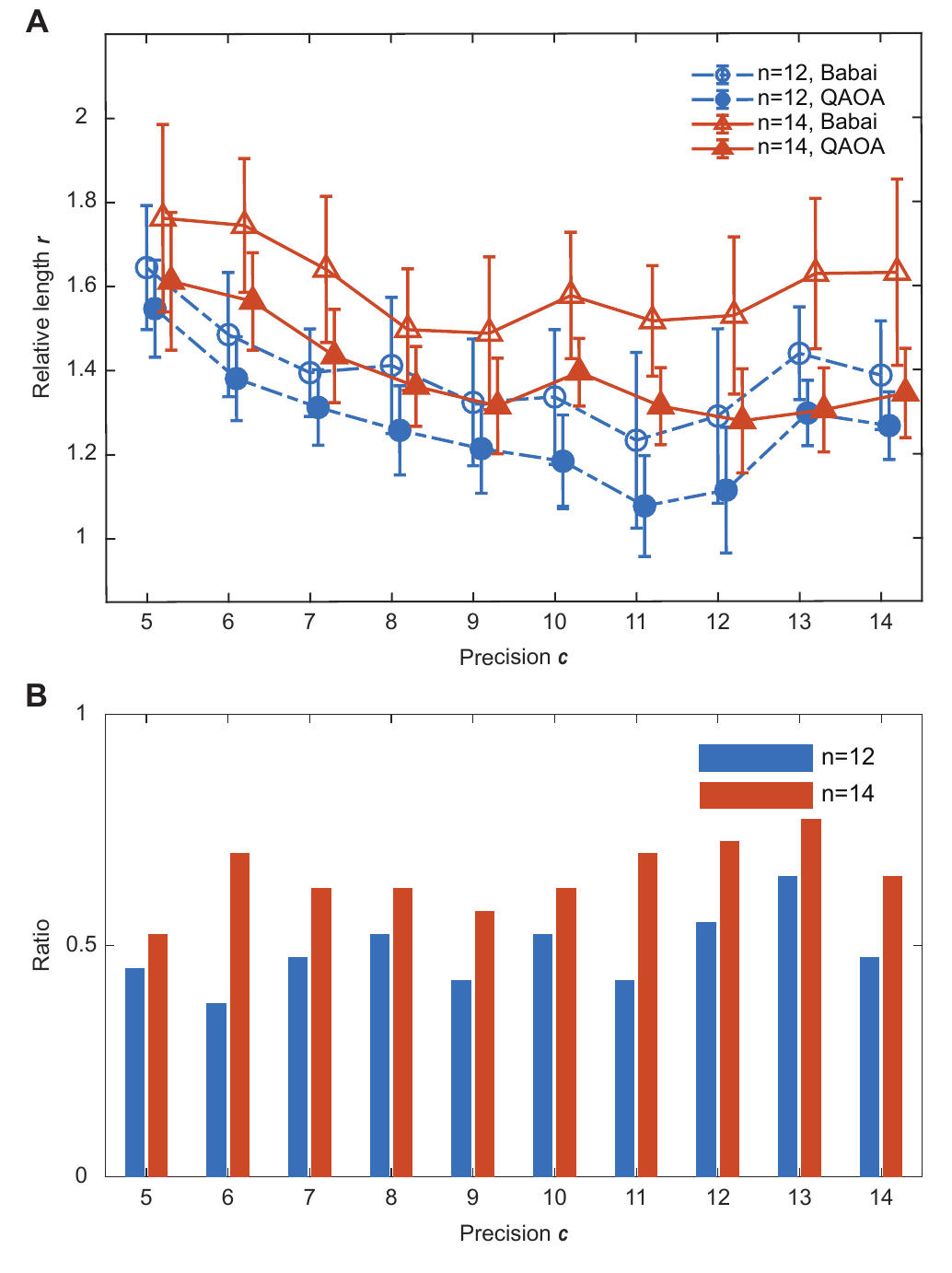}
\caption{Performance of the quantum optimizer with the increasing precision of the lattice. \textbf{A}, Relative length results for the quantum and classical methods. The horizontal axis represents precision parameter $c$,  which is positively correlated with the determinant of the lattice. \textbf{B}, Advantage ratio for the 40-random samples, blue (orange) bars for the $n=12$ ($n=14$) case. We can see from the picture that the quantum optimization results is better than that of Babai's in average cases, especially when the determinant (positively correlated with $c$) of the lattice is large.}\label{prec1}
\end{figure}

\begin{figure}[t]
\centering
\includegraphics[width=0.48\textwidth]{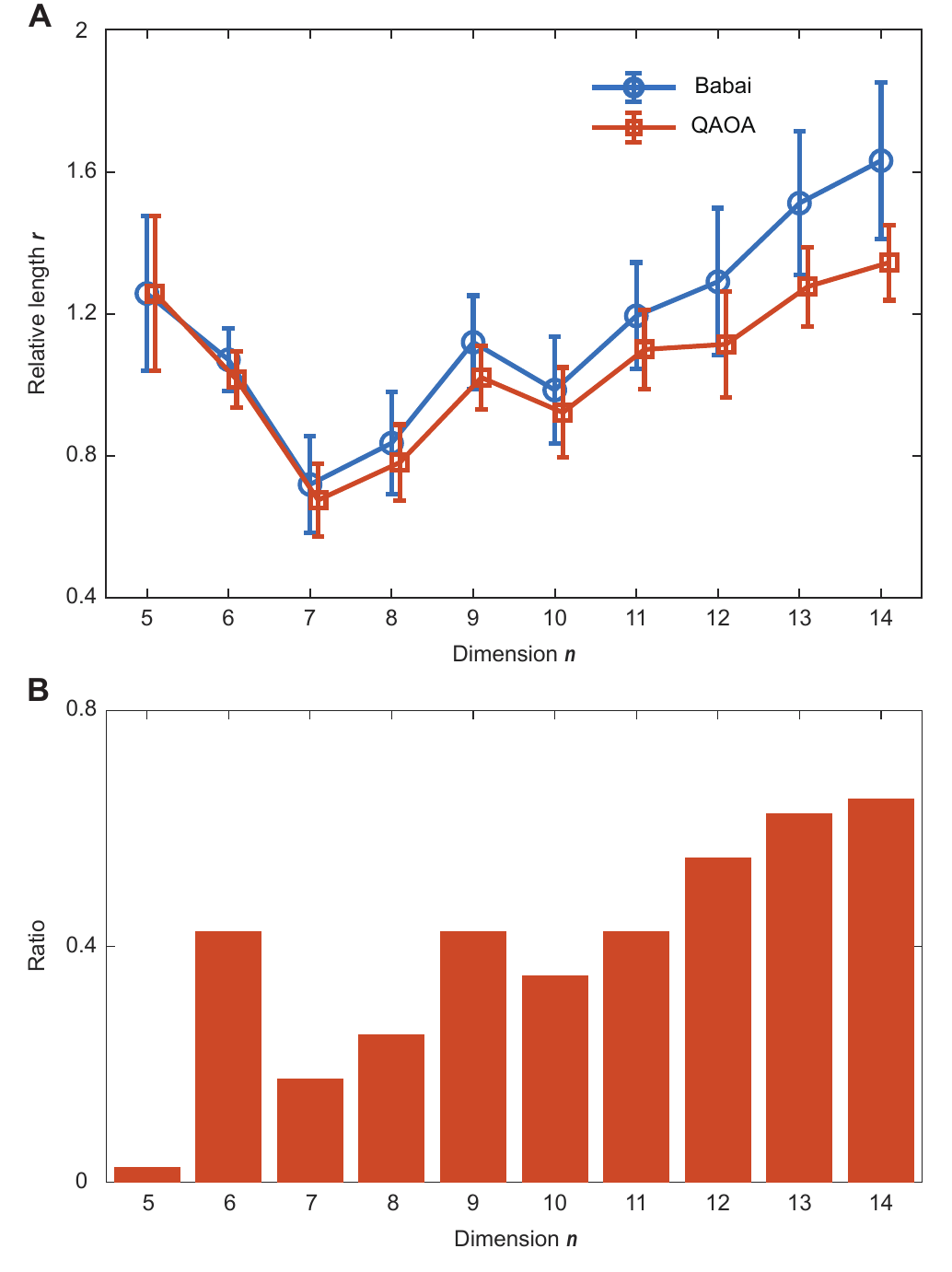}
\caption{Performance of the quantum optimizer with the increasing dimension of the lattice. \textbf{A}, Relative length results for the quantum and classical methods. The horizontal axis represents lattice dimension $n$. The results are averaged over 40  random samples with $c=n$. The error bars give a confidence interval under a unit standard deviation. \textbf{B}, Advantage ratio for the 40-random samples. We can find that the relative distance gap between the quantum and classical results becomes more significant as the dimension of the lattice grows.}\label{evidence2}
\end{figure}

\subsection{Quantum advantage and  lattice precision  }
We further study the advantage of the quantum optimizer with increasing precision parameter $c$ of the lattice. In Fig.~\ref{prec1}\textbf{A},  we present the numerical results when the parameter $c$ increases from $5$ to $14$ in the dimension of $ n = 12, 14 $ separately. The results are averaged by 40 randomly generated CVP samples for each set of parameters $\{n, c\}$. Among them, the circles (triangles) represent the calculation results of $ n = 12$ ($ n=14 $). The solid (hollow) symbols represent the results of quantum (classical). The error bars give a confidence interval under a unit standard deviation. In both the case of $ n = 12 $ and $ n = 14 $, a shorter vector is obtained after quantum optimization. Taking the $ n = 14 $ case as an example, we can see that the quality gap between the results of Babai's algorithm and the quantum algorithm gradually increases with the increase of the parameter $c$, which indicates that the vector quality after quantum optimization is higher than that of Babai's algorithm in an average sense when the determinant of the lattice grows. In addition, we have counted the advantage sample ratio for the quantum results over the 40 random samples, shown by the blue and orange bars in Fig.~\ref{prec1}\textbf{B}. We found that the ratio of quantum advantages at $ n = 12 $ is about $0.5$, and this proportion increases to $0.65$  when $ n = 14 $. The results indicate that quantum advantage becomes more significant when the lattice dimension increases. This results will be further demonstrated in the following part.

\subsection{Quantum advantage and  lattice dimension}
Here we study the the relation between the advantages of quantum optimizer versus the dimension of lattice up to $n=14$.  For each dimension $n$, the results are averaged over 40  random samples with $c=n$. As can be found in Fig.~\ref{evidence2}\textbf{A}, the vector quality after quantum optimization is higher than the result of the classic Babai algorithm in the average sense. That is, we can find a shorter vector through quantum optimization. The quality gap between the quantum and classical results becomes more significant as the dimension of the lattice grows, which means the advantage of the quantum method becomes more significant in the larger system. At the same time, we have counted the ratio of the quantum advantage over the 40 random samples, showing the results in Fig.~\ref{evidence2}\textbf{B}. The advantage ratio is highlighted when the dimension increases, which is consistent with the different trends of the vector quality curves in Fig.~\ref{evidence2}\textbf{A}. Both results indicate that the advantages of quantum methods will become more and more obvious when the dimension of the lattice is increased.
 
\section{The resource estimation for RSA-2048}

\subsection{Introduction}
How many quantum resources does it take to factor a 2048-bit RSA integer? This part we focus on the specific quantum resources required to factor a 2048-bit RSA integer based on the SQIF algorithm. The quantum resources considered mainly include the number of physical qubits and the depth of the QAOA circuit with single layer. Usually, quantum circuits cannot be directly executed on quantum computing devices, as their design does not consider the qubit connectivity characteristics or the topology construction of actual physical systems. The execution process often requires additional quantum resources such as ancilla qubits and extending circuit depths. We discuss the quantum resources required to factor real-life RSA numbers in terms of complete graph topology (Kn),  2-dimensional lattice topology (2DSL), and one-dimensional chain topology (LNN), respectively. We demonstrate with specific schemes that the embedding process needs no extra qubits overhead.
Furthermore, the circuit depth of QAOA with a single layer is linear to the dimension $n$ of the quantum system for all the three topology systems. As a result, we consume a sublinear quantum resources to factor integers using the SQIF algorithm. Taking RSA-2048 as an example, the number of qubits required is $n= 2*2048/\text{log}2048\sim 372$. The quantum circuit depth of QAOA with single layer is $1118$ in Kn topology system, $1139$ in 2DSL system and $1490$ in the simplest LNN system, which is achievable for the NISQ devices in the near future or even today.

\subsection{Problem description}

\begin{figure*}
\centering
\includegraphics[width=0.95\textwidth]{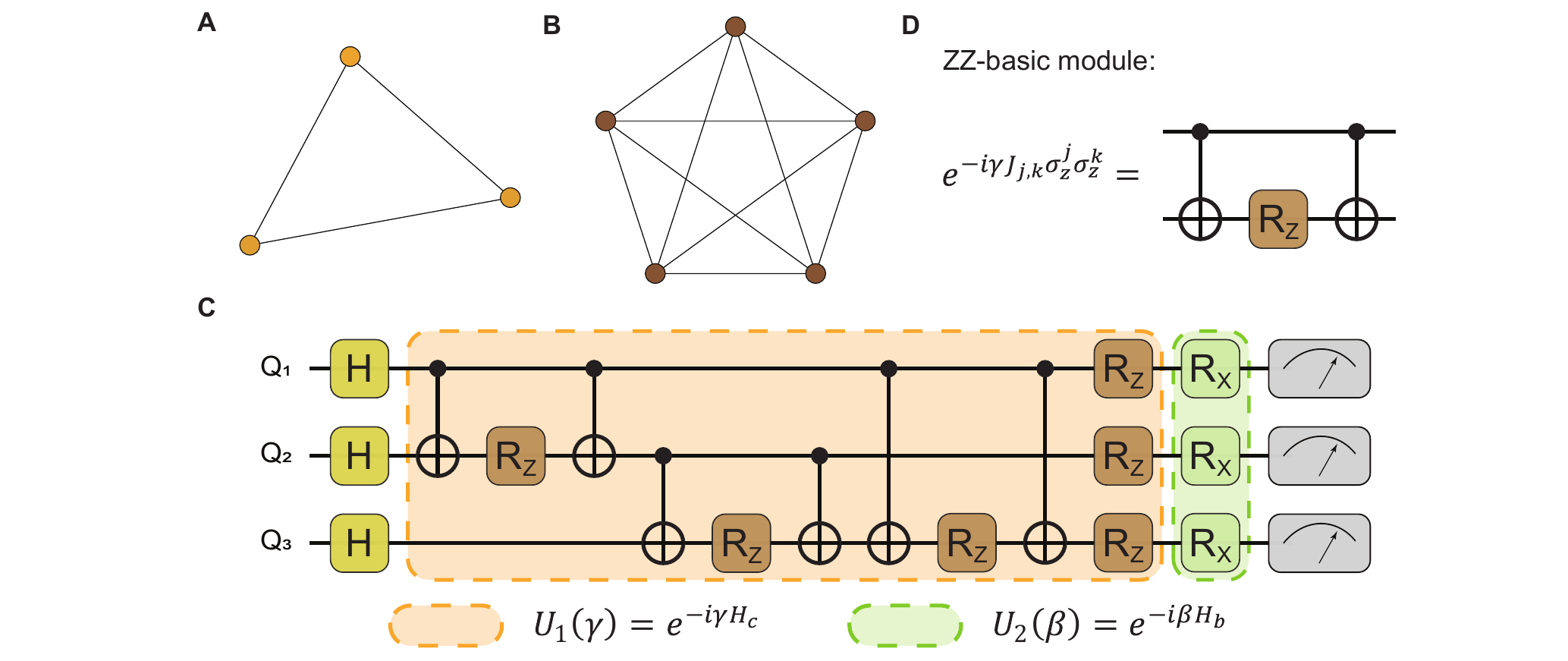}
\caption{Qubits connectivity and the typical quantum circuit of QAOA. \textbf{A} represents the qubits connectivity for the 3-qubit case, which is a $K_3$ graph. \textbf{B} for the 5-qubit case, which is a $K_5$ graph. \textbf{C}. A typical QAOA circuit with single layer for the 3-qubit QUBO Hamiltonian. It manly consists of two  $U_1(\gamma)$ and $U_2(\beta)$, which correspond to the evolution operator of the problem Hamiltonian $H_c$ and mixing Hamiltonian $H_b$.  \textbf{D}. ZZ-basic module. It is consist of two CNOT gates and a single qubit Rz rotation arranged together as a sandwich.}\label{cir1}
\end{figure*}

First, we review the construction of the problem Hamiltonian $Hc$. Using the single-qubit encoding rules, the corresponding Hamiltonian $Hc$ could be taken as a 2-dimensional Ising model of the following form:
\begin{equation}
Hc=\sum_{i=1}^{n}{h_i\sigma_z^i}+\sum_{i,j=1}^{n}{J_{i,j}\sigma_z^i\sigma_z^j},
\end{equation}
where the parameters $h_i, J_{i,j}$ are determined by the coefficients of the primary and quadratic terms of the quadratic unconstrained binary optimization (QUBO) problem . The summation symbol on the right side of the second equation above traverses all subscript combinations. If we regard each quadratic term as an edge in an undirected graph, all the ZZ-terms $\{\sigma_z^i\sigma_z^j\}_{i<j}$  will form an $n$-order complete graph Kn. Namely, the connectivity topology of the logical qubits is a Kn-graph. Take the 3-qubit case and 5-qubit case in the main text as examples, the qubit topology of the problem Hamiltonian is a 3-order and 5-order complete graph, as shown in Fig.~\ref{cir1}\textbf{A, B}, respectively.

A typical QAOA circuit for the Kn-type Hamiltonian is shown in Fig.~\ref{cir1}\textbf{C}.  Two types of unitaries are involved in the quantum circuit of single-layer QAOA iteration.  $U_1(\gamma)$  is the evolution operator of the problem Hamiltonian $H_c$. $U_2(\beta)$ is the evolution operator of the mixing Hamiltonian $H_b$, which consists of single qubit rotations around the x-axis. $U_1(\gamma)$ can be implemented by the single qubit Rz-rotations and the two-qubit blocks, which are corresponding to the local terms (primary) and ZZ-terms of the problem Hamiltonian $H_c$. We mainly focus on the two-qubit blocks which is specifically defined as 
\begin{equation}
 ZZ_{j,k}(\gamma )=e^{-i\gamma J_{j,k}\sigma_z^j\sigma_z^k}.
\end{equation}
The $ZZ_{j,k}(\gamma )$ unitary  can be realized  by the combination of two CNOT gates and one Rz gate like a sandwich, as shown in Fig.~\ref{cir1}\textbf{D}. In the following discussions, we'll take the gate-combination as a basic module of depth 3 without considering its compilation in a specific physical system. The actual physical system often compiles this module according to its native universal quantum gates, and this often increases at most $O(1)$ additional quantum operations and circuit depths.

Since the circuit depth of the single-qubit operations in $U_1(\gamma)$ and $U_2(\beta)$ is at most 2 in a physical system, we will not discuss them later. We focus on the embedding problem of the quadratic terms in $U_1(\gamma)$ into the  physical system.  Specifically, to estimate the overhead of qubits and the circuit depth after embedding a group of ZZ-terms which form a Kn-type circuit. This problem will be referred to as a Kn-type embedding problem in the following part. 

\subsection{Circuit depth  under complete graph topology}

We first consider an ideal situation that any two qubits can interact directly. Namely, the qubits connectivity of the  physical system is a complete graph.  In this scenario, the Kn-type embedding problem does not need additional qubits or quantum SWAP gates. Therefore, the depth of the quantum circuit can be made optimal. 

The $n$-node complete graph Kn contains $ n(n-1)/2 $ edges, which means the QAOA circuit contains $O(n^2)$ ZZ-basic modules. The depth of the single-layer QAOA circuit is $O(n^2)$ without optimization. Since the ZZ-terms in operator $U_1(\gamma)$ are mutual commutes, we are free to rearrange the order of all two-qubit interactions to minimize the depth of the embedded circuit. Here we introduce an optimization scheme based on the maximum matching theory in an undirected graph, which will reduce the depth of the circuit to $O(n)$. The scheme is optimal when considering the ZZ-term as a basic module.

\begin{definition}\label{matching}
{Matching and maximum matching:} Denote an undirected graph by $G(V, E)$,  and $M\subseteq E(G)$ is a subset of edges, satisfying: $\forall (e_i,e_j)\in M$, $e_i,e_j$ are not adjacent in $G$. Edges in $M$ do not share a common vertex. Then $M$ is said to be a matching of $G$. For each edge $e=(u, v)$ in the matching $M$, we call the edge $e$, or the vertices $u, v$ matched by $M$. Each vertex in the graph is either not matched by $M$ or only matched by one edge in $M$. If there is no matching $M'$ in $G$ such that $|M'|>|M|$, then $M$ is said to be a maximum matching in $G$. If every vertex in $G$ is matched by $M$, then $M$ is said to be a perfect matching of $G$.
\end{definition}
According to the definition, a perfect matching must be a maximum matching. The number of maximum matchings in a complete graph can be answered by the following lemma.
\begin{lemma}\label{lemma_graph}
(Maximum matching in the complete graph)
There are $2n-1$ perfect matchings with non-repeated edges in an even-order complete graph $K_{2n}$. There are $2n-1$ maximum matchings with non-repeated edges in an odd-order complete graph $K_{2n-1}$. 
\end{lemma}
\begin{proof}
For an even-order complete graph $K_{2n}$, let the vertices of a $(2n-1)$-regular polygon be $v_1,v_2,...,v_{2n-1}$, and add a vertex $ v_{2n}$ at the center. Take any vertex $v_i, 1\le i\le 2n-1$, construct a matching $M_i$ including edge $(v_i, v_{2n})$ and all the edges perpendicular to it. It is easy to prove that $M_i$ is a perfect match for any $i$, and there is no common edge between $M_i$ and $M_j$ when $ i \ne j$. Therefore, an even-order complete graph $K_{2n}$ has $2n-1$ perfect matchings with non-repeated edges. Since there are total $n(2n-1)$ different edges in $K_{2n}$, and the $2n-1$ different perfect matchings already matched $n(2n-1)$  non-repeated edges, there are no other matching with non-repeated edges.
For the odd-order complete graph $K_{2n-1}$, we need to add an auxiliary vertex $v_{2n}$, then the situation is turned to the even-order case. The difference is that we need to remove the edge $(v_i, v_{2n})$ in each perfect match. As a result, we get $2n-1$ maximum matchings for the odd-order case. This completes the proof.
\end{proof}
\begin{proposition}
If the qubits connection of the  physical system forms a complete graph, then the Kn-type Hamiltonian can be embedded in the physical system without additional qubits, and the depth of the embedded quantum circuit is $O(n)$.
\end{proposition}
\begin{proof}
According to the definition of matching in Definition \ref{matching}, there is no common vertex between the edges that belong to the same matching. So the corresponding two-qubit interactions can be performed on the quantum circuit simultaneously and in parallel.  Suppose each ZZ-term is compiled using the ZZ-basic module shown in Fig.~\ref{cir1}\textbf{D}, then the circuit depth of the ZZ-terms in the same matching is 3. According to Lemma \ref{lemma_graph}, when $n$ is even, there are n-1 non-repeated perfect matchings in a complete graph, and these perfect matchings infiltrate all edges in graph Kn. Therefore, we can construct a quantum circuit with $n-1$ layers according to the $n-1$ perfect matchings. And each layer executes $n/2$ ZZ-basic modules without common qubits in parallel. As the circuit depth of each layer is 3,  the total circuit depth is $3(n -1)$. When $n$ is odd, the complete graph Kn has $n$ maximum matchings with non-repeated edges by Lemma \ref{lemma_graph}. Each maximum matching infiltrates $(n-1)/2$ edges, so these maximum matchings infiltrate all edges in Kn. Similarly, a quantum circuit with $n$ layers can be constructed, and each layer executes $(n-1)/2$ ZZ-terms  without common qubits in parallel, with a total circuit depth of $3n$. In summary, the Kn-type embedding process can be implemented   without additional qubits and the depth of the embedded quantum circuit is $O(n)$. This completes the proof.
\end{proof}
The proof of Lemma \ref{lemma_graph} gives the exact construction of each perfect matching of the complete graph Kn, which means an exact construction scheme of the embedded quantum circuit with depth $O(n)$ is given. The scheme is optimal when considering the ZZ-term as a basic module. In this situation, the number of ZZ-terms in a perfect matching is maximized, namely, the quantum gate operations have the highest parallelism. Since all the maximum (perfect) machines cover all the edges non-repeatedly, the scheme is optimal.

The qubit connectivity is a valuable resource in NISQ devices. Usually, it is difficult to achieve large-scale fully connected topology for actual quantum systems. However, it is easier to realize in some special quantum systems such as the trapped ions system~\cite{pagano2020quantum}, the optical quantum system, and the system with large quantum memory~\cite{gouzien2021factoring}.

\subsection{Circuit depth under linear chain topology and lattice topology}

The linear chain topology is  one of the most common  structures, which can be realized relatively easily in real quantum systems. This topology, also known as linear nearest neighbor (LNN) architecture, in which the qubits are arranged on a line and only the nearest neighbor couplings are  available. In this section, we focus on discussing the resource of quantum gate  and circuit depth when embedding  a Kn-type Hamiltonian into a one-dimensional linear chain system. The results for the lattice system can be made as a corollary.

The embedding problem from arbitrary types of Hamiltonian topology into LNN has been widely studied~\cite{takahashi2007quantum,kutin2006shor,cheung2007translation,hirata2009efficient,saeedi2011synthesis,wille2016look,farghadan2017quantum}, and some  mature methods have been formed. In 2007, Donny Cheung et al. studied the overhead of mutual conversions between various topology models based on the graph-theoretic model~\cite{cheung2007translation}. They pointed out that mapping arbitrary circuits to LNN would require at most $O(n)$ extra depth overhead based on the linearity of parallel sorting. However, no specific conversion scheme is given for mapping Kn-type Hamiltonian to LNN. In 2009,  Yuichi Hirata proposed an efficient method to map arbitrary quantum circuits to LNN based on the idea of bubble sorting~\cite{hirata2009efficient}.  In 2021, the researchers of Google applied the circuit of parallel bubble sorting to complete the embedding from $K_{17}$ to LNN and conducted related experiments on  Sycamore superconducting quantum processor~\cite{harrigan2021quantum}. Overall, mapping Kn to LNN requires an additional $O(n^2)$ SWAP operations and an additional $O(n)$ circuit depth overhead based on the parallel bubble sorting circuit. In the following, we will give proof of this conclusion independently based on  the parallel bubbling algorithm. 

To embed a fully connected graph Kn into LNN, additional SWAP operations are required to swap the positions of the qubits, i.e. to permute (or sort) the vertices of the graph into some specifical order. How to reduce the overhead of the SWAP gates is the crucial  issue. According to Ref.~\cite{hirata2009efficient}, the swap network of bubble sort is optimal to fulfill the corresponding permutation job, and the following lemma holds.
\begin{lemma}\label{permutation}
Let $x_1,x_2,...,x_n$ be the initial order of $n$ qubits under the LNN architecture. Consider a permutation to change the order into $x_{j_1},x_{j_2},...,x_{j_n}$, the least SWAP gates overhead is equivalent to the number of swap operations in the bubble sort.
\end{lemma}
Lemma \ref{permutation} indicates that the bubble sort algorithm is optimal for qubits order permutation in the case that only the nearest neighbor qubits are coupled. Therefore, the quantum SWAP circuit for exchanging qubits is equivalent to the classical swap circuit for bubble sort. Bubble sort is a  sort algorithm for a given dataset. The algorithm starts from the head of the dataset, compares the first two elements, and if the first element is greater than the second, swaps them. This process is performed for each pair of adjacent elements in the dataset until the tail of the dataset is reached. The whole process repeats until the last round with no swap happening. Since any two elements are compared only once during the process of bubble sort, the average and worst-case running time of bubble sort are both $O(n^2)$. 

Consider the worst case which would be useful for our analysis later. Let the initial order of $n$ qubits be $1, 2,...,n$, and now we want to sort them reversely as $n,n-1,...,1$. According to Lemma \ref{permutation}, the number of swap gates used in the bubble sort algorithm is the least for this purpose. According to the rules of bubble sort, any two qubits are swapped and only swapped once, and a total $n(n-1)/2$ swaps are required, which made an exact cover of the edges in the complete graph Kn. Therefore, the classical swap network of bubble sort that implements the reverse order permutation just corresponds to the embedding of Kn-type Hamiltonian to LNN.

The parallel bubble sort algorithm is a parallel version of the bubble sort algorithm which can reduce the running time to $O(n)$. The main idea of parallel bubble sort is to compare all adjacent pairs of input data at the same time and then iterate alternately between odd and even phases. A pseudocode description of this algorithm can be found in Algorithm~\ref{pbub}. Assuming the size of the input data  is $n$, the parallel bubble sort will perform $n$ iterations of the main loop. For each iteration, it is divided into odd and even phases according to the parity of the main loop. When $n$ is an odd number, both the odd and even phases perform $(n-1)/2$ compare-swap operations, which can be performed in parallel; When $n$ is even, the odd and even phases perform $n/2$  and $n/ 2-1$ compare-swap operations, respectively. The following lemma holds for the parallel bubble sort algorithm.

\begin{algorithm} 
    \caption{parallel bubble sort} \label{pbub}
    \SetAlgoNoLine
    \KwIn{ Data, which is a dataset with n elements }
    \KwOut{ Data, with reversed order }
    \For{i from 1 to $n$ }{
        Flag=mod(i,2),\quad \# a flag for odd and even phases.\\
        \For{j from 1 to $\lfloor{(n-1+\text{Flag})/2}\rfloor$}{
            \If {Data[2j-Flag]$<$ Data[2j+1-Flag]}{
                swap Data[2j-Flag] and Data[2j+1-Flag]
            }
        }
    }
\end{algorithm}  

\begin{figure*}
\centering
\includegraphics[width=0.95\textwidth]{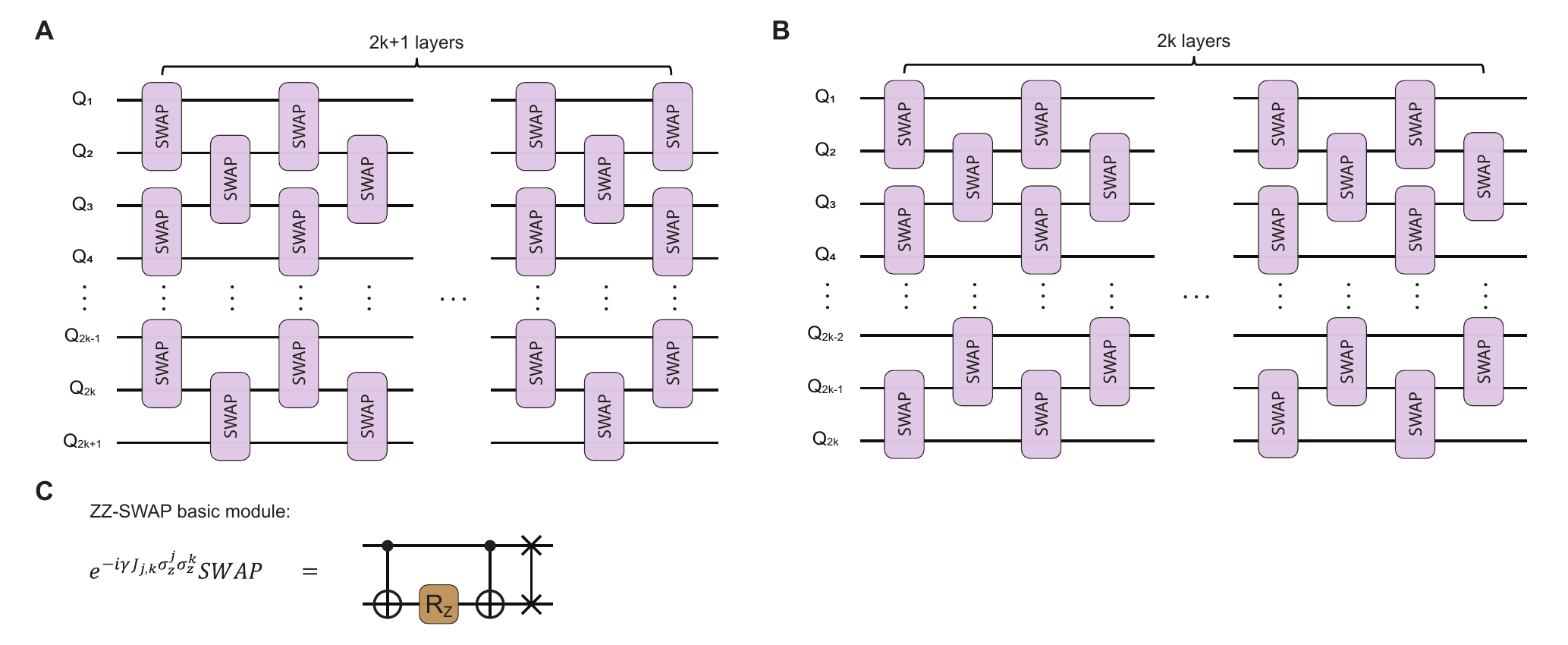}
\caption{ Circuit of parallel bubbling swap network implementing reverse order sorting. \textbf{A} for the odd qubits case and \textbf{B} for the even qubits case. Both cases have the same layers of outer loops to $n$, which leads to an $O(n)$ depth circuit. \textbf{C}. ZZ-SWAP basic module. It is a 4-depth quantum circuit constructed by a SWAP operation after a ZZ-basic module.}\label{Parallel} 
\end{figure*}

\begin{lemma}\label{pbubble} 
Let $n$ be the size of the input data and $ p\le n/2$ be the number of processors, then the time complexity of the algorithm is $O(n^2/2p)$. The  algorithm achieves the minimum when $p=\lfloor n/2\rfloor$ with at most $n$ iterations.
\end{lemma}
If the number of processors is less than $n/2$, a single processor will perform about $\lfloor n/2p\rfloor $  the compare-swap operations for the inner loop of each phase. In this case, the complexity of the algorithm is $O(n^2/2p)$. Here, the optimal number of processors is $\lfloor n/2\rfloor$. Each processor in the inner loop will perform at most one compare-swap operation. The outer loop of the algorithm can complete the entire bubble sort task with at most $n$ iterations. In quantum computing, it is often assumed that the quantum device has enough control processors to enable  two-qubit operations executed in parallel if they do not share the same qubit. Therefore, the following conclusion is established.
\begin{proposition}
The Kn-type circuit can be embedded in the LNN physical system without additional qubits, and the depth of the embedded quantum circuit is $O(n)$.
\end{proposition}
\begin{proof}
 Let the initial order of the $n$ qubits LNN system be $1,2,...,n$, and now it will be rearranged into $n,n-1,...,1$. The number of swaps required by the bubble sort algorithm is $n(n-1)/2$.
The swap network of bubble sort covers the edges of the complete graph Kn exactly, which made an embedding from Kn to LNN. This process can be fulfilled by the parallel bubbling circuit $\Gamma$ with $n/2$ processors. Then the circuit needs at most $n$ loops according to Lemma \ref{pbubble}, and each loop executes $\lfloor n/2\rfloor$  swap operations in parallel. The specific parallel swap network that implements this process can be found in Fig.~\ref{Parallel}. Hence the embedded circuit in the LNN system can be constructed by replacing the swap gate in $\Gamma$ with the ZZ-SWAP basic module (shown in Fig.~\ref{Parallel}\textbf{C}). Since the depth of the ZZ-SWAP basic module is 4, the depth of the embedded circuit is $4n$. The whole embedding process can be implemented without  additional qubits. This completes the proof.
\end{proof}
In addition, since the ZZ-basic module corresponding to each edge in Kn becomes the ZZ-SWAP basic module, an additional overhead including $n(n-1)/2$ SWAP operations and $n$-depth circuit are required after embedding. Meanwhile, the order of the qubits will be reversed after the execution of the circuit, and the qubit order will be restored by iterating the QAOA circuit again.

The lattice topology system is also referred as a 2-dimensional square lattice (2DSL), in which the qubits are arranged as a two-dimensional lattice, and only the adjacent interactions are allowed. Since a one-dimensional chain can be directly embedded into the 2-dimensional lattice  (find a one-dimensional path in 2DSL), the embedded circuit depth of the Kn type  Hamiltonian will not exceed the LNN case. Hence the following corollary is established.

\begin{corollary}
The embedded circuit depth for Kn-type Hamiltonian into a 2-dimensional square lattice (2DSL) system is $O(n)$  without additional qubits.
\end{corollary}
In summary, the complete graph topology is the ideal topology, and the circuit for the Kn type Hamiltonian  can be embeded optimally in depth $O(n)$ without any additional quantum resources. For the LNN topology,  additional $O(n)$ circuit depth is required. Since the embedded circuit scheme based on parallel bubble sorting shares the same depth overhead $O(n)$,  it is also optimal for both the LNN and 2DSL systems in the meaning of the '$O$' symbol. Moreover, Ref.~\cite{cheung2007translation}  provides a more efficient method to embed a Kn-type Hamiltonian into a 2DSL system  with $O(\sqrt n)$ additional circuit  depth, which takes advantage of the convenience of two-dimensional lattice structure.

\subsection{Resource estimation for RSA-2048}

Here we discuss the quantum resources needed to challenge real-life RSA numbers based on the results above. Since  no additional qubits in the process of circuit embedding consumed, the number of qubits required to factor an $m$ bit integer is $n=2m/\text{log}m$ (here we take the precision parameter $c=1$). The embedded circuit depths for the Kn-type Hamiltonian are $3n$ and $4n$ in the complete graph system and the LNN system, respectively. The circuit depth for a 2DSL system can be optimized to $3n+\sqrt n$, according to Donny Cheung et al.~\cite{cheung2007translation}.  The results are obtained without considering the native compilation of the ZZ-basic module (or ZZ-SWAP basic module) in a specific physical system. Taking RSA-2048 as an example, the number of qubits is about $2*2048/\text{log}2048\sim 372$, and the circuit depth of the single layer QAOA is $\sim 3+2=1118$ in completely connected systems, which includes 1-depth single qubit Rz operations and 1-depth single qubit Rx operations. It is $\sim 4n+2= 1490$ and $\sim 3n+\sqrt n+2=1139$ in a one-dimensional chain system and a 2-dimensional square lattice system, respectively. The quantum resources required for different lengths of RSA numbers are shown in Table~\ref{resource}.

\begin{table}[ht]
\begin{center}
\caption{ Quantum resource estimation for RSA numbers. The principal quantum resources mentioned are the number of qubits, and the quantum circuit depth of QAOA with one layer in three typical topologies, including an all-to-all connected system (Kn), 2d-lattice system (2DSL), and one-dimensional chain system (LNN). The results are obtained without considering the native compilation of the ZZ-basic module (or ZZ-SWAP basic module) in a specific physical system.}\label{resource} 
\begin{tabular}{@{}lcccc@{}}
\toprule
RSA number & Qubits  & Kn-depth & 2DSL-depth & LNN-depth \\
\hline
RSA-128 & 37 & 113 &121  & 150 \\
RSA-256 & 64 & 194 & 204 & 258 \\
RSA-512 & 114 & 344 & 357 & 458 \\
RSA-1024 & 205 & 617 & 633 & 822 \\
RSA-2048 & 372 & 1118 & 1139 & 1490 \\

\botrule
\end{tabular}
\end{center}
\end{table}

\begin{table}[ht]
\begin{center}
\caption{Touch-size of RSA numbers for some famous real quantum devices. The results are given according to the qubits connectivity and basic logical gate groups of the quantum devices by using our algorithm. The last column gives the basic depth condition needing to satisfy for the devices. The circuit depth in systems with ``others" topology type will be calculated according to the LNN model.}\label{device} 
\begin{tabular}{@{}cccccc@{}}
\toprule
system & devices  & qubits & topology & touch-size  & depth-least \\
\hline
\multirow{5}{*}{\makecell[c]{supercon\\-ducting \\qubits}}
& Sycamore        & 53     & 2DSL     & 201       & 170 \\
& Eagle      & 127    & others   & 581       & 510        \\
& Aspen-M & 80     & others   & 334       & 322        \\
& Zuchongzhi2    & 66     & 2DSL     & 264       & 210       \\
& Tianmu-1        & 36     & 2DSL     & 124       & 118        \\

\multirow{2}{*}{\makecell[c]{trapped \\ ions}}
& Maryland      & 40     & Kn       & 142      & 122        \\
& IonQ          & 79     & Kn       & 329      & 239        \\

\botrule
\end{tabular}
\end{center}
\end{table}
We have also analyzed the scale of RSA-numbers, namely the touch-size that existing quantum computing devices can reach under some ideal conditions, in which the claimed qubits are all relatively ideal or with high fidelity. The results are given according to the qubits connectivity of the quantum devices by using the SQIF algorithm. The quantum processors considered mainly including Sycamore, Eagle, Aspen-M, Zuchongzhi2, Tianmu-1, and trapped-ion devices from Maryland and Ion Q. All the devices are released publicly or could be visited through quantum cloud platforms. Meawhile, we give the ``depth-least" results for the devices to try the touch-size RSA numbers , which represent the necessary depth condition and estimated by the depth of single layer QAOA circuit. Specifically, if the quantum processor is a two-dimensional grid topology, the circuit depth is calculated according to the 2DSL model.  For the topology type of ``others",  the depth will be calculated according to the LNN model. The detailed  results are shown in Table \ref{device} .
 
We can find from Table \ref{device} that the touch-size of NISQ devices is close to the real life RSA numbers today. Such as the 127 qubits ibm-eagle machine, whose touch-size is 581 and the least circuit depth necessary is 510, which means if all the qubits work well and reaches a considerable fidelity after 510 circuit depth, it can be used to try factoring RSA-581. However, the touch-size is an ideal basic situation, the QAOA usually works more than one layer and deeper circuit required. Besides, the quantum speedup is unknown, it is still a long way to break RSA quantumly.



\end{document}